\documentclass[aps,pra,amsmath,amssymb,floatfix,twocolumn,amsmath,superscriptaddress,twocolumn,nofootinbib,tighten,letterpaper]{revtex4-1}
\usepackage{multirow}
\usepackage{bbold}
\usepackage{subfigure}
\usepackage{color}
\usepackage[table]{xcolor}
\usepackage{mathrsfs}
\usepackage{hyperref}
\usepackage{bm}
\usepackage{stackrel}

    \DeclareMathOperator{\sech}{sech}
		\renewcommand\vec[1]{\ensuremath\boldsymbol{#1}} 
\usepackage{amsfonts, relsize}
\usepackage{graphics}
\usepackage{graphicx}
\usepackage{amssymb}
\usepackage{amsbsy}
\usepackage{amsmath}
\usepackage{array}
\usepackage{fontenc}
\usepackage{textcomp}
\usepackage{rotating}
\usepackage{comment}

\usepackage{booktabs}

\definecolor{RowColor}{rgb}{0.88,1,0.9}

\begin{document}
\title{\bf Interacting spin-3/2 fermions in a Luttinger (semi)metal:\\ competing phases and their selection in the global phase diagram}

\author{Andr$\acute{\mbox{a}}$s L. Szab$\acute{\mbox{o}}$}
\affiliation{Max-Planck-Institut f\"{u}r Physik komplexer Systeme, N\"{o}thnitzer Str. 38, 01187 Dresden, Germany}

\author{Roderich Moessner}
\affiliation{Max-Planck-Institut f\"{u}r Physik komplexer Systeme, N\"{o}thnitzer Str. 38, 01187 Dresden, Germany}

\author{Bitan Roy}\thanks{Corresponding author: bitan.roy@lehigh.edu}
\affiliation{Max-Planck-Institut f\"{u}r Physik komplexer Systeme, N\"{o}thnitzer Str. 38, 01187 Dresden, Germany}
\affiliation{Department of Physics, Lehigh University, Bethlehem, Pennsylvania, 18015, USA}

\date{\today}
\begin{abstract}
We compute the effects of electronic interactions on gapless spin-3/2 excitations that in a noninteracting system emerge at a bi-quadratic touching of Kramers degenerate valence and conduction bands in three dimensions, also known as a Luttinger semimetal. This model can describe the low-energy physics of HgTe, gray-Sn, 227 pyrochlore iridates and half-Heuslers. For the sake of concreteness we only consider the short-range components of the Coulomb interaction (extended Hubbardlike model). By combining mean-field analysis with a renormalization group (RG) calculation (controlled by a ``small" parameter $\epsilon$, where $\epsilon=d-2$), we construct multiple cuts of the global phase diagram of interacting spin-3/2 fermions at zero and finite temperature and chemical doping. Such phase diagrams display a rich confluence of competing orders, among which rotational symmetry breaking nematic insulators and time reversal symmetry breaking magnetic orders (supporting Weyl quasiparticles) are the prominent candidates for excitonic phases. We also show that even repulsive interactions can be conducive to both mundane $s$-wave and topological $d$-wave pairings. The reconstructed band structure (within the mean-field approximation) inside the ordered phases allows us to organize them according to the energy (entropy) gain in the following (reverse) order: $s$-wave pairing, nematic phases, magnetic orders and $d$-wave pairings, at zero chemical doping. However, the paired states are energetically superior over the excitonic ones for finite doping. The phase diagrams obtained from the RG analysis show that for sufficiently strong interactions, an ordered phase with higher energy (entropy) gain is realized at low (high) temperature. In addition, we establish a ``selection rule" between the interaction channels and the resulting ordered phases, suggesting that repulsive short-range interactions in the magnetic (nematic) channels are conducive to the nucleation of $d$-wave ($s$-wave) pairing among spin-3/2 fermions. We believe that the proposed methodology can shed light on the global phase diagram of various two and three dimensional interacting multi-band systems, such as Dirac materials, doped topological insulators and the like.  
\end{abstract}

\maketitle

\section{Introduction}

The discovery of new phases of matter, and the study of the transitions between them, form the core of
condensed matter and materials physics~\cite{chaikin}. Much experimental ingenuity is devoted to 
realising and controlling different tuning parameters in the laboratory--such as temperature, pressure, magnetic field, 
chemical composition--to drive a system from one phase to the other~\cite{chaikin}. 
Most simply, the appearance of different phases can often be appreciated from the competition between energy and entropy. For example, with decreasing temperature transitions from water vapor to liquid to ice are intimately tied with the reduction of entropy or gain in energy. Similarly,  in a metal, entropy of gapless excitations on the Fermi surface is exchanged  for condensation energy as a superconducting gap opens~\cite{tinkham}.

\begin{figure}[t!]
\includegraphics[width=.96\linewidth]{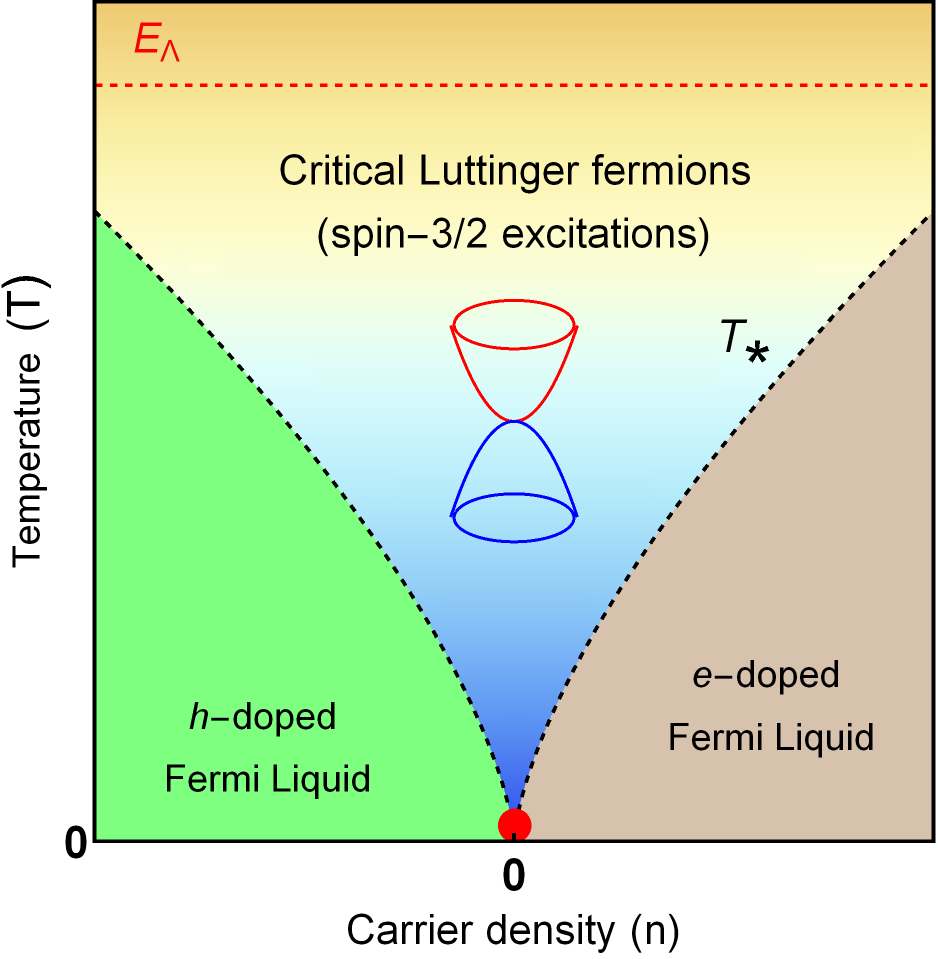}
\caption{A schematic phase diagram of noninteracting Luttinger fermions. The red dot represents the Luttinger semimetal fixed point separating electron- and hole-doped Fermi liquid of spin-3/2 fermions. The shaded sector corresponds to the quantum critical regime associated with the $z=2$ fixed point in $d=3$ (the red dot), controlling scaling properties in this regime (see Sec.~\ref{Sec:Intro_noninteracting}). Above a nonuniversal high energy cutoff $E_\Lambda$ (red dashed line), spin-3/2 fermions lose their jurisdiction. The crossover temperature $T_\ast$ (black dashed lines) above which critical Luttinger fermions are operative is given by $T_\ast \sim |n|^{z/d}$, with $z=2$ and $d=3$, where $n$ is the carrier density measured from the bi-quadratic band touching point. In this work we demonstrate the role of electron-electron interactions on the quantum critical regime (the shaded region), at finite temperature and chemical doping (see Sec.~\ref{Sec:Intro_Interacting} for summary of our results).    
}~\label{Fig:CriticalFan_Noninteracting}
\end{figure}

With increasing complexity of quantum materials attained in recent decades, the richness of the global phase diagram of strongly correlated materials has amplified enormously. And various prototypical representatives, such as cuprates, pnictides, heavy fermion compounds display concurrence of competing orders, among which spin- and charge-density-wave, superconductivity, nematicity are the most prominent ones. Besides establishing the existence  of -- and, hopefully, eventually utilizing in technological applications -- these phenomena, an obvious challenge is to discover any simplifying perspective, or at least heuristic classification scheme, for predicting and classifying them: {do there exist any organising principles among multiple competing orders that can shed light on the global phase diagram of strongly correlated materials?} Restricting ourselves to a specific but remarkably rich metallic system, we here give a partially affirmative answer to this question. We study a collection of strongly interacting spin-3/2 fermions in three dimensions that in the normal phase display a bi-quadratic touching of Kramers degenerate valence and conduction bands at an isolated point in the Brillouin zone, see Fig.~\ref{Fig:CriticalFan_Noninteracting}. This system is also known as Luttinger (semi)metal~\cite{luttinger, murakami-zhang-nagaosa}.

Such peculiar quasiparticle excitations can be found for example in HgTe~\cite{hgte}, gray-Sn~\cite{gray-sn-1, gray-sn-2}, 227 pyrochlore iridates (Ln$_2$Ir$_2$O$_7$, where Ln is the lanthanide element)~\cite{Savrasov, Balents1, Exp:Nakatsuji-1, Exp:Nakatsuji-2, Exp:armitage} and half-Heusler compounds (such as LnPtBi, LnPdBi)~\cite{Exp:cava, Exp:felser, binghai}. Bi-quadratic band touching has recently been observed in the normal state of Pr$_2$Ir$_2$O$_7$~\cite{Exp:Nakatsuji-1} and Nd$_2$Ir$_2$O$_7$~\cite{Exp:Nakatsuji-2} via angle resolved photo emission spectroscopy (ARPES). While most of the iridium-based oxides support all-in all-out arrangement of 4d Ir-electrons~\cite{takagi, krempa-1, krempa-2, nagaosa, Balents3, yamaji, troyer, tokura-1, tokura-2}, a singular member of this family, namely Pr$_2$Ir$_2$O$_7$, possibly resides at the brink of a metallic spin-ice or 3-in 1-out ordering and supports a large anomalous Hall conductivity~\cite{AHE-1, AHE-2, AHE-3, goswami-roy-dassarma, gang-chen}. While these materials harbor various competing magnetic orders (due to comparable Hubbard and spin-orbit interactions), smallness of the Fermi surface prohibits the onset of superconductivity at the lowest achievable temperature. On the other hand, half-Heusler compounds accommodate both anti-ferromagentism and unconventional superconductivity~\cite{Exp:Takabatake, Exp:Paglione-1a, Exp:Bay2012, Exp:visser-1, Exp:Taillefer, Exp:Zhang, Exp:visser-2, Exp:Paglione-1, Exp:Paglione-2, Exp:TbPdBi}. A transition between them can be triggered by tuning the de Gennes factor~\cite{Exp:Paglione-1}. Moreover, the superconducting YPtBi (with transition temperature $T_c=0.78$K) supports gapless BdG quasiparticles at low temperatures~\cite{Exp:Paglione-2}. Therefore, besides its genuine fundamental importance, our quest focuses on a timely issue due to growing material relevance of interacting spin-3/2 fermions, which has triggered a recent surge of theoretical works~\cite{krempa-1, krempa-2, nagaosa, Balents3, yamaji, troyer, goswami-roy-dassarma, gang-chen, balents-kim, Herbut-1, lai-roy-goswami, herbut-1a, Herbut-2, kharitonov, arago, BJ-Yang,  Fang2015, Yang16, brydon-1, brydon-2, Herbut-3, savary-1, roy-nevidomskyy, savary-2, sato,hosur, CXLiu-1, CXLiu-2, brydon-3, sunbinLee}.

The effects of electronic interactions on spin-3/2 fermions are addressed within the framework of an extended Hubbardlike model, composed of only the short-range components of repulsive Coulomb interactions.~\footnote{In this work we neglect the long-range tail of the Coulomb interaction. When the chemical potential is pinned at the band touching point, long-range Coulomb interaction can give rise to an infrared stable non-Fermi liquid fixed point~\cite{Abrikosov1, Abrikosov2, balents-kim}, which, may however be unstable toward the formation of an excitonic nematic phase~\cite{kohn, Herbut-1, herbut-1a}. By contrast, at finite temperature due to the thermal screening, yielding finite mass to photon, it becomes short-ranged in nature~\cite{kapusta}. Otherwise, purely long-range interaction in doped Luttinger systems yields $s$-wave and $p$-wave pairings~\cite{krempa-longrange-1, krempa-longrange-2}. However, these phases have not been observed in any Luttinger material so far.} Here we aim to construct a minimal interacting model that captures the competition among various experimentally observed ordered states, e.g. magnetic and superconducting, in different compounds, where the normal state quasiparticles are described by three-dimensional spin-3/2 Luttinger fermions at low energies. We show that an extended Hubbard model, composed of all symmetry allowed momentum-independent four-fermion interactions serves this purpose, at least qualitatively.

Due to the vanishing density of states (DoS) in a Luttinger system (namely, $\varrho(E) \sim \sqrt{E}$), sufficiently weak short-range interactions are irrelevant perturbations. Therefore, any ordering takes place at finite coupling. We here employ a renormalization group (RG) analysis to construct various cuts of the global phase diagram at zero as well as finite chemical doping [see Figs.~\ref{Fig:FiniteT_Summary}, ~\ref{Fig:swave_IsotropicLSM} and ~\ref{Fig:dwaves_IsotropicLSM}], and combine it with mean-field analysis to gain insight into the organizing principle among distinct broken-symmetry phases (BSPs)~\footnote{All cuts of the global phase diagram are obtained from a RG analysis, accounting for interaction effects on Luttinger fermions. Such an RG analysis can only predict the phase boundaries between the Luttinger (semi)metal and various BSPs in an unbiased fashion. However, our RG analysis is not applicable deep inside a BSP (depicted as gray shaded regions in Figs.~\ref{Fig:FiniteT_Summary}, ~\ref{Fig:swave_IsotropicLSM}, ~\ref{Fig:dwaves_IsotropicLSM}, ~\ref{Fig:GlobalPD_Intro_FS}, ~\ref{Fig:T1uMagnetg4smallalpha} and ~\ref{Fig:GlobalPD_Intro_noFS}) and cannot capture order-order transitions. When a cut of the phase diagram accommodates two competing phases, as in Fig.~\ref{Fig:FiniteT_Summary}, ~\ref{Fig:swave_IsotropicLSM}, ~\ref{Fig:dwaves_IsotropicLSM}, ~\ref{Fig:GlobalPD_Intro_FS}, and ~\ref{Fig:GlobalPD_Intro_noFS}, inside the ordered phase typically there exists a regime of coexistence (as the corresponding order-parameters never fully commute with each other, see Sec.~IIB3 and Sec.~\ref{SubSec:SelectionRule}), with two pure phases on either side of it. Such coexistence can be captured by standard mean-field analysis, which goes beyond the scope of the present discussion.}. A gist of our findings can be summarized as follows.

\begin{figure}[t!]
\includegraphics[width=.95\linewidth]{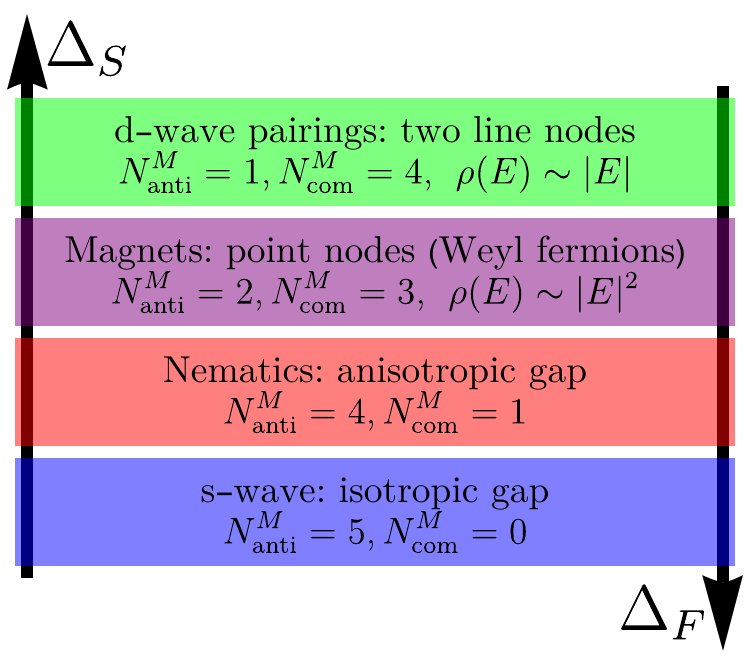}
\caption{ Hierarchy of various dominant orders in an interacting Luttinger semimetal according to the gain of the condensation energy ($\Delta_F$) and entropy ($\Delta_S$) inside the ordered phases. The condensation energy (entropy) gain increases in the direction of the $\Delta_F (\Delta_S)$ arrow. If an order-parameter matrix ($M$) anti-commutes (commutes) with $N^{M}_{\rm anti}(N^{M}_{\rm com})$ number of  matrices appearing in the Luttinger model [see Eq.~(\ref{Eq:Luttinger_Hamiltonian_Intro})], then $\Delta_F \sim N^{M}_{\rm anti}$ and $\Delta_S \sim N^{M}_{\rm com}$ (qualitatively), see Sec.~\ref{SubSec:energy_Entropy} for detailed discussion. At finite chemical doping superconducting orders are always energetically superior over the excitonic ones. In the global phase diagram, phases with higher gain in condensation energy (entropy) appear at low (high) temperature, see for example Figs.~\ref{Fig:FiniteT_Summary}, ~\ref{Fig:swave_IsotropicLSM} and ~\ref{Fig:dwaves_IsotropicLSM}. We also display the scaling of density of states [$\varrho(E)$] at low-energies in the presence of both point and line nodes. Notice that with increasing $N^{M}_{\rm anti}$ ($N^{M}_{\rm com}$) the stiffness of the spectral gap (amount of gapless quasiparticles), determining the scaling of the DoS at low-energy, increases.  
}~\label{Fig:energyentropy}
\end{figure}

1. By computing the reconstructed band structure (within the mean-field approximation) we organize dominant BSPs according to the condensation energy and entropy gains. Results are summarized in Fig.~\ref{Fig:energyentropy}. We note that while the stiffness (uniform or anisotropic) of the band gap measures the condensation energy gain, the amount of gapless quasiparticles (resulting in power-law scaling of DoS at low energies) measures the entropy inside the ordered phase. All cuts of the global phase diagram (obtained from a RG analysis) show that the low (high) temperature phase yields larger condensation energy (entropy) gain, see Fig.~\ref{Fig:FiniteT_Summary}.

2. The quasiparticle spectra inside the $s$-wave paired state and two nematic phases (belonging to the $T_{2g}$ and $E_g$ representations of the cubic or $O_h$ point group) are fully gapped. Hence, nucleation of these three phases leads to the maximal gain of condensation energy, and  they appear as the dominant BSPs at zero temperature, as shown in Fig.~\ref{Fig:FiniteT_Summary}.

3. At finite temperature condensation energy gain competes with the entropy, and phases with higher entropy are realized at higher temperatures. Onset of any nematicity results in an anisotropic gap, in contrast to the situation with an $s$-wave pairing. Thus former orderings are endowed with larger (smaller) entropy (condensation energy), and are found at higher temperatures in Figs.~\ref{Subfig:FiniteT_g1}, \ref{Subfig:FiniteT_g2}. By contrast, the dominant magnetic orders (belonging to the $A_{2u}$ and $T_{1u}$ representations) produce gapless Weyl quasiparticles and result in $\varrho(E) \sim |E|^2$ scaling of the DoS at low energies. Hence, these two magnetic orders carry larger entropy than the nematic phases or the $s$-wave pairing, and can only be found at finite temperature, see Figs.~\ref{Subfig:FiniteT_g3}, \ref{Subfig:FiniteT_g4}. Luttinger semimetal (LSM)-$A_{2u}$ magnetic order (results from the all-in all-out state in pyrochlore lattice~\cite{Savrasov, Balents3}) transition at finite temperature is consistent with the experimental observation in Nd$_2$Ir$_2$O$_7$~\cite{takagi}, while the $T_{1u}$ magnetic ordering (yielding a 3-in 1-out ordering in pyrochlore lattice and supporting anomalous Hall effect~\cite{goswami-roy-dassarma}) can be germane for Pr$_2$Ir$_2$O$_7$~\cite{AHE-1,AHE-2,AHE-3}~\footnote{ARPES measurements in Nd$_2$Ir$_2$O$_7$~\cite{Exp:Nakatsuji-2} and Pr$_2$Ir$_2$O$_7$~\cite{Exp:Nakatsuji-1} suggest that the LSM in these compound is isotropic.}.

4. Local four fermion interactions in the nematic channels are conducive for $s$-wave pairing (at zero and finite chemical doping), see Fig.~\ref{Subfig:FiniteT_g1}, ~\ref{Subfig:FiniteT_g2} and \ref{Fig:swave_IsotropicLSM} , whereas short-range magnetic interactions give birth to topological $d$-wave pairing (only at finite chemical doping), see Fig.~\ref{Fig:dwaves_IsotropicLSM}. We further elaborate such an emergent ``selection rule" in Sec.~\ref{Sec:Intro_Interacting}. Confluence of magnetic order and $d$-wave pairing (resulting in gapless BdG quasiparticles, found in YPtBi~\cite{Exp:Paglione-2}) is in (qualitative) agreement with the global phase diagram of LnPdBi~\cite{Exp:Paglione-1}.

The theoretical approach outlined in this work is quite general and can be extended to address the effects of electronic interactions in various strongly correlated multi-band systems, among which two-dimensional Dirac and quadratic fermions (respectively relevant for monolayer and bilayer graphene)~\cite{graphene-RMP, Balatsky}, three-dimensional doped topological, crystalline and Kondo insulators~\cite{TI-RMP-1, TI-RMP-2, TKI-Review}, Weyl materials~\cite{Weyl-RMP}, twisted bilayer graphene~\cite{tblg-1, tblg-2, tblg-3}, are the most prominent and experimentally pertinent ones. The proposed organization principle (Sec.~IIB1) and the selection rule (Sec.~IIB3) among competing orders are consistent with recently reported confluence between (1) charge-density-wave and $f$-wave superconductor~\cite{scherrerCDWfwave} and (2) quantum spin Hall insulator and $s$-wave superconductor~\cite{assaadQSHIswave} in doped monolayer graphene, respectively demonstrated from non-perturbative functional RG and quantum Monte Carlo simulations. We discuss these cases in Appendix~\ref{appendix:Dirac}. In the future we will systematically study these systems in details. Even though competing orders in correlated insulators and standard Fermi liquids (such as the one realized in a square lattice system) have been investigated in terms of the intertwinded orders~\cite{fradkin-kivelson-tranquada, fernandes-orth-schmalian}, the organizing principle [see Sec.~IIB1] and selection rules [see Sec.~IIB3] for gapless fermionic topological multi-band systems with or without strong spin-orbit coupling remains vastly unexplored so far.

\subsection{Outline}

The rest of the paper is organized as follows. In the next section we present an extended summary of our main results. The low-energy description of the Luttinger model and its symmetry properties are discussed in Sec.~\ref{Sec:Luttinger_Model}. In Sec.~\ref{Sec:BSPs} we discuss the reconstructed band structure inside various excitonic and superconducting phases. In Sec.~\ref{Sec:e-e_Interaction} we introduce the interacting model for spin-3/2 fermions and analyze the propensity toward the formation of various orderings within a mean-field approximation. Sec.~\ref{Sec:RG} is devoted to a renormalization group (RG) analysis of interacting Luttinger fermions at zero and finite temperatures and chemical doping. We summarize the main results and highlight some future outlooks in Sec.~\ref{Sec:Conclusion}. Additional technical details are relegated to appendices.

\section{Extended Summary}~\label{Sec:ExtendedSummary}

Our starting point is a collection of spin-3/2 fermions for which the normal state is described by a bi-quadractic touching of Kramers degenerate valence and conduction bands. The corresponding Hamiltonian operator is~~\cite{luttinger, murakami-zhang-nagaosa} 
\begin{equation}~\label{Eq:Luttinger_Hamiltonian_Intro}
\hat{h}_{\rm L} ({\bf k})=-\frac{k^2}{2m} \left[ \cos\alpha \sum^{3}_{j=1} \hat{d}_j \Gamma_j +\sin\alpha \sum^{5}_{j=4} \hat{d}_j \Gamma_j\right] -\mu,
\end{equation}
where $\hat{d}_j$s are five $d$-wave harmonics in three dimensions, $\Gamma_j$s are five mutually anticommuting four dimensional Hermitian matrices, and $m/[\cos \alpha \; (\sin \alpha)]$ is the effective mass for gapless excitations in the $T_{2g}(E_g)$ orbitals in a cubic environment. Momentum $k$ and chemical potential $\mu$ are measured from the band touching point. Additional details of this model are discussed in Sec.~\ref{Sec:Luttinger_Model} and Appendix~\ref{Append:Luttingerdetails}. The mass anisotropy parameter ($\alpha$) lies within the range $0 \leq \alpha \leq \frac{\pi}{2}$~\cite{goswami-roy-dassarma}. But, for the sake of concreteness we restrict our focus on the isotropic system with $\alpha=\frac{\pi}{4}$. For discussion on the evolution of phase diagrams with varying $\alpha$, see Secs.~\ref{SubSec:susceptibility} and ~\ref{Sec:RG}, and Figs.~\ref{Fig:GlobalPD_Intro_FS} and ~\ref{Fig:GlobalPD_Intro_noFS}. The LSM is realized as an \emph{unstable} fixed point at $\mu=0$, the red dot in Fig.~\ref{Fig:CriticalFan_Noninteracting}. This fixed point is characterized by the dynamic scaling exponent $z=2$, determining the relative scaling between energy and momentum according to $E \sim |{\bf k}|^z$. The chemical potential is a \emph{relevant} perturbation at this fixed point, with the scaling dimension $[\mu]=2$. Hence, the correlation length exponent at this fixed point is $\nu=1/2$. Therefore, the LSM can be envisioned as a quantum critical point (QCP) separating electron-doped (for $\mu>0$, the brown region) and hole-doped (for $\mu<0$, the green region) Fermi liquid phases, as shown in Fig.~\ref{Fig:CriticalFan_Noninteracting}. Our discussion is focused on the quantum critical regime (the shaded region).

The crossover temperature ($T_\ast$) separating the quantum critical regime accommodating gapless spin-3/2 excitations from the Fermi liquid phases can be estimated in the following way 
\begin{equation}
T_\ast \sim \frac{\hbar^2}{2m} \times \frac{1}{\xi^2} \sim \frac{\hbar^2}{2m} \: |n|^{2/3}, 
\end{equation}  
where $\xi$ is a characteristic length scale and $|n|\sim |\mu|/\xi^d$ is the carrier density. Two critical exponents ($z$ and $\nu$) and the dimensionality of the system ($d=3$) determine the scaling of various thermodynamic (such as specific heat, compressibility) and transport (such as dynamic conductivity) quantities in this regime.

\begin{figure*}
\subfigure[]{
\includegraphics[width=0.40\linewidth]{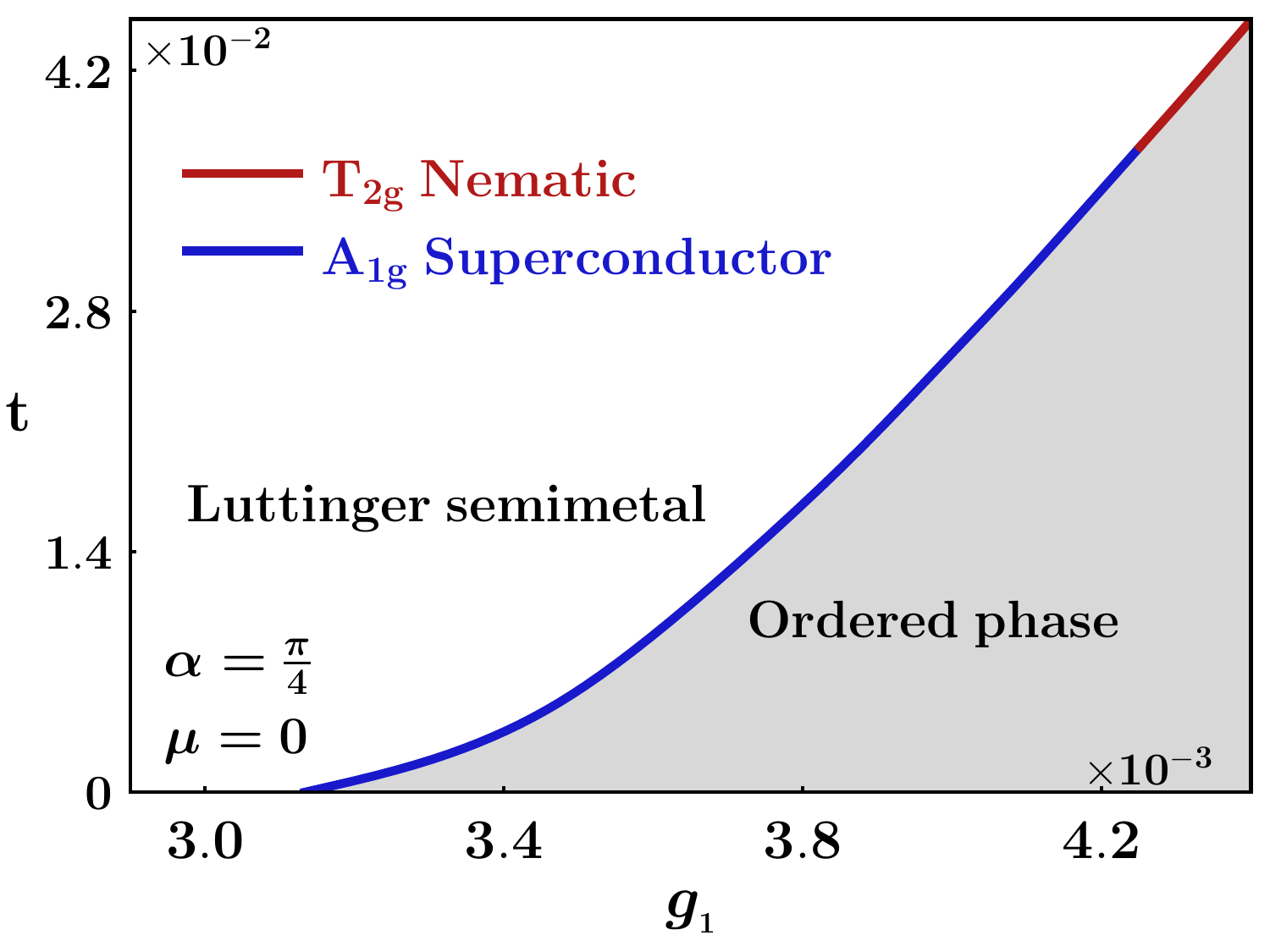}~\label{Subfig:FiniteT_g1}
}\hspace{0.5cm}
\subfigure[]{
\includegraphics[width=0.40\linewidth]{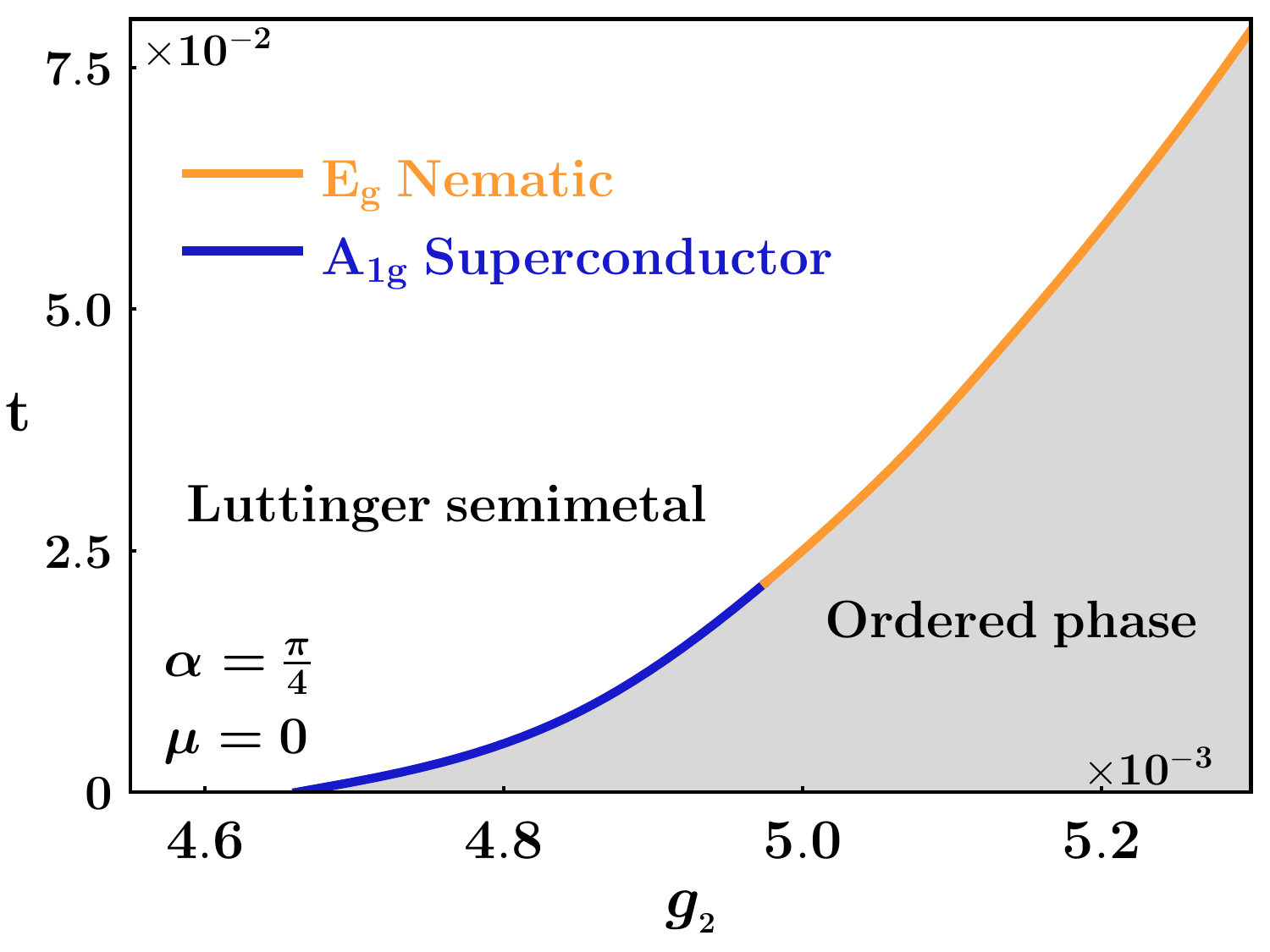}~\label{Subfig:FiniteT_g2}
}
\subfigure[]{
\includegraphics[width=0.40\linewidth]{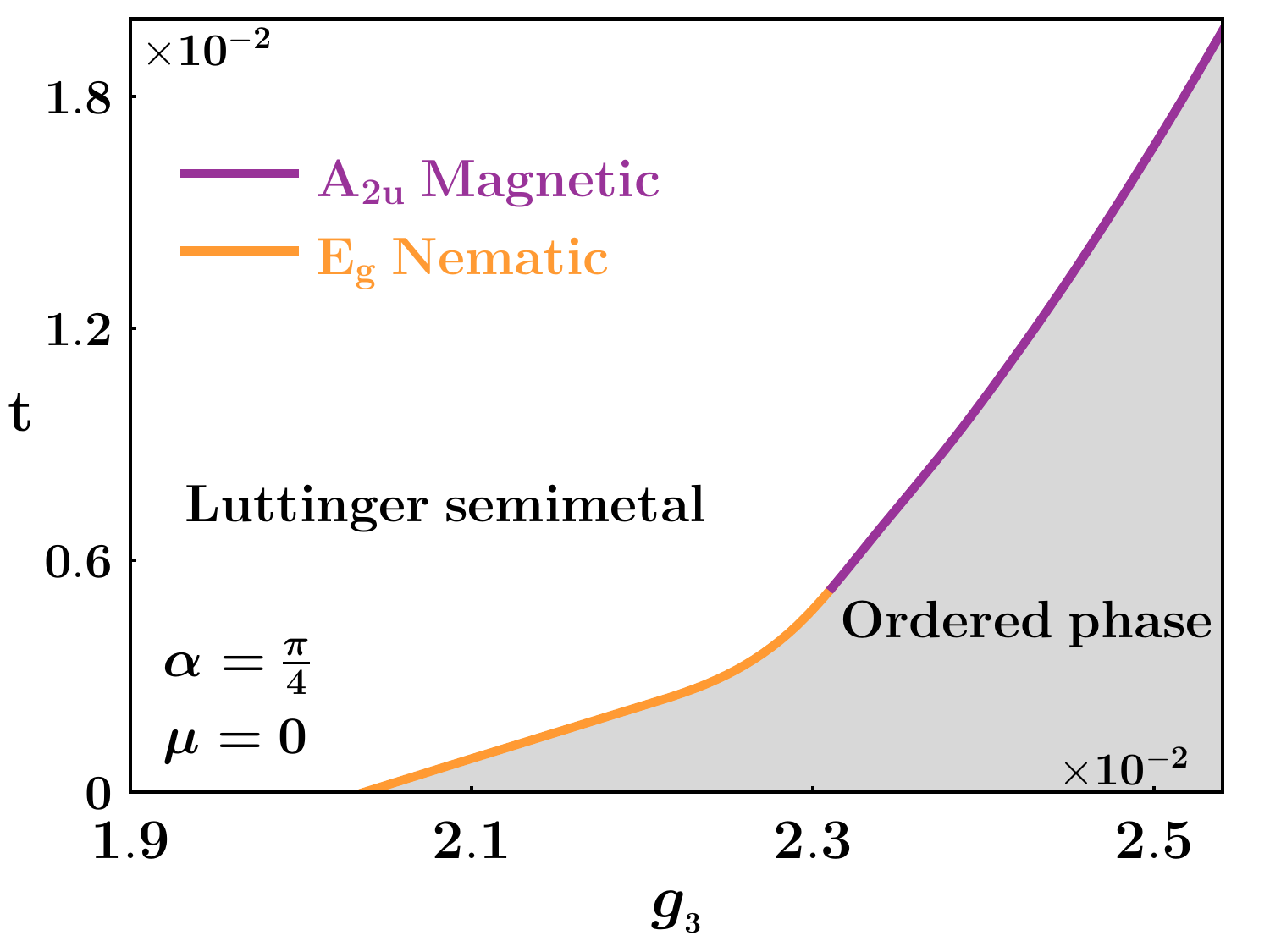}~\label{Subfig:FiniteT_g3}
}\hspace{0.5cm}
\subfigure[]{
\includegraphics[width=0.40\linewidth]{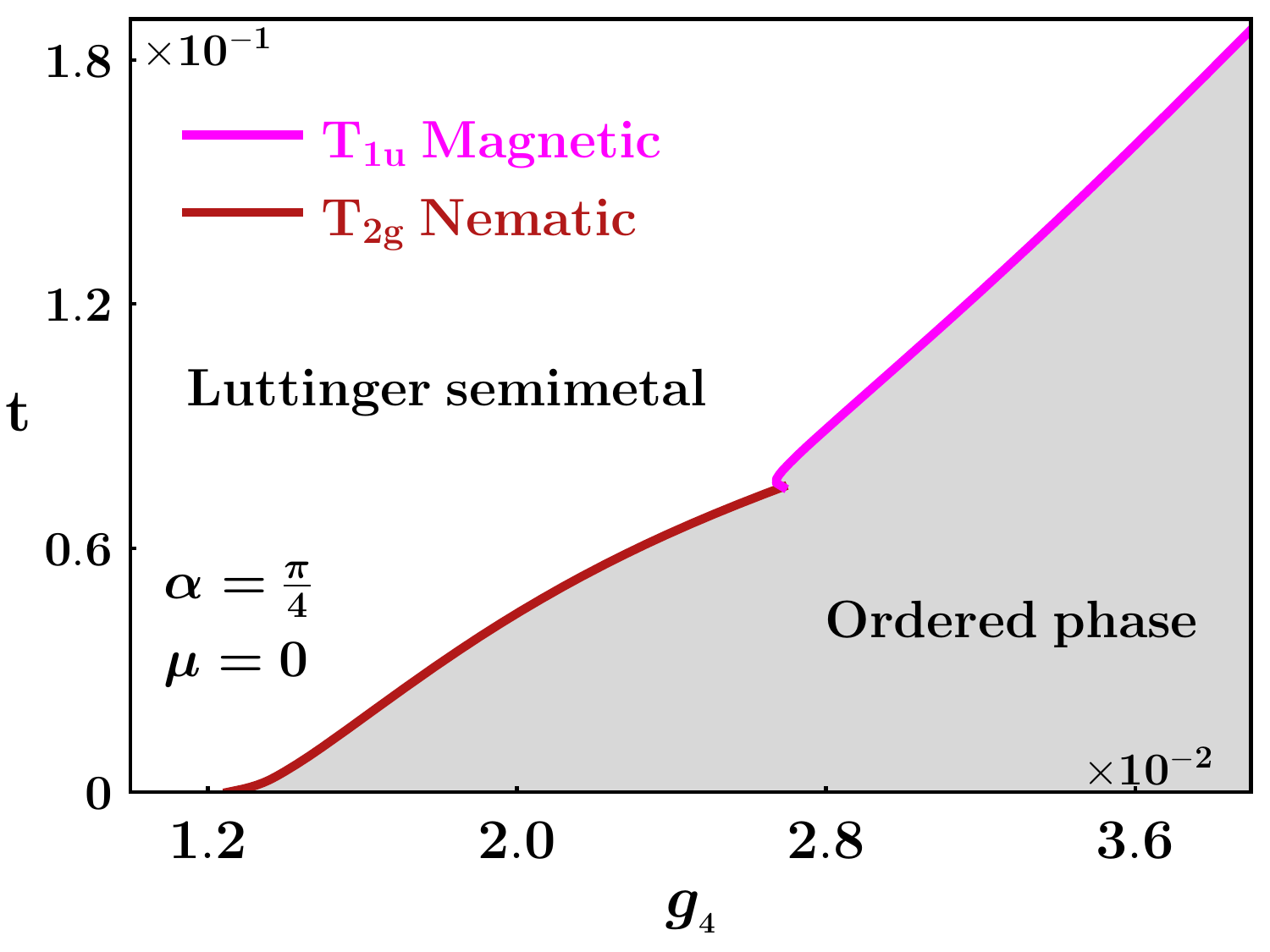}~\label{Subfig:FiniteT_g4}
}
\caption{ Various cuts of the finite temperature ($t$) phase diagram of an interacting isotropic Luttinger semimetal. The interaction couplings ($g_{_i}$) are measured in units of $\epsilon$ (also in Figs.~\ref{Fig:swave_IsotropicLSM}, ~\ref{Fig:dwaves_IsotropicLSM}, ~\ref{Fig:GlobalPD_Intro_FS}, ~\ref{Fig:T1uMagnetg4smallalpha} and ~\ref{Fig:GlobalPD_Intro_noFS}). But nature of the order states in all cuts of the phase diagram do not depend on the value of $\epsilon>0$. Panels (a) and (b) respectively depict onset of $T_{2g}$ and $E_g$ nematic orders at higher temperatures as we tune interactions in these two channels. Recall while the $s$-wave superconductor yields fully gapped spectra, both nematic phases produce anisotropic gaps [see Sec.~\ref{SubSec:band-topology}]. Hence, the energy-entropy competition [see Sec.~\ref{SubSec:energy_Entropy}] favors $s$-wave pairing (nematic phases) at low (high) temperature. When we tune the strength of (c) $A_{2u}$ and (d) $T_{1u}$ magnetic interactions, $E_g$ and $T_{2g}$ nematic orders set in at low temperature, respectively, and corresponding magnetic orders nucleate only at higher temperature, since both magnetic orders produce gapless Weyl fermions (less gain of condensation energy, but higher entropy). See Sec.~\ref{SubSec:FiniteTRG} for details of the renormalization group analysis at finite $t$. The white regions represent Luttinger semimetal without any ordering. The gray shaded region to the ordered phase (see footnote~2), and its boundaries with Luttinger semimetal, occupied by various ordered phases are shown in different colors..       
}~\label{Fig:FiniteT_Summary}
\end{figure*}

\subsection{Critical scaling in noninteracting system}~\label{Sec:Intro_noninteracting}

The free-energy density (up to an unimportant temperature ($T$) independent constant) inside the critical regime is given by (setting $k_B=1$) 
\begin{equation}~\label{Eq:freeenergy_Intro}
f= T^{5/2} \; \frac{(2 m)^{\frac{3}{2}}}{4 \pi^{\frac{3}{2}}} \left[ {\rm Li}_{\frac{5}{2}} \left(- e^{\mu/T} \right) + {\rm Li}_{\frac{5}{2}} \left(- e^{-\mu/T} \right) \right],  
\end{equation}  
where ${\rm Li}$ represents the polylogarithimic function. The specific heat of this system is given by 
\begin{equation}~\label{Eq:specificheat_Intro}
C_V= -T \frac{\partial^2 f}{\partial T^2} \approx T^{3/2} \: \frac{(2 m)^{\frac{3}{2}}}{32 \pi^{\frac{3}{2}}} \left[ 15 a -b \; \frac{\mu^2}{T^2}\right],
\end{equation}
for $\mu/T \ll 1$ (ensuring that the system resides inside the critical regime), where $a \approx 1.7244$ and $b \approx 0.6049$. From the above expression of the free-energy density we can also extract the scaling of compressibility ($\kappa$), given by 
\begin{equation}~\label{Eq:compressibility_Intro}
\kappa= - \frac{\partial^2 f}{\partial \mu^2} \approx \sqrt{T} \: \frac{(2 m)^{\frac{3}{2}}}{2 \pi^{\frac{3}{2}}} \left[ b + 6 c \; \frac{\mu^2}{T^2} \right],
\end{equation}
where $c \approx 0.00989$. Therefore, the presence of finite chemical doping does not alter the leading power-law scaling of physical observables, such as $C_V \sim T^{3/2}$, $\kappa \sim \sqrt{T}$, but only provides \emph{subleading} corrections, which are suppressed by a parametrically small quantity $\mu/T$.~\footnote{The scaling of specific heat and compressibility is determined by the dimensionality ($d$) and dynamic scaling exponent ($z$) according to $C_V \sim T^{1+d/z}$ and $\kappa \sim T^{d/z}$, respectively.} Also note that 
\begin{equation}~\label{Eq:universalratio_Intro}
\frac{C_V/T}{\kappa} \approx 5.37611
\end{equation} 
is a universal ratio, capturing the signature of a $z=2$ quantum critical point in $d=3$. A detailed derivation of this analysis is shown in Appendix~\ref{Append:freeenergy}. Qualitatively similar sub-leading corrections are also found in the scaling of the dynamic conductivity, which we discuss now.

Gauge invariance mandates that the conductivity ($\sigma$) must scale as $\sigma \sim \xi^{2-d}$. Hence for a collection of $z=2$ quasiparticles (such as the Luttinger fermions), $\sigma \sim \sqrt{T}$ or $\sqrt{\omega}$ in three spatial dimensions. Indeed we find that the Drude (Dr) component of the dynamic conductivity in the Luttinger system is given by [see Appendix~\ref{Append:dynamic conductivity}] 
\begin{equation}~\label{Eq:Drude}
\sigma_{\rm Dr} (\omega,T) = e^2 \; \delta\left( \frac{\omega}{T}\right) \sqrt{m \; T} \:\: F_{\rm Dr} \left( \frac{\mu}{T}\right),
\end{equation}
where $F_{Dr} (x)$ is a monotonically increasing universal function of its argument [see Eq.~(\ref{Eq:Drude_Function}) and Fig.~\ref{Fig:Drude_LSM}] and $\omega$ is the frequency. On the other hand, the inter-band (IB) component of the optical conductivity reads as 
\begin{equation}~\label{Eq:Inter-band}
\sigma_{\rm IB}(\omega,T)= e^2 \sqrt{m \; \omega} \:\: \sum_{\tau=\pm} \; \tanh \left( \frac{\omega + 2 \tau \mu}{4 T} \right).
\end{equation} 
Hence, inter-band component of the optical conductivity vanishes as $\sqrt{\omega}$ as $\omega \to 0$ and the LSM can be identified as a \emph{power-law insulator}.

Therefore, even when the chemical doping is finite there exists a wide quantum critical regime, shown in Fig.~\ref{Fig:CriticalFan_Noninteracting}, where the scaling of thermodynamic and transport quantities are essentially governed by $z=2$ quasiparticles, and the chemical potential provides only sub-leading corrections. Next we highlight the imprint of finite temperature and chemical doping on the global phase diagram of interacting spin-3/2 fermions.

\subsection{Electron-electron interactions in a Luttinger (semi)metal}~\label{Sec:Intro_Interacting}

In this work we compute the effects of electron-electron interactions on Luttinger fermions, occupying the critical regime of the noninteracting fixed point, see Fig.~\ref{Fig:CriticalFan_Noninteracting}. In this regime any short-range or local four-fermion interaction ($\lambda$) is an \emph{irrelevant} perturbation, since 
\begin{equation}
[\lambda]=z-d=2-3=-1, \nonumber 
\end{equation}
due to the vanishing DoS, namely $\varrho(E) \sim \sqrt{E}$. We use a RG analysis, tailored to address the effects of electronic interactions on Luttinger fermions in $d=3$, constituting a $z=2$ band structure, to arrive at various cuts of the global phase diagram of this system. If, on the other hand, temperature and chemical doping are such that the system resides inside a Fermi liquid phase, the notion of $z=2$ nodal quasiparticles becomes moot and our RG analysis loses its jurisdiction.~\footnote{ Note that in the presence of a Fermi surface the DoS is constant and the interaction coupling $\lambda$ is dimensionless. Consequently a Fermi liquid becomes unstable toward the formation of a superconductor (often non-$s$-wave) even in the presence of repulsive electron-electron interactions, following the spirit of the Kohn-Luttinger mechanism~\cite{Sankar-RG-RMP, kohn-Luttinger, KL-Chubukov-Review, zanchi-schulz, hlubina, raghu-kivelson} and the superconducting transition temperature ($T_c$) mimics the BCS-scaling law $T_c \sim \exp[-1/\lambda]$.} Furthermore, we augment the RG analysis with an organizing principle based on the competition between energy and entropy. To this end we rely on the computation of the reconstructed band structure inside the ordered phases within the mean-field approximation. Subsequently, we also promote a ``selection rule" among neighboring phases in the global phase diagram, originating purely from their algebraic or symmetry properties. For the sake of simplicity we concentrate on the isotropic system ($\alpha=\frac{\pi}{4}$) in the following three subsections. Nonetheless, our results hold (at least qualitatively) for any arbitrary value of $\alpha$, as summarized in the last subsection.

\begin{center}
{\bf IIB1. Organizing principle: Emergent Topology \& Energy-Entropy}
\end{center}

Let us first promote an organizing principle among BSPs according to their contribution to the energy and entropy gain and anticipate their presence in the global phase diagram. In what follows we highlight the reconstructed band structure inside the dominant ordered phases within a mean-field or Hartree-Fock approximation, which by construction undermines the ordered parameter fluctuations. The emergent band topology is computed by diagonalizing an effective single-particle Hamiltonian, composed of the noninteracting Luttinger Hamiltonian and corresponding order parameter, see Sec.~\ref{SubSec:band-topology} for details.

Perhaps it is natural to anticipate that at zero temperature strong electronic interactions favor the phases that produce the largest spectral gap, as the onset of these ordered states offers maximal gain of condensation energy. In a LSM there are three candidate BSPs that yield fully gapped quasiparticle spectra: (a) an $s$-wave superconductor, producing a uniform and isotropic gap, and (b) two nematic orders (belonging to the $T_{2g}$ and $E_g$ representations), producing anisotropic gaps. As shown in Fig.~\ref{Fig:FiniteT_Summary}, only these three phases can be found in an isotropic and interacting LSM at zero temperature.

The energy-entropy competition, leading to an organizing principle among competing phases at finite temperature, can be appreciated from the scaling of the DoS or the stiffness (uniform or anisotropic) of the spectral gap. As mentioned above, the $s$-wave pairing and nematic orders respectively produce uniform and anisotropic gaps, whereas two magnetic orders, belonging to the $A_{2u}$ and $T_{1u}$ representations, respectively produce eight~\cite{Balents3} and two~\cite{goswami-roy-dassarma} isolated simple Weyl nodes, around which the DoS vanishes as $\varrho(E) \sim |E|^2$ for sufficienlty low energies. On the other hand, each copy of the $d$-wave pairings accommodates two nodal loops for which the low-energy DoS scales as $\varrho(E) \sim |E|$ [see Sec.~\ref{SubSec:energy_Entropy}]~\cite{brydon-1, Herbut-3, roy-nevidomskyy, savary-2, sunbinLee}. Since we are interested in energy or temperature scales much smaller than the ultraviolet cutoff or bandwidth ($|E| \ll 1$), the structure of the spectral gap (isotropic or anisotropic) and power-law scaling of DoS carry sufficient information to organize the ordered phases according to their contribution to condensation energy and entropy, summarized in Fig.~\ref{Fig:energyentropy}. In brief, existence of more gapless points (resulting in higher DoS near $E=0$) yields larger entropy, while a more uniform gap leads to higher gain in condensation energy. From various cuts of the global phase diagram at finite temperature, see Fig.~\ref{Fig:FiniteT_Summary}, we note the following common feature: \emph{Among competing orders, the one with maximal gain in condensation energy appears at low temperature, while the phase with higher entropy is realized at higher temperature}, in accordance with the general principle of energy-entropy competition. Since the DoS in a LSM scales as $\varrho(E) \sim \sqrt{E}$ (maximal entropy), it can always be found at sufficiently high (weak) enough temperature (interactions).

\begin{figure}[t!]
\includegraphics[width=4.25cm,height=3.75cm]{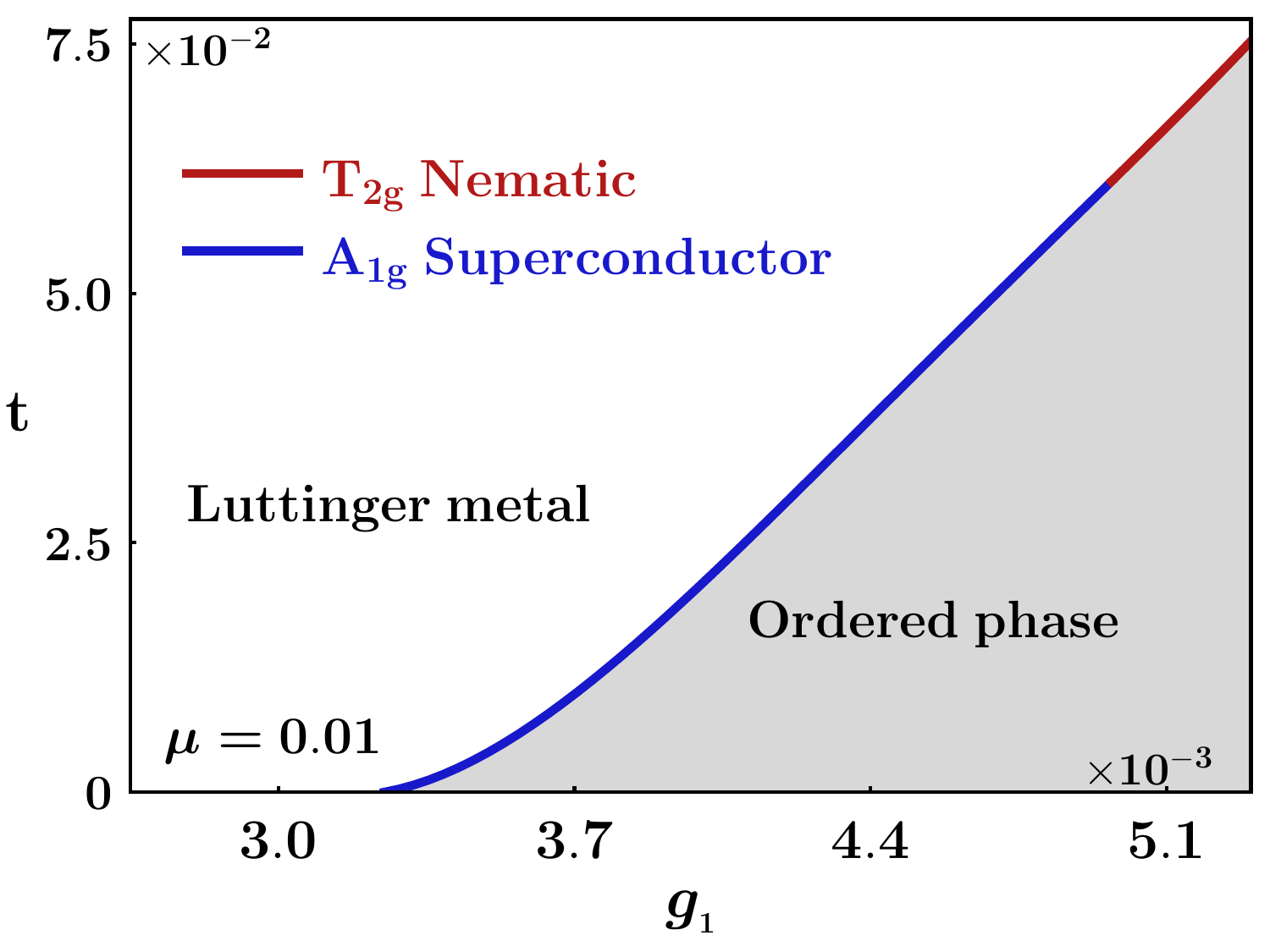}%
\includegraphics[width=4.25cm,height=3.75cm]{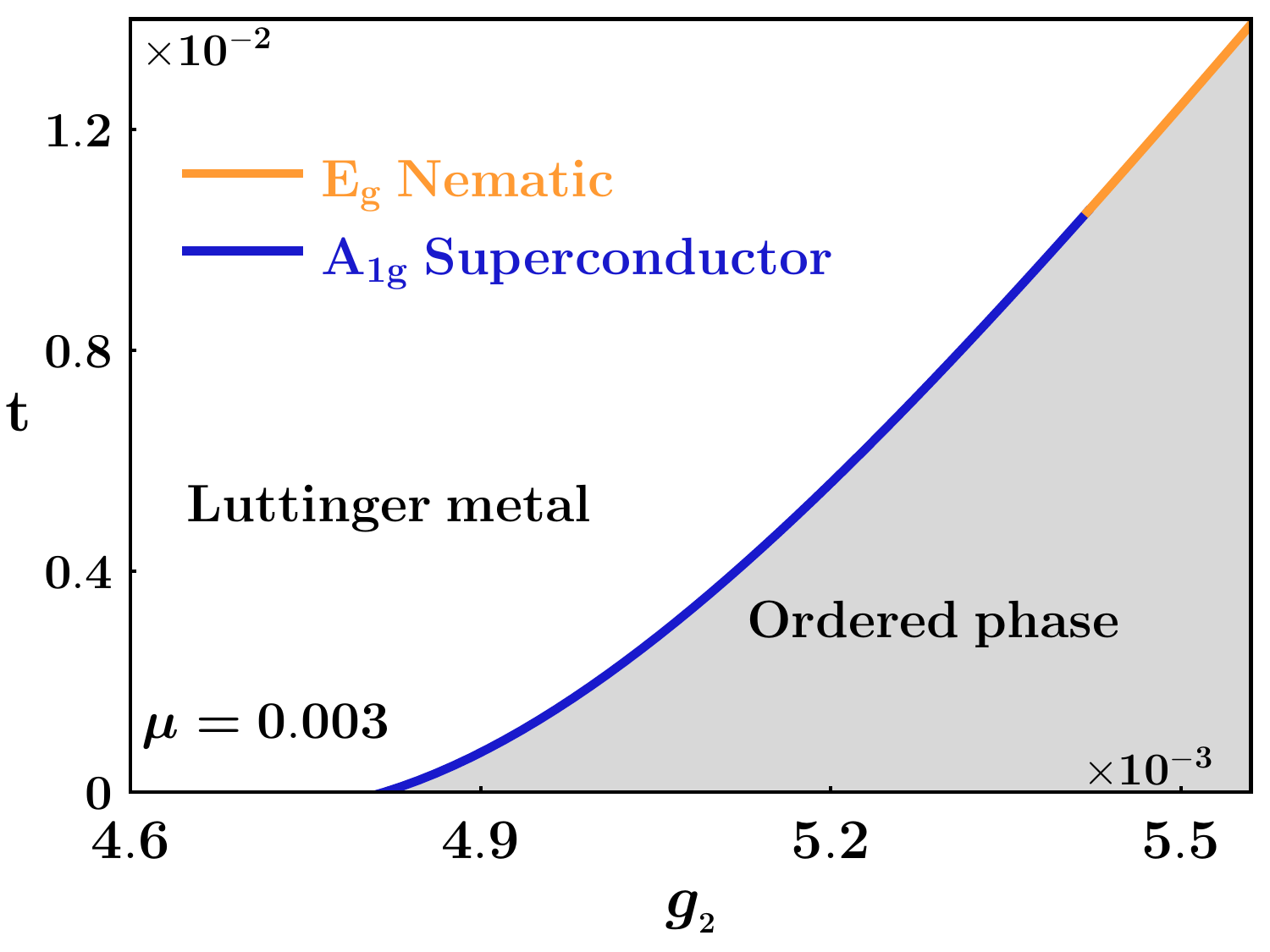}
\caption{ Realization of $s$-wave pairing from repulsive electronic interactions in the (a) $T_{2g}$ and (b) $E_g$ nematic channels, for finite $\mu$. In the presence of chemical doping the $s$-wave pairing occupies a larger portion of the phase diagrams and the threshold couplings for the onset of two nematic orders get pushed toward stronger couplings [compare with Figs.~\ref{Subfig:FiniteT_g1} and \ref{Subfig:FiniteT_g2}]. Hence, nematic interactions favor the onset of $s$-wave pairing in an isotropic Luttinger system ($\alpha=\frac{\pi}{4}$). Here coupling constants are measured in units of $\epsilon$. The gray shaded regions represent ordered states (see footnote~2), and its boundaries with Luttinger metal, occupied by various ordered phases are shown in different colors.
}~\label{Fig:swave_IsotropicLSM}
\end{figure}

\begin{figure}[t!]
\includegraphics[width=4.25cm,height=3.75cm]{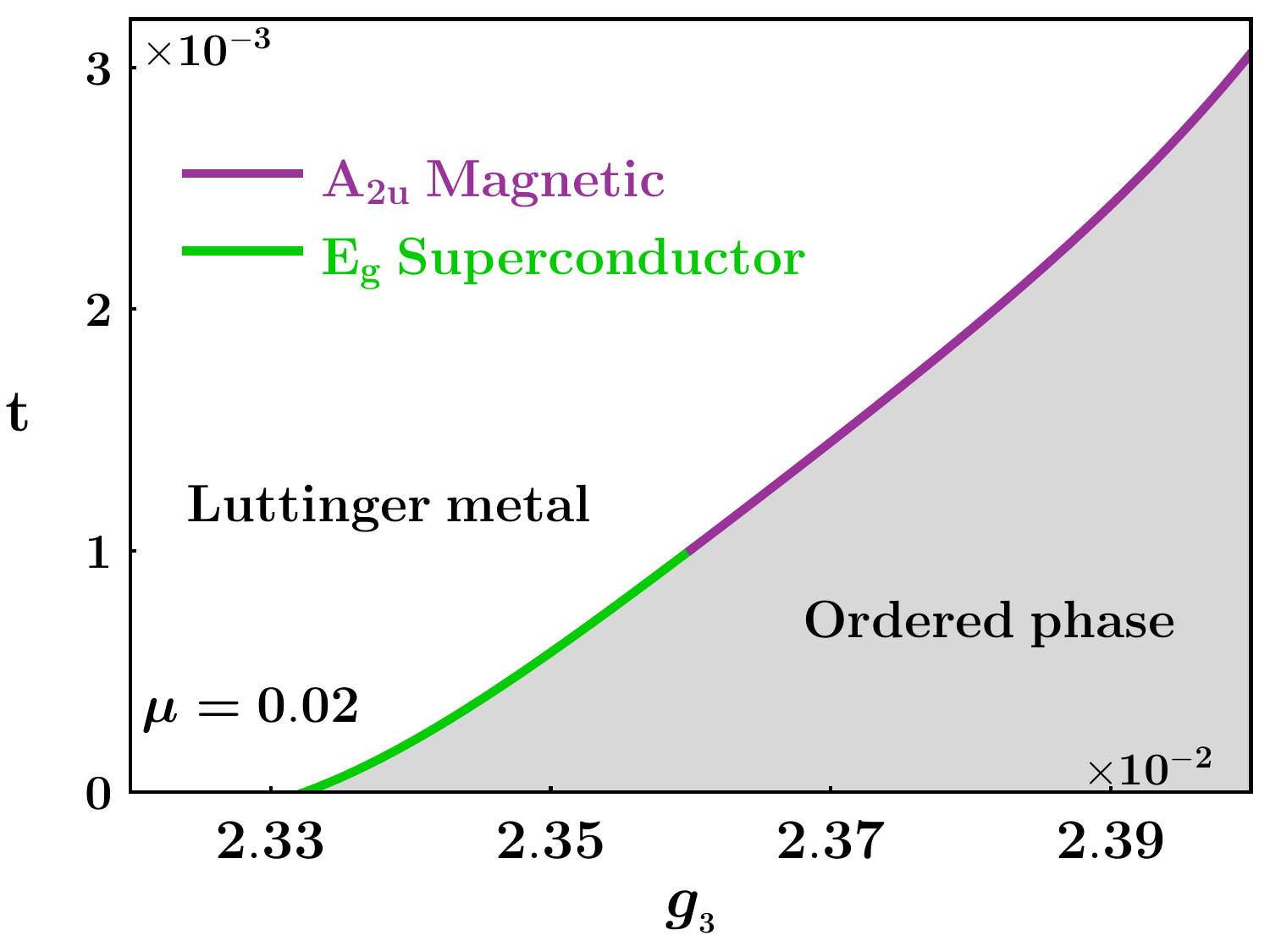}%
\includegraphics[width=4.25cm,height=3.75cm]{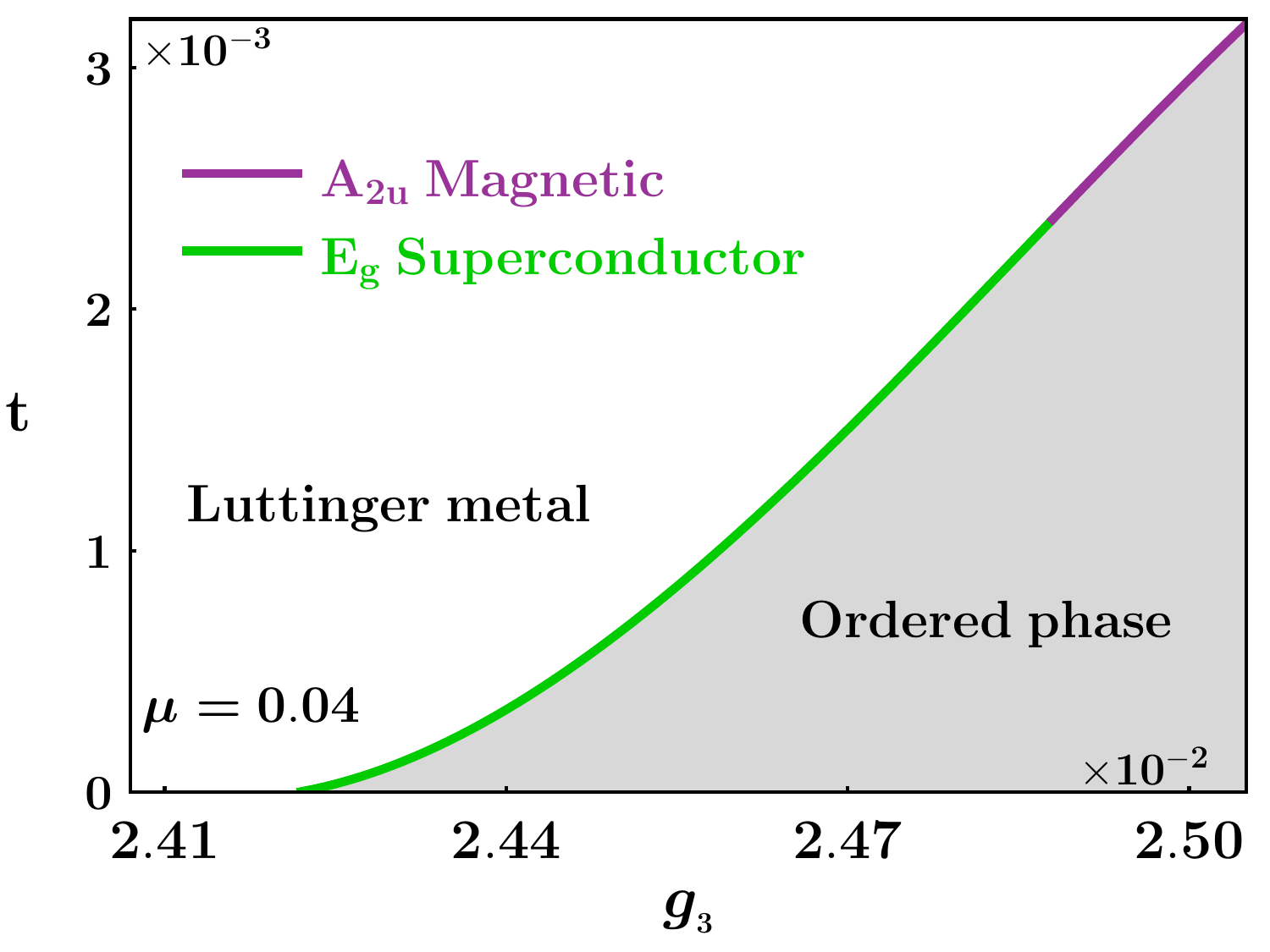}
\includegraphics[width=4.25cm,height=3.75cm]{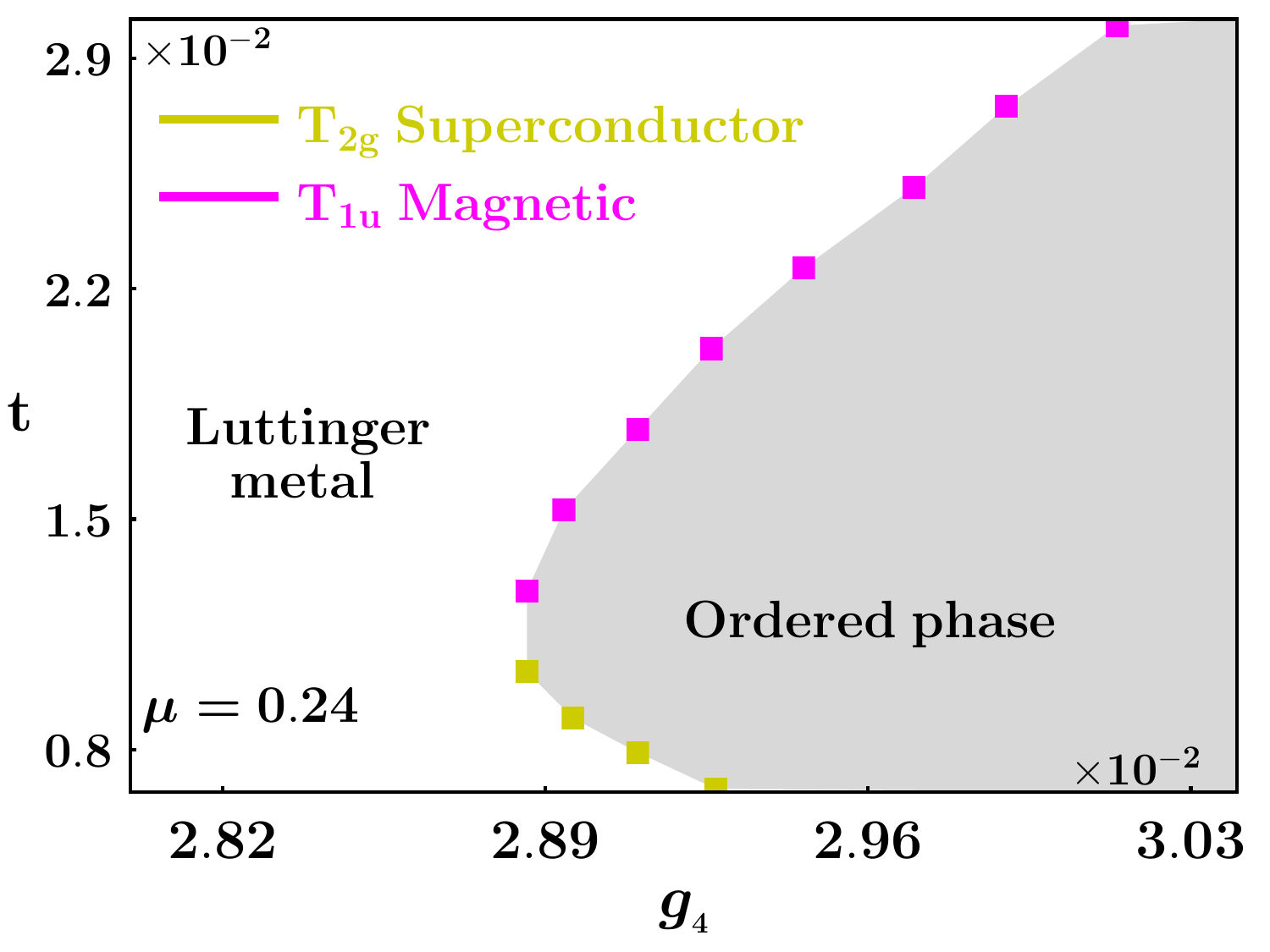}%
\includegraphics[width=4.25cm,height=3.75cm]{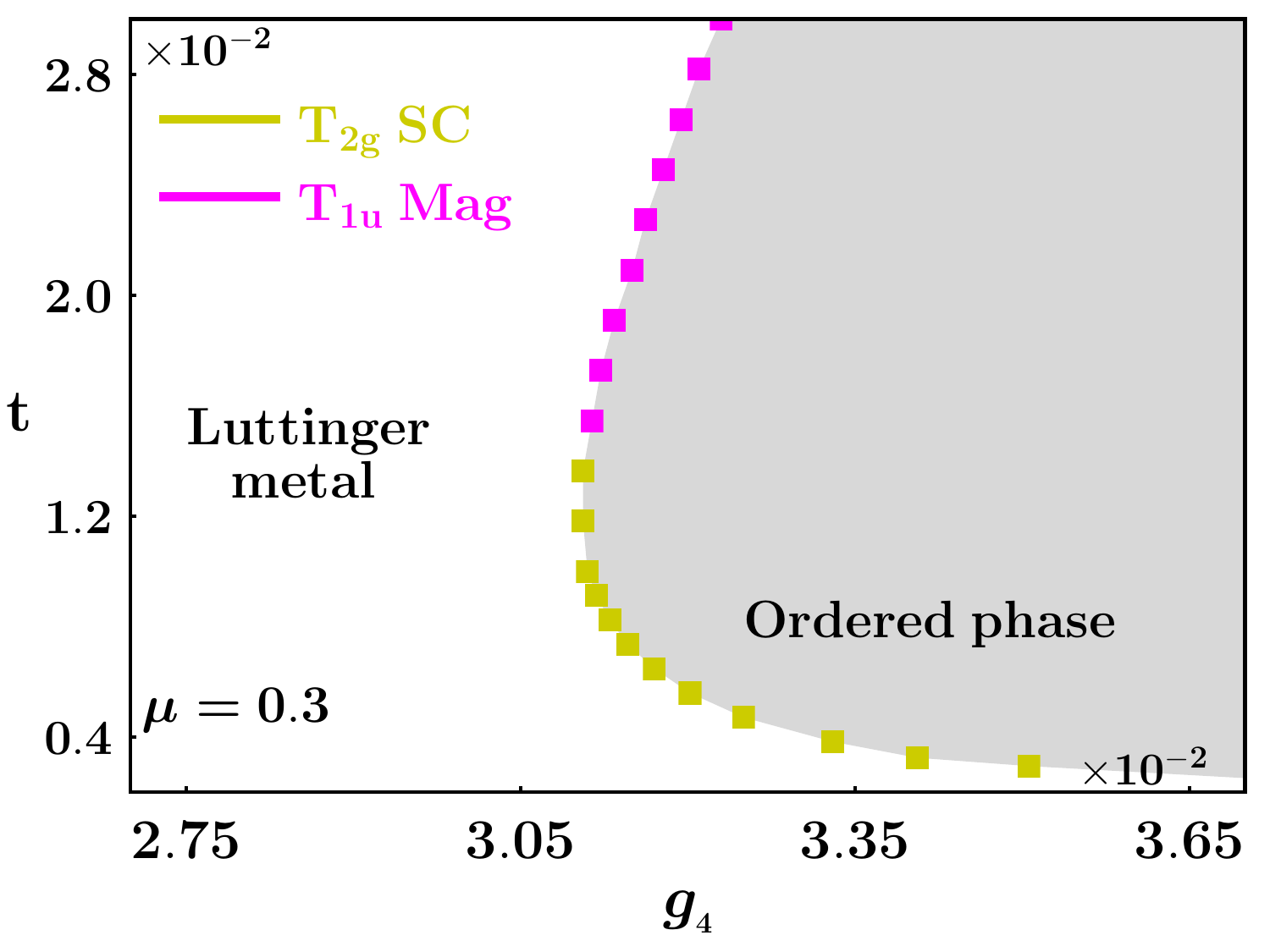}
\caption{Top row: Onset of $d$-wave pairing (belonging to the $E_{g}$ representation) at low temperature from magnetic interaction in the $A_{2u}$ channel, namely $g_{_3}$, in the presence of finite chemical doping $\mu$ (cf. Fig.~\ref{Subfig:FiniteT_g3} for $\mu=0$). The $A_{2u}$ magnetic order gets pushed toward stronger coupling with increasing $\mu$. Bottom row: Similar phase diagrams depicting the onset of $d$-wave pairing (belonging to the $T_{2g}$ representation) from strong repulsive magnetic interaction in the $T_{1u}$ channel, when $\mu >0$ (compare it with Fig.~\ref{Subfig:FiniteT_g4} for $\mu=0$). With increasing chemical potential the onset of the $T_{1u}$ magnetic order takes place at stronger coupling. Hence, magnetic interactions are conducive for the nucleation of $d$-wave pairings in correlated Luttinger metal. Here coupling constants are measured in units of $\epsilon$ and the gray shaded regions represent ordered states (see footnote~2). Note that the shape of the phase boundaries in the lower panel suggests that the LSM-ordered phase transition at low temperatures and finite chemical doping is possibly first-order in nature. Even though the RG methodology is tailored to capture continuous transitions, the possibility of a first-order transition is extremely sparse. The boundaries between Luttinger metal and various ordered phases are shown in different colors.   
}~\label{Fig:dwaves_IsotropicLSM}
\end{figure}

A similar conclusion can also be arrived at when the chemical potential is placed away from the band touching point. At finite chemical doping all particle-hole (two nematic and two magnetic) orders produce a Fermi surface (according to the Luttinger theorem~\cite{Luttinger_Theorem}) and hence a finite DoS. By contrast, any superconducting order at finite doping maximally gaps the Fermi surface. Therefore, at finite chemical doping superconducting orders are energetically superior to the excitonic orders, and they can be realized at sufficiently low temperature even in the presence of repulsive electronic interactions. Concomitantly, the particle-hole orders are pushed to the higher temperature and interaction regime, see Figs.~\ref{Fig:swave_IsotropicLSM} and ~\ref{Fig:dwaves_IsotropicLSM}.

Even though we gain valuable insights into the organization of various BSPs in the global phase diagram of strongly interacting spin-3/2 fermions from the competition between energy and entropy inside the ordered phases (guided by emergent topology of reconstructed band structure), the phase diagrams shown in Figs.~\ref{Fig:FiniteT_Summary},~\ref{Fig:swave_IsotropicLSM} and ~\ref{Fig:dwaves_IsotropicLSM} are obtained from an unbiased RG analysis, which systematically accounts for quantum fluctuations beyond the saddle point or mean-field approximation. Next we highlight the key aspects of the RG analysis.

\begin{center}
{\bf IIB2. Methodology: Renormalization Group }
\end{center}

The RG analysis we pursue in this work is controlled by a ``small" parameter $\epsilon$, measuring the deviation from the lower critical two spatial dimensions ($d=2$) of the theory, where local four-fermion interactions are \emph{marginal}, with $\epsilon=d-z=d-2$, and hence $[\lambda]=-\epsilon$. Both temperature and chemical potential (bearing the scaling dimension of energy) are \emph{relevant} perturbations at the $z=2$ fixed point, with scaling dimension $[T]=[\mu]=z=2$. The RG flow equations are cast in terms of the dimensionless coupling constants ($g_{_i}$s), temperature ($t$) and chemical potential ($\tilde{\mu}$), defined as 
\begin{equation}~\label{eq:dimensionlessparameters}
g_{_i}=\frac{m \lambda_i \Lambda^{\epsilon}}{4 (2 \pi)^3}, \:\:\:
t=\frac{2 m T}{\Lambda^2}, \:\:\: \tilde{\mu}=\frac{2 m \mu}{\Lambda^2},
\end{equation}
which can be inferred from their bare scaling dimensions. Here $\Lambda$ is the ultraviolet momentum cutoff.

The leading order RG analysis can be summarized in terms of the following set of coupled flow equations
\begin{eqnarray}~\label{Eq:RG_Introduction}
&&\frac{d t}{d \ell}= z t, \quad
\frac{d \mu}{d \ell}=z \mu, \nonumber \\
&&\frac{d g_{_i}}{d \ell}=-\epsilon g_{_i} + \sum_{j,k} g_{_j} g_{_k} \: H_{jk} \left(\alpha, t, \mu \right),
\end{eqnarray}    
where $\ell$ is the logarithm of the RG scale. For notational simplicity we take $\tilde{\mu} \to \mu$ in the above flow equations, and $H_{jk} \left(\alpha, t, \mu \right)$ are functions of the mass anisotropy parameter $\alpha$, $t$ and $\mu$. The RG flow equations for $g_i$s are obtained by systematically accounting for quantum corrections to the quadratic order in the $g_{_i}$s. The relevant Feynman diagrams are shown in Fig.~\ref{Fig:FeynDiag_Interaction}. A more detailed discussion of the RG analysis is presented in Sec.~\ref{Sec:RG} and Appendix~\ref{Append:RGdetails}. Some salient features of the RG analysis are the followings.

\begin{enumerate}

\item Temperature ($t$) and chemical potential ($\mu$) provide two infrared cutoffs~\cite{chakravarty, vafek, roy-juricic-twistedBLG}, respectively given by 
\begin{equation}~\label{Eq:Infraredcutoffs_Intro}
\ell^t_\ast=\frac{1}{z} \; \ln\left(\frac{1}{t(0)} \right), \:\:\: 
\ell^\mu_\ast=\frac{1}{z} \; \ln\left(\frac{1}{\mu(0)} \right)
\end{equation}
for the flow of quartic couplings $\left\{ g_{_i} \right\}$, where $t(0)$ and $\mu(0)$ represent the bare values ($<1$). Ultimately $\ell_\ast =\min \left( \ell^t_\ast, \ell^\mu_\ast\right)$ stops the flow of four-fermion interactions. At zero temperature and chemical doping the system is devoid of any such natural infrared cutoff, implying $\ell_\ast \to \infty$.

\item Any weak local four-fermion interaction is an irrelevant perturbation and all orderings (realized when $g_{_i} (\ell_\ast) \to \infty$) take place at finite coupling $g_{_i} \sim \epsilon$ through quantum phase transitions (QPTs). Such QPTs are controlled by quantum critical points (QCPs) and all transitions are continuous in nature. The universality class of the transition is determined by two critical exponents, given by 
\begin{equation}~\label{Eq:Exponents_Intro}
\nu^{-1}=\epsilon + {\mathcal O} (\epsilon^2) \quad \text{and} \quad z=2 + {\mathcal O} (\epsilon), 
\end{equation}
and for the physically relevant situation $\epsilon=1$.        

\end{enumerate}

\begin{figure*}[t!]
\subfigure[]{\includegraphics[width=0.425\linewidth]{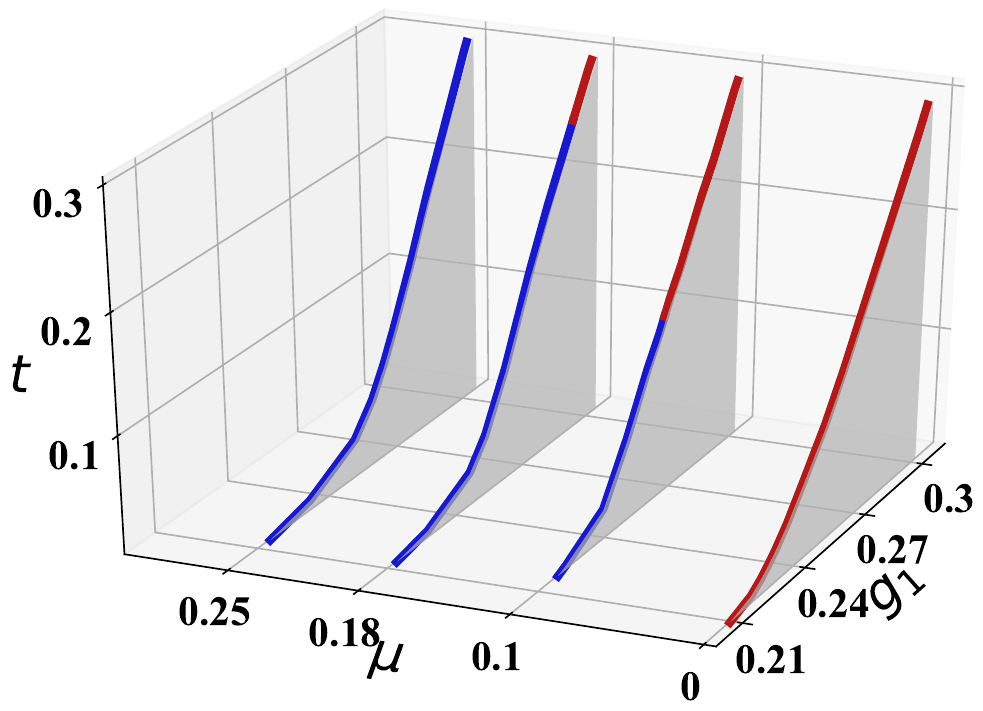}~\label{Subfig:T2gswaveMU}}
\hspace{0.5cm}
\subfigure[]{\includegraphics[width=0.47\linewidth]{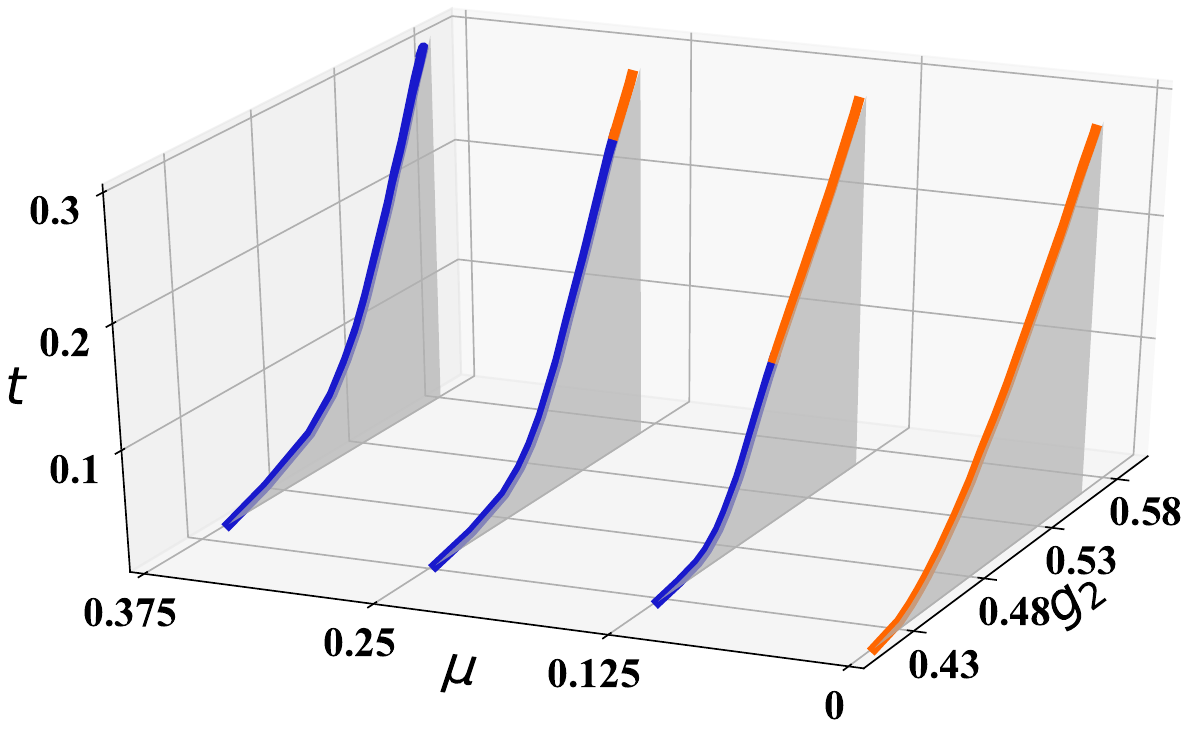}~\label{Subfig:EgswaveMU}} 
\vspace{0.1cm}
\subfigure[]{\includegraphics[width=0.475\linewidth]{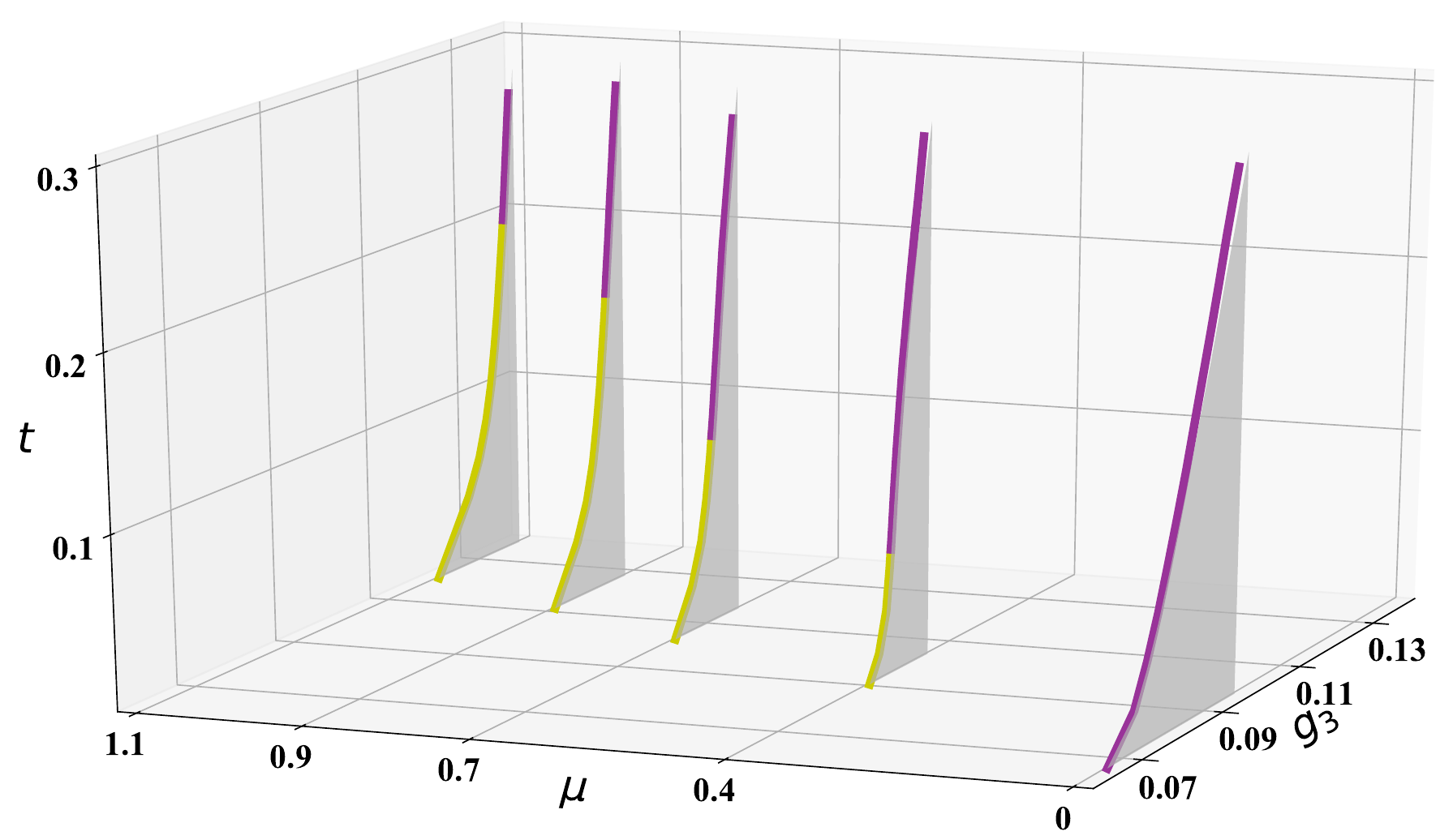}~\label{Subfig:A2udwaveMU}}
\hspace{0.5cm}
\subfigure[]{\includegraphics[width=0.40\linewidth]{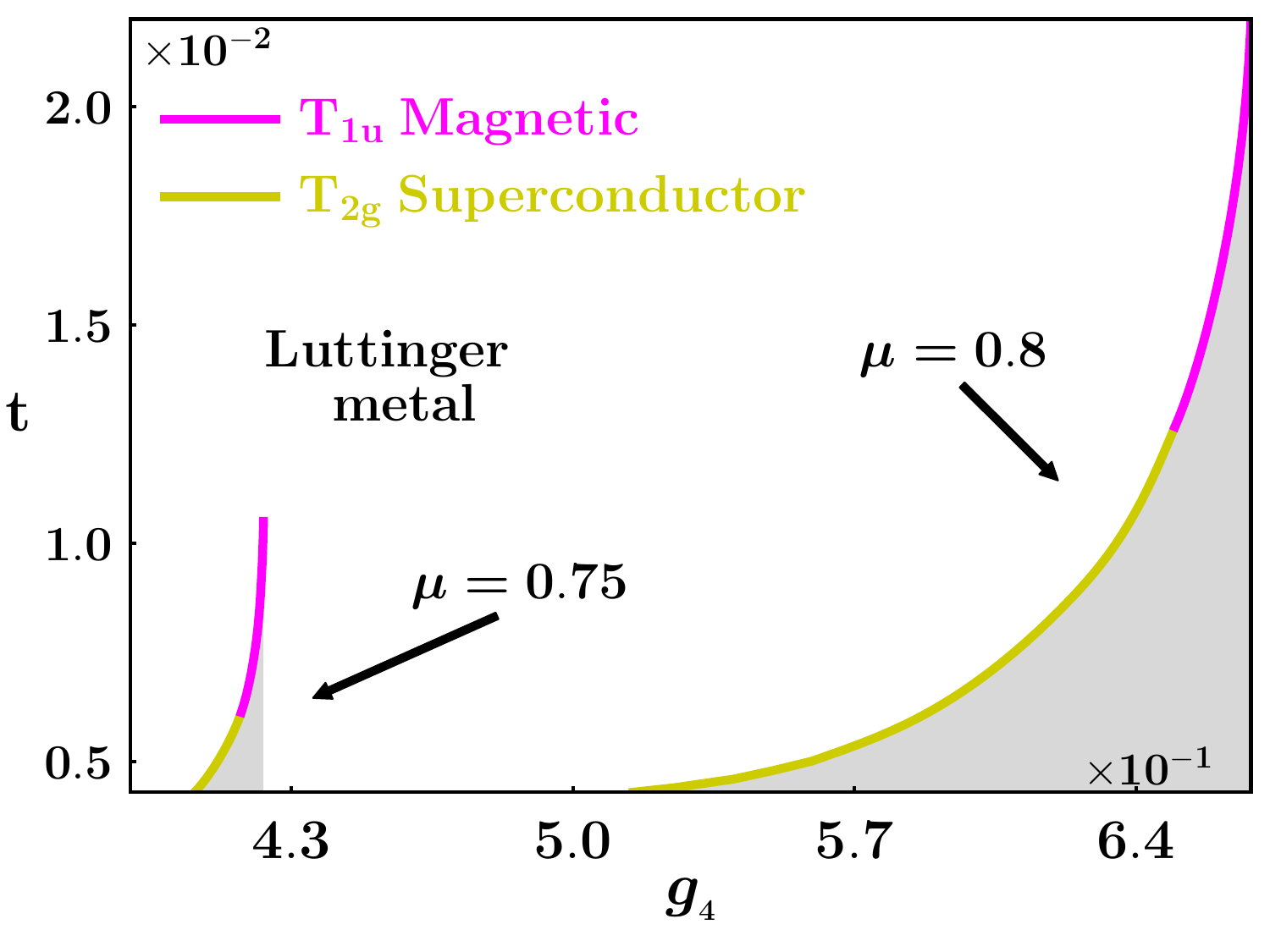}~\label{Subfig:T1udwaveMU}}
\caption{Specific cuts of the phase diagram of an interacting, but anisotropic Luttinger metal ($\alpha \neq \frac{\pi}{4}$), showing the appearance of various superconducting phases at finite chemical doping ($|\mu| >0$) from repulsive electron-electron interactions, accommodating only excitonic orders for $\mu=0$. Panels (a) and (b) show appearance of a conventional $s$-wave pairing (blue lines) from strong repulsive interaction in the $T_{2g}$ and $E_g$ nematic channels for $\alpha=1.5$ and $\alpha=0.1$, respectively [see also Figs.~\ref{Subfig:T2gswave} and ~\ref{Subfig:Egswave}]. The phase boundaries of two nematic phases with Luttinger (semi)metal are respectively denoted by red and golden yellow lines. Nucleation of various topological $d$-wave pairings, belonging to the $E_g$ (dark yellow lines) and $T_{2g}$ (dark green lines) representations, from strong repulsive magnetic interactions in the $A_{2u}$ and $T_{1u}$ channels, are respectively shown in panel (c) and (d) for $\alpha=1.5$ and $\alpha=0.1$ [see also Figs.~\ref{Subfig:A2uEg} and ~\ref{Subfig:T1uT2g}]. The phase boundaries of $A_{2u}$ and $T_{1u}$ magnetic phases with Luttinger (semi)metal are shown by purple and magenta lines. Due to a large separation of the interaction strength $g_{_4}$ required for any ordering at $\mu=0$ and $|\mu|>0$, we display the $\mu=0$ cut of the phase diagram from panel (d) in Fig.~\ref{Fig:T1uMagnetg4smallalpha}. The region at weaker interaction and higher temperature is occupied by correlated Luttinger metal (white regions), without any long-range ordering. In panels (a) and (b) $100 \; g_{_i} \to g_{_i}$ and in (c) $10 \; g_{_i} \to g_{_i}$. Throughout the coupling constants are measured in units of $\epsilon$ and the ordered states are displayed as the gray shaded regions (see footnote~2). 
}~\label{Fig:GlobalPD_Intro_FS}
\end{figure*}

\begin{figure}[t!]
\includegraphics[width=7cm,height=5cm]{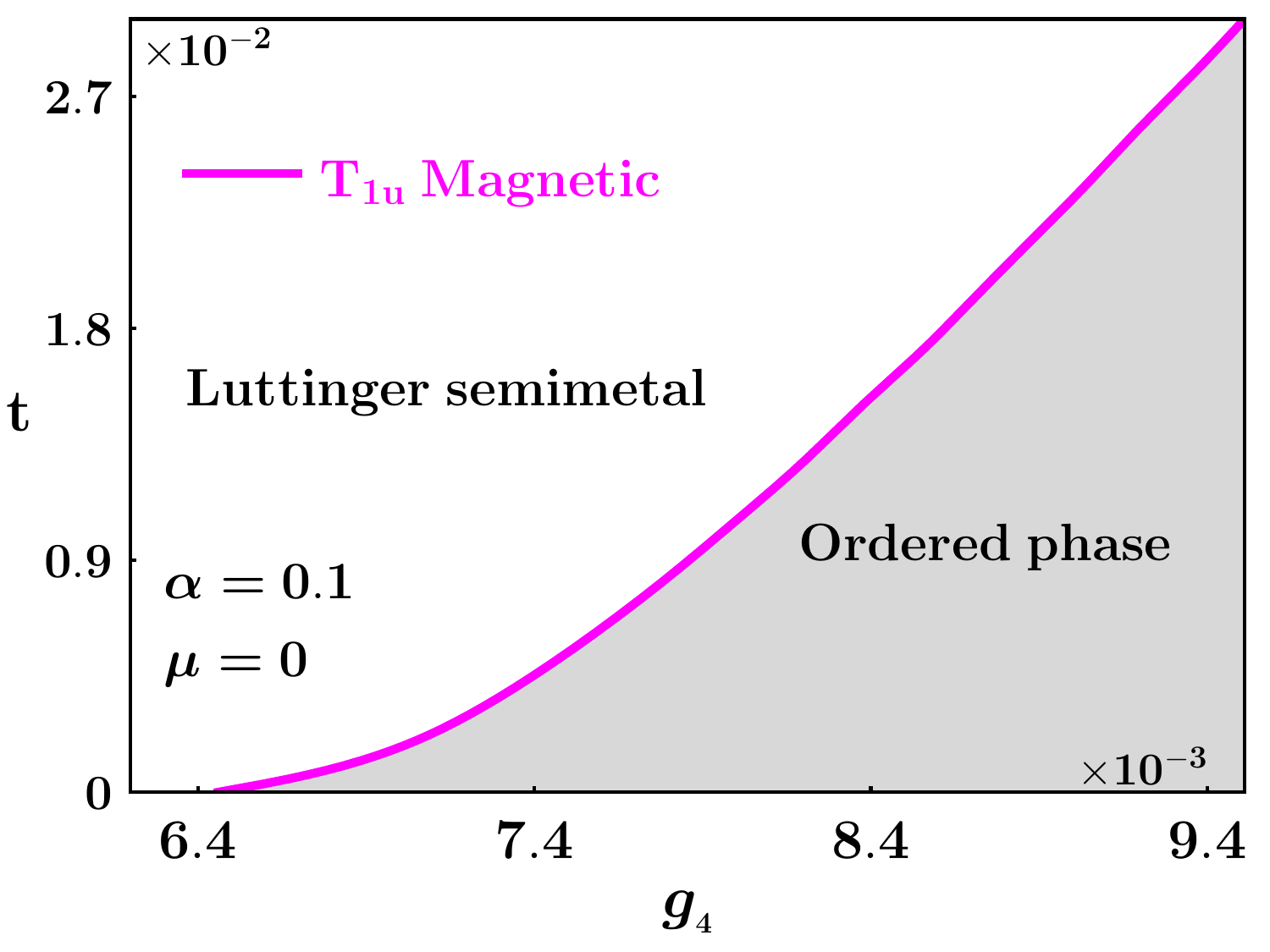}
\caption{The phase diagram of an interacting Luttinger semimetal for $\alpha=0.1$ and $\mu=0$, obtained by tuning the strength of the magnetic interaction in the $T_{1u}$ channel ($g_{_4}$), measured in units of $\epsilon$. The shaded (white) region represents the ordered phase (Luttinger semimetal).
}~\label{Fig:T1uMagnetg4smallalpha}
\end{figure}

Using the RG analysis we arrive at various cuts of the global phase diagram of interacting spin-3/2 fermions at (1) zero chemical doping [see Fig.~\ref{Fig:FiniteT_Summary}] and (2) for finite-$\mu$ [see Figs.~\ref{Fig:swave_IsotropicLSM} and \ref{Fig:dwaves_IsotropicLSM}]. In all these cuts of the phase diagram (as well as the one shown in Figs.~\ref{Fig:GlobalPD_Intro_FS}, ~\ref{Fig:T1uMagnetg4smallalpha} and ~\ref{Fig:GlobalPD_Intro_noFS}) coupling constants are always measured in units of $\epsilon$ and most importantly the nature of the ordered states is completely impervious to the exact value of $\epsilon$.

The universality class of the QPT leaves its signature on the scaling of the transition temperature ($t_c$). Note that $t_c \sim |\delta_i|^{\nu z}$~\cite{sondhi-RMP, sachdev-book}, where $\delta_i=\left(g_{_i}-g^\ast_{_i}\right)/g^\ast_{_i}$ is the reduced distance from the critical point, located at $g^\ast_{_i}$. Hence, $t_c \sim |\delta|^2$ for $\nu=1$ and $z=2$, obtained from the leading order $\epsilon$ expansion, after setting $\epsilon=1$, irrespective of the choice of the coupling constant and the resulting BSP (see Fig.~\ref{Fig:ScalingTc}). We discuss this issue in detail in Sec.~\ref{SubSec:FiniteTRG}. Even though ultimately we are interested in three-dimensional interacting Luttinger materials for which $\epsilon=1$, the RG methodology employed here follows the general spirit of the $\epsilon$ expansion, succinctly employed in the past for bosonic $\Phi^4$ theory and fermionic Gross-Neveu model for which as well $\epsilon=1$~\cite{zinn-justin:book, zinn-justin-moshe-moshe}.

\begin{center}
{\bf IIB3. Competing Orders \& Selection Rule}
\end{center}

The correspondence between a given interaction coupling and the resulting phases can be appreciated by formulating the whole theory in the basis of an eight component Nambu-doubled spinor $\Psi_{\rm Nam}$, introduced in Sec.~\ref{SubSec:Nambu-doubling}. In this basis the Luttinger Hamiltonian $\hat{h}_{\rm L} ({\bf k}) \to \eta_3 \hat{h}_{\rm L} ({\bf k})$. Pauli matrices $\{ \eta_\mu \}$ operate on the Nambu or particle-hole index. Any four-fermion interaction takes the form $g_{_{\rm I}} \; ( \Psi^\dagger_{\rm Nam}\; \hat{\rm I}\; \Psi_{\rm Nam} )^2$ and an order parameter ($\Delta_{\rm O}$) couples to a fermion bilinear according to $\Delta_{\rm O} \; ( \Psi^\dagger_{\rm Nam} \; \hat{{\rm O}} \; \Psi^\dagger_{\rm Nam} )$, where $\hat{\rm I}$ and $\hat{\rm O}$ are eight dimensional Hermitian matrices. We argue that when $g_{_{\rm I}}$ is sufficiently strong, it can support \emph{only} two types of ordered phases, for which~\footnote{\label{footnote-selectionrule}If $\hat{\rm O}$ and $\hat{\rm I}$ are multi-component vectors of $M_{\rm O}$ and $M_{\rm I}$ elements, respectively, then condition (2) is satisfied when at least $\lceil \frac{M_{\rm O}}{2} \rceil$ matrices anti-commute with $\lceil \frac{M_{\rm I}}{2} \rceil$ matrices, where $\lceil \cdots \rceil$ is the ceiling function. } 
\begin{equation}~\label{Eq:SelectionRule_Intro}
\text{either} \:\:\: (1) \: \hat{\rm O} \equiv \hat{\rm I} \:\:\: \text{or} \:\:\:  (2) \: \{ \hat{\rm O}, \hat{\rm I} \}=0. 
\end{equation} 
This outcome can be appreciated in the following way.

When an interaction coupling $g_{_{\rm I}}$ diverges toward $+\infty$ under coarse graining (indicating onset of a BSP), it provides \emph{positive} scaling dimension to an order parameter field $\Delta_{\rm O}$ only when one of the above two conditions is satisfied. We substantiate this argument by considering the relevant Feynman diagrams [see Fig.~\ref{Fig:FeynDiag_Susceptibility}] in Sec.~\ref{SubSec:SelectionRule}. Even though we arrive at this ``selection rule" among competing orders from a leading order RG calculation, such a simple argument relies on internal symmetries among competing orders (breaking different symmetries) and is expected to hold at the non-perturbative level. We now support this claim by focusing on a specific example.

Let us choose a particular four-fermion interaction (in the $T_{2g}$ nematic channel) 
\begin{equation}
g_{_1} \sum^3_{j=1} \left( \Psi^\dagger_{\rm Nam} \: \eta_3 \; \Gamma_j \: \Psi_{\rm Nam}\right)^2. \nonumber             
\end{equation} 
From the phase diagrams shown in Figs.~\ref{Subfig:FiniteT_g1} for zero and finite temperature and \ref{Fig:swave_IsotropicLSM}(left) for finite chemical doping, we find that when this coupling constant is sufficiently strong, it supports two distinct BSPs. 

\begin{enumerate}

\item A nematic order following the $T_{2g}$ representation, for which $\hat{\rm O}= \eta_3 \{\Gamma_1,\Gamma_2,\Gamma_3 \}$. In this case selection rule (1) from Eq.~(\ref{Eq:SelectionRule_Intro}) is satisfied, since $\hat{\rm I}= \eta_3 \{\Gamma_1,\Gamma_2,\Gamma_3 \}$, and hence $\hat{\rm O} \equiv \hat{\rm I}$.

\item An $s$-wave superconductor following the trivial $A_{1g}$ representation, for which $\hat{\rm O}= \{\eta_1, \eta_2\} \Gamma_0$, where $\Gamma_0$ is a four-dimensional identity matrix. The onset of $s$-wave pairing follows from selection rule (2) in Eq.~(\ref{Eq:SelectionRule_Intro}), since $\{ \hat{\rm O}, \hat{\rm I} \}=0$.   

\end{enumerate}

\noindent
Moreover we realize that the $T_{2g}$ nematic order and the $s$-wave pairing together constitute an $O(5)$ vector, $\left\{\eta_3 \Gamma_1, \eta_3 \Gamma_2, \eta_3 \Gamma_3, \eta_1 \Gamma_0, \eta_2 \Gamma_0 \right\}$, of five mutually anti-commuting matrices, reflecting the enlarged internal symmetry between these two orders. Following the same spirit we arrive at the following observations.

\begin{enumerate}

\item Four fermion interaction in the $E_g$ nematic channel ($g_{_2}$) supports (a) a nematic order transforming under the $E_g$ representation [satisfying selection rule (1)] and (b) an $s$-wave superconductor [satisfying selection rule (2)], as shown in Figs.~\ref{Subfig:FiniteT_g2} and ~\ref{Fig:swave_IsotropicLSM}(right). One can construct an $O(4)$ vector by combining these two order-parameters [see Eq.~(\ref{Eq:mass0})].

\item Four fermion interaction in the $A_{2u}$ channel ($g_{_3}$) supports (a) a magnetic order transforming under the $A_{2u}$ representation [satisfying selection rule (1)], and (b) $E_g$ nematic order and $d$-wave pairings [satisfying selection rule (2)], as shown in Figs.~\ref{Subfig:FiniteT_g3} and ~\ref{Fig:dwaves_IsotropicLSM}(top). Notice, we can construct multiple copies of composite SU(2) order parameters, by combining the $A_{2u}$ order with $E_g$ nematic or $d$-wave pairing, see Figs.~\ref{Fig:Triangles_Excitons}(a) and \ref{Fig:Triangles_Pairing1}.  

\item Four fermion interaction in the $T_{1u}$ magnetic channel ($g_{_4}$) supports (a) a magnetic order transforming under the $T_{1u}$ representation [satisfying selection rule (1)], and (b) $T_{2g}$ nematic and $d$-wave pairings [satisfying selection rule (2)], as shown in Figs.~\ref{Subfig:FiniteT_g4} and ~\ref{Fig:dwaves_IsotropicLSM}(bottom). Combining the magnetic order with $T_{2g}$ nematic or $d$-wave pairing we can construct multiple copies of composite SU(2) vector, see Figs.~\ref{Fig:Triangles_Excitons}(b), ~\ref{Fig:Triangles_Excitons}(c) and \ref{Fig:Triangles_Pairing2}.  

\end{enumerate}  

\noindent
A more detailed discussion supporting these scenarios is presented in Sec.~\ref{SubSec:SelectionRule}. Therefore, combining the energy-entropy competition (obtained from the reconstructed band topology within a mean-field approximation) and an unbiased (controlled by a small parameter $\epsilon$) RG analysis with the selection rule we gain valuable insights into the nature of broken symmetry phases, competing orders and quantum critical phenomena in the global phase diagram of strongly interacting spin-3/2 fermions.

\begin{center}
{\bf IIB4. Anisotropic Luttinger (semi)metal}
\end{center}

So far, we centered our focus on the isotropic system [realized for $\alpha=\frac{\pi}{4}$ in Eq.~(\ref{Eq:Luttinger_Hamiltonian_Intro})]. Note that for $\alpha=\frac{\pi}{4}$ the system enjoys an enlarged (but artificial) SO(3) rotational symmetry. However, in a cubic environment $\alpha \neq \frac{\pi}{4}$ in general. Nonetheless, all the central results we quoted in the last three subsections hold (at least qualitatively) for any arbitrary value of $\alpha$: $0 \leq \alpha \leq \frac{\pi}{2}$. The discussion on the role of the mass anisotropy parameter $\alpha$ on the global phase diagram of interacting spin-3/2 fermions is rather technical, which we address in depth in Secs.~\ref{SubSec:susceptibility}, \ref{SubSec:QPTinLSM}, \ref{SubSec:LuttingerMetalRG}. We here only quote some key results, which nicely corroborate with the rest of the discussion from this section. 

\begin{enumerate}

\item We identify the mass anisotropy parameter as a valuable non-thermal tuning parameter, and for suitable choices of this parameter one can find (a) $T_{2g}$ nematic and $A_{2u}$ magnetic order respectively for strong enough $g_{_1}$ and $g_{_3}$ couplings (when $\alpha \to \frac{\pi}{2}$), see Figs.~\ref{Subfig:T2gswaveMU} and ~\ref{Subfig:A2udwaveMU}, (b) $E_g$ nematic and $T_{1u}$ magnetic orders for strong enough $g_{_2}$ and $g_{_4}$ (when $\alpha \to 0$), as shown in Figs.~\ref{Subfig:EgswaveMU} and ~\ref{Fig:T1uMagnetg4smallalpha} at zero temperature and chemical doping, respectively. These outcomes are in agreement with selection rule (1).

\item At finite chemical doping (a) an $s$-wave pairing emerges from repulsive electronic interaction in the $T_{2g}$ channel [see Fig.~\ref{Subfig:T2gswaveMU}] as well as $E_g$ channel [see Fig.~\ref{Subfig:EgswaveMU}], (b) $d$-wave pairings belonging to the $E_g$ and $T_{2g}$ representations respectively appear for repulsive interaction in the $A_{2u}$ channel [see Fig.~\ref{Subfig:A2udwaveMU}] and $T_{1u}$ channel [see Fig.~\ref{Subfig:T1udwaveMU}]. These outcomes are in accordance with selection rule (2), as we argued previously for an isotropic Luttinger system.    

\end{enumerate}   

We now proceed to a detailed discussion on each component of our work, starting from the noninteracting Luttinger model.


\section{Luttinger model}~\label{Sec:Luttinger_Model}

We begin the discussion with the Luttinger model describing a bi-quadratic touching of Kramers degenerate valence and conduction bands at an isolated point (here chosen to be the $\Gamma=(0,0,0)$ point, for convenience) in the Brillouin zone. In this section, we first present the low-energy Hamiltonian and discuss its symmetry properties (Sec.~\ref{SubSec:Hamiltonian}). Subsequently, we introduce the corresponding imaginary time ($\tau$) or Euclidean action and the notion of the renormalization group (RG) scaling (Sec.~\ref{SubSec:Lagrangian}). Finally, we define an eight-component Nambu-doubled basis that allows us to capture all, including both particle-hole or excitonic and particle-particle or superconducting, orders within a unified framework (Sec.~\ref{SubSec:Nambu-doubling}).

    \subsection{Hamiltonian and Symmetries}~\label{SubSec:Hamiltonian}

The Hamiltonian operator describing a bi-quadratic touching of Kramers degenerate valence and conduction bands in three dimensions is given by~\cite{luttinger, murakami-zhang-nagaosa} 
\begin{equation}~\label{Eq:Luttinger_Model}
\hat{h}_{\mathrm L} ({\bf k})=-k^2 \left[ \mathlarger{\sum^{3}_{j=1}} \frac{\hat{d}_j ( \hat{{\bf k}} )}{2m_1} \Gamma_j 
+ \mathlarger{\sum^{5}_{j=4}} \frac{\hat{d}_j ( \hat{{\bf k}})}{2m_2} \Gamma_j\right]-\mu \Gamma_0,
\end{equation}
where $\Gamma_0$ is the four-dimensional identity matrix. Chemical potential $\mu$ and momenta ${\bf k}$ are measured from the band touching point. Here, $\hat{{\bf d}}( \hat{{\bf k}})$ is a five-dimensional unit vector that transforms under the $l=2$ representation under the orbital $SO(3)$ rotations. Hence $\hat{{\bf d}}( \hat{{\bf k}})$ is constructed from the $d$-wave form factors or spherical harmonics $Y^{m}_{l=2}(\theta, \phi)$, as shown in Appendix~\ref{Append:Luttingerdetails}. The corresponding four-component spinor basis is given by 
\begin{equation}~\label{Eq:spinor}
\Psi^\top_{\bf k}= \left( c_{{\bf k},+\frac{3}{2}}, c_{{\bf k},+\frac{1}{2}}, c_{{\bf k},-\frac{1}{2}}, c_{{\bf k},-\frac{3}{2}} \right),
\end{equation}  
where $c_{{\bf k},j}$ is the fermionic annihilation operator with momenta ${\bf k}$ and spin projection $j=\pm 3/2$ and $\pm 1/2$.
The five mutually anti-commuting $\Gamma$ matrices are defined as 
\begin{eqnarray}~\label{Eq:Gammarepresentation}
\Gamma_1 = \kappa_3 \sigma_2, \:\: \Gamma_2=\kappa_3 \sigma_1, \:\: \Gamma_3=\kappa_3 \sigma_0,  \nonumber \\
\Gamma_4 = \kappa_1 \sigma_0, \:\: \Gamma_5 = \kappa_3 \sigma_3.
\end{eqnarray}
Two sets of two dimensional Pauli matrices $\{ \kappa_\nu \}$ and $\{ \sigma_\nu \}$ respectively operate on the sign (${\rm sgn}[j]$) and magnitude ($|j| \in \{ 1/2, 3/2 \}$) of the spin projections, where $\nu=0,1,2,3$. To close the Clifford algebra of all four-dimensional Hermitian matrices we also define ten commutators according to $\Gamma_{jk}=\left[ \Gamma_j, \Gamma_k \right]/(2i)$, with $j>k$ and $j,k=1, \cdots, 5$. All sixteen four-dimensional matrices can be expressed in terms of the products of spin-3/2 matrices (${\mathbf J}$), as also shown in Appendix~\ref{Append:Luttingerdetails}.

The energy spectra for Luttinger fermions are given by $\pm E_s({\bf k})-\mu$, where for $s=\pm 1$
\begin{equation}
E_s({\bf k})= \frac{k^2}{2m} \sqrt{ \cos^2\alpha \sum^{3}_{j=1}  \hat{d}^2_j + \sin^2\alpha  \sum^{5}_{j=4}\hat{d}^2_j },
\end{equation}
reflecting the quadratic band touching for $\mu=0$, which is protected by the cubic symmetry. For brevity we drop the explicit dependence of $\{ \hat{d}_j \}$ on $\hat{\bf k}$. 

Notice that the independence of $E_s({\bf k})$ on $s$ manifests the Kramers degeneracy of the valence and conduction bands, ensured by (1) the time-reversal ($\mathcal T$) and (2) the parity or inversion ($\mathcal P$) symmetries. Specifically, under the reversal of time, ${\bf k} \to -{\bf k}$ and $\Psi_{\bf k} \to \Gamma_1 \Gamma_3 \Psi_{-{\bf k}}$ and hence ${\mathcal T}= \Gamma_1 \Gamma_3 {\mathcal K}$, where ${\mathcal K}$ is the complex conjugation, yielding ${\mathcal T}^2=-1$ (reflecting Kramers degeneracy of bands). Under the inversion ${\mathcal P}: {\bf k} \to -{\bf k}$ and $\Psi_{\bf k} \to \Psi_{-{\bf k}}$.

The ``average" mass $m$ and the mass anisotropy parameter $\alpha$ are respectively given by~\cite{goswami-roy-dassarma} 
\begin{equation}~\label{Eq:anisotropicparameter}
m=\frac{m_1 m_2}{m_1+m_2}, \:\: \alpha=\tan^{-1} \left( \frac{m_2}{m_1}\right).
\end{equation}
Note that $\{ \hat{d}_j \}$ for $j=1,2,3$ and $j=4,5$ belong to the $T_{2g}$ (three component) and $E_g$ (two component) representations of the cubic or octahedral ($O_h$) point group, and $m_1$ and $m_2$ are effective masses in these two orbitals, respectively. The mass anisotropy parameter $\alpha$ allows us to smoothly interpolate between (1) the $m_1 \to \infty$ limit when the dispersion of the $T_{2g}$ orbital becomes flat, yielding $\alpha \to 0$ and (2) $m_2 \to \infty$ when the $E_g$ orbital becomes non-dispersive, leading to $\alpha \to \frac{\pi}{2}$. For $\alpha=\frac{\pi}{4}$ or $m_1=m_2$, the Luttinger model enjoys an enlarged spherical symmetry. Any $\alpha \neq \frac{\pi}{4}$ captures a \emph{quadrupolar distortion} in the system (still preserving the cubic symmetry). In what follows, we treat $\alpha$ as a \emph{non-thermal} tuning parameter to explore the territory of interacting Luttinger fermions.

The connection between the spin projections ($j=\pm 3/2$ and $\pm 1/2$) and the bands can be appreciated most economically by taking ${\bf k}=(0,0,k)$. For such a specific choice of momentum axis, the Luttinger Hamiltonian takes a \emph{diagonal} form, given by 
\begin{equation}
\hat{h}_{\rm L}(k \hat{z})=\frac{k^2}{2m_2} {\rm Diag.} \left[ -1,1,1,-1\right]-\mu.
\end{equation} 
From the above expression we can immediately infer that the pseudospin projections on the valence and conduction bands are respectively $|j|=3/2$ and $1/2$. 

    \subsection{Lagrangian and Scaling}~\label{SubSec:Lagrangian}

The imaginary time ($\tau$) Euclidean action corresponding to the non-interacting Luttinger model is given by 
\begin{equation}~\label{Eq:action_nonint}
S_0 = \int d\tau d^d{\bf x} \: \Psi^\dagger (\tau,{\bf x}) \; \hat{h}_{\rm L} ({\bf k} \to -i \nabla) \; \Psi (\tau,{\bf x}). 
\end{equation}
The action remains invariant under the following rescaling of space-(imaginary)time coordinates and the fermionic field
\begin{equation}~\label{Eq:rescaling}
{\bf x} \to e^{\ell} \; {\bf x}, \:\: \tau \to e^{z \ell} \; \tau, \:\: \Psi \to e^{-d \ell/2} \; \Psi,
\end{equation}
where $z$ is the dynamic scaling exponent, measuring the relative scaling between energy and momentum according to $E({\bf k}) \sim |{\bf k}|^z$. For Luttinger fermions $z=2$. The parameter $\ell$ is the logarithm of the RG scale. In what follows in Secs.~\ref{Sec:e-e_Interaction} and \ref{Sec:RG}, we use the above scaling ansatz while addressing the effects of electronic interactions in this system. Under the above rescaling of parameters, the temperature ($T$) and chemical potential ($\mu$) scale as 
\begin{equation}
T \to e^{-z \ell} T, \:\: \mu \to e^{-z \ell} \mu. 
\end{equation} 
Therefore, the scaling dimension of these two quantities is $[T]=[\mu]=z$ (same as that of energy). Throughout, we use the \emph{natural unit}, in which $\hbar=k_B=1$.   

         \subsection{Nambu doubling}~\label{SubSec:Nambu-doubling}

To facilitate the forthcoming discussion we here introduce an eight-component Nambu-doubled spinor basis (suitable to capture both exitonic and superconducting orders within a unified framework) according to 
\begin{equation}~\label{Eq:Nambu_Basis}
\Psi_{\rm Nam}= \left[ \begin{array}{c}
\Psi_{\bf k} \\
\Gamma_1 \Gamma_3 \left( \Psi^\dagger_{-{\bf k}} \right)^\top
\end{array}
\right],
\end{equation}
where $\Psi_{\bf k}$ is a four-component spinor, see Eq.~(\ref{Eq:spinor}). In the lower block of $\Psi_{\rm Nam}$ we absorb the unitary part of the time-reversal operator ${\mathcal T}$, ensuring that the eight-component Nambu spinor ($\Psi_{\rm Nam}$) transforms the same way as the original four component spinor $\Psi_{\bf k}$ under the $SU(2)$ pseudospin rotation.   
In this basis the eight-dimensional Luttinger Hamiltonian takes a simple form
\begin{equation}~\label{Eq:LuttingerModel_Nambu}
\hat{h}^{\rm Nam}_{\mathrm L} ({\bf k})= \eta_3 \; \hat{h}_{\rm L} ({\bf k}),
\end{equation}
and the time-reversal operator becomes ${\mathcal T}_{\rm Nam}=\eta_0 \Gamma_1 \Gamma_3 {\mathcal K}$. The newly introduced set of Pauli matrices $\{\eta_\nu \}$ operates on the Nambu or particle-hole indices, with $\nu=0,1,2,3$. Therefore, by construction while the excitonic orders assume \emph{block-diagonal} form, all superconducting orders are \emph{block-off-diagonal} in the Nambu subspace. Note that $\hat{h}^{\rm Nam}_{\mathrm L} ({\bf k})$ \emph{commutes} with the number operator $\hat{N}=\eta_3 \Gamma_0$. 

\section{Broken symmetry phases}~\label{Sec:BSPs}

Next we discuss possible BSPs in this system. We introduce various possible excitonic and superconducting orders in the Nambu basis ($\Psi_{\rm Nam}$) in two subsequent sections. Finally, we discuss the reconstructed band structure and emergent topology inside the ordered phases.

    \subsection{Particle-hole or excitonic orders}~\label{SubSec:particle-hole_Nambu}

The effective single-particle Hamiltonian in the presence of all possible momentum-independent or local or intra-unit cell excitonic orders is given by 
\begin{equation}~\label{Eq:excitonicOP_Hamil}
H^{{\rm exc}}_{{\rm local}}= \int d^3{\bf r} \:\: \left( \Psi^\dagger_{\rm Nam} \; \hat{h}^{{\rm exc}}_{{\rm local}} \; \Psi_{\rm Nam} \right),
\end{equation}
where 
\allowdisplaybreaks[4]
\begin{eqnarray}~\label{Eq:excitonicOP}
\hat{h}^{{\rm exc}}_{{\rm local}} &=& \overbrace{\Delta_0 \eta_3 \Gamma_0}^{\rm Density} \quad
+ \overbrace{\eta_3 \left[ \sum^{3}_{j=1} \Delta^{j}_1 \Gamma_j + \sum^{5}_{j=4} \Delta^{j}_2 \Gamma_j \right]}^{\rm Nematic}  \\
&+& \underbrace{\eta_0 \left[ \Delta_3 \Gamma_{45} + \sum^{3}_{j=1} \Delta^{j}_{4} \Gamma_{45} \Gamma_j + \sum^{3}_{j=1} \sum^{5}_{k=4} \Delta^{jk}_5 \Gamma_{jk} \right]}_{\rm Magnetic}. \nonumber  
\end{eqnarray}
The ordered phases can be classified according to their transformation under the cubic ($O_h$) point group symmetry. Regular fermionic density ($\Delta_0$) does not break any symmetry (hence does not correspond to any ordering) and transforms under the trivial $A_{1g}$ representation. A three-component \emph{nematic} order-parameter, constituted by $\vec{\Delta}_1=\left( \Delta^{1}_1, \Delta^{2}_1, \Delta^{3}_1 \right)$, transforms under the $T_{2g}$ representation. By contrast, a two-component nematic order transforming under the $E_g$ representation is captured by $\vec{\Delta}_2=\left( \Delta^1_2, \Delta^2_2\right)$. Both of them break only the cubic symmetry, but preserve time-reversal and inversion symmetries. The ordered phase represents either a time-reversal invariant insulator or a Dirac semimetal, about which more in a moment [see Sec.~\ref{SubSec:band-topology}]. Since five $\Gamma$ matrices transform as components of a rank-2 tensor under SO(3) rotations, the two nematic phases represent \emph{quadrupolar} orders, see Appendix~\ref{Append:Luttingerdetails}.

All ordered phases shown in the second line of Eq.~(\ref{Eq:excitonicOP}) break time-reversal symmetry and represent different magnetic orders. For example, $\Delta_3$ corresponds to an \emph{octupolar} order (since $\Gamma_{45} \sim J_x J_y J_z$), transforming under the singlet $A_{2u}$ representation. In a pyrochlore lattice of 227 iridates such an ordered phase represents the ``\emph{all-in all-out}" arrangement of electronic spin between two adjacent corner-shared tetrahedra~\cite{Savrasov, Balents3}. By contrast, ``\emph{two-in two-out}" or ``\emph{spin-ice}" magnetic orderings on a pyrochlore lattice are represented by a three-component vector $\vec{\Delta}_4=\left( \Delta^1_4, \Delta^2_4, \Delta^3_4 \right)$ (accounting for six possible two-in two-out arrangements in a single tetrahedron). Since $\Gamma_{45} \Gamma_{j} \sim 7 J_j - 4 J^3_j$ such an ordered phase contains a linear superposition of dipolar and octupolar moments, and transforms under the $T_{1u}$ representation~\cite{goswami-roy-dassarma}. Any other magnetic ordering can be represented by a six component vector $\Delta^{jk}_{5}$ with $j=1,2,3$ and $k=4,5$. No physical realization of such multi-component magnetic ordering in any material is currently known, and we do not delve into the discussion on such ordering for the rest of the paper.

            \subsection{Particle-particle or superconducting orders}~\label{SubSec:particle-particle_Nambu}

The effective single particle Hamiltonian in the presence of all possible momentum-independent or local or intra-unit cell superconducting orders reads~\cite{brydon-1, Herbut-3, savary-1, roy-nevidomskyy}
\begin{eqnarray}~\label{Eq:superconductingOP_Hamil}
H^{{\rm pair}}_{{\rm local}}= \int d^3{\bf r} \:\: \left( \Psi^\dagger_{\rm Nam} \: \hat{h}^{{\rm pair}}_{{\rm local}} \: \Psi_{\rm Nam} \right),
\end{eqnarray}
where 
\allowdisplaybreaks[4]
\begin{eqnarray}~\label{Eq:superconductingOP}
\hat{h}^{{\rm pair}}_{\rm local} &=& \left( \eta_1 \cos \phi +\eta_2 \sin \phi \right)  \bigg[ \overbrace{\Delta^{\rm p}_{A_{1g}} \Gamma_0}^{\rm s-wave} \nonumber \\
&+& \underbrace{\sum^3_{j=1} \Delta^{{\rm p},j}_{T_{2g}} \Gamma_j + \sum^5_{j=4} \Delta^{{\rm p},j}_{E_g} \Gamma_j }_{\rm d-wave}\bigg], 
\end{eqnarray} 
and $\phi$ is the global $U(1)$ superconducting phase. Any pairing proportional to $\eta_1 (\eta_2)$ preserves (breaks) time-reversal symmetry (recall that the time-reversal operator in the Nambu basis is ${\mathcal T}_{\rm Nam}=\eta_0 \Gamma_1 \Gamma_3 {\mathcal K}$). Here, $\Delta^{\rm p}_{A_{1g}}$ is the amplitude of the $s$-wave pairing, transforming under the $A_{1g}$ representation. The $s$-wave pairing breaks only the global $U(1)$ symmetry, but preserves the cubic symmetry. On the other hand, $\Delta^{{\rm p},j}_{T_{2g}}$ captures the amplitude of three $d$-wave pairings (for $j=1,2,3$) transforming under the $T_{2g}$ representation, and $\Delta^{{\rm p},j}_{E_g}$ for $j=4,5$ represents the amplitude of two $d$-wave pairings belonging to the $E_g$ representation. Notice $\{ \Gamma_j, \; j=1, \cdots, 5 \}$ can be expressed in terms of the product of two spin-3/2 matrices, and all five $d$-wave pairings break the cubic symmetry, while introducing a lattice distortion or electronic nematicity in the system. Hence, they stand as representatives of \emph{quadrupolar nematic superconductors}.


        \subsection{Reconstructed band structure and emergent topology}~\label{SubSec:band-topology}

\begin{figure}[t!]
\subfigure[]{\includegraphics[width=4cm,height=4cm]{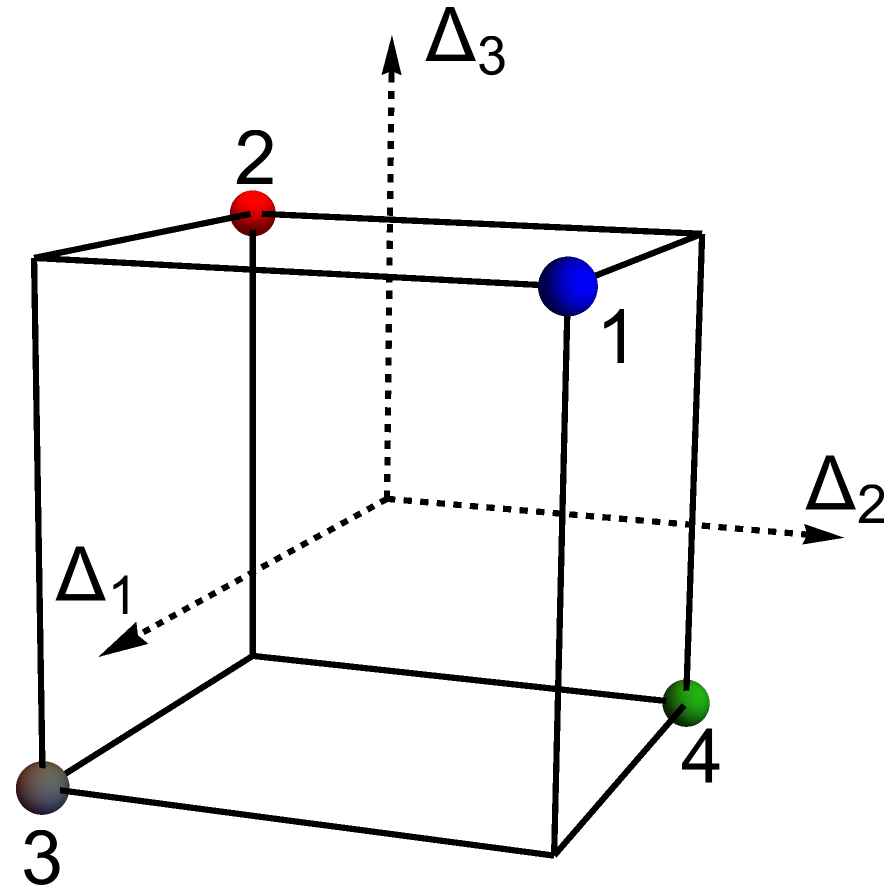}~\label{Fig:T2g_PhaseLocking}}
\subfigure[]{\includegraphics[width=4cm,height=4cm]{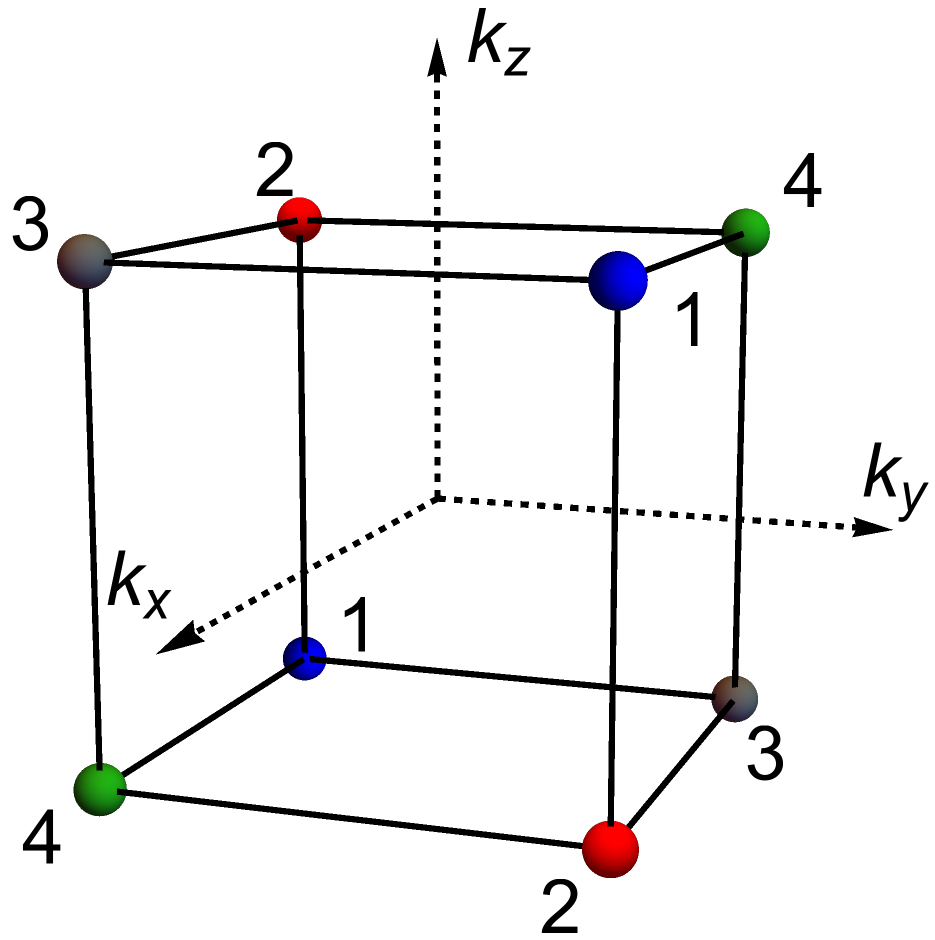}~\label{Fig:T2g_Diracpoints}}
\caption{ (a) Four phase lockings (blue, red, green and black dots) of the three component $T_{2g}$ nematic order, see Eq.~(\ref{Eq:T2gPhaselocking_DSM}), that yield gapless phase (topological Dirac semimetal) inside the ordered state. Corresponding locations of two Dirac points in momentum space are shown in panel (b). For any other generic phase locking, the ordered phase is a time-reversal symmetry preserving insulator. Dirac points are located along the body-diagonals ($C_{3v}$ axes in a cubic system).
}~\label{Fig:T2g-Topology}
\end{figure}

Next we consider the reconstructed band structure inside different BSPs which provides valuable information regarding the emergent topology inside ordered phases. The onset of any ordering discussed in the previous sections destabilizes the bi-quadratic touching and gives rise to either gapped or gapless quasiparticles (see below). Furthermore, this exercise will allow us to appreciate the energy-entropy competition among different orderings [see Sec.~\ref{SubSec:energy_Entropy}], which ultimately plays a decisive role in the organization of various phases in the global phase diagram of interacting Luttinger fermions.

1. $T_{2g}$ nematicity: The three component order-parameter for the $T_{2g}$ nematic phase gives birth to gapless quasiparticles for the following \emph{four} configurations
\begin{eqnarray}~\label{Eq:T2gPhaselocking_DSM}
\vec{\Delta}_1=\frac{|\Delta_1|}{\sqrt{3}} \bigg\{ 
\underbrace{\left(+,+,+ \right)}_{1},
\underbrace{\left(-,-,+ \right)}_{2},
\underbrace{\left(+,-,- \right)}_{3},
\underbrace{\left(-,+,- \right)}_{4} 
\bigg\}. \nonumber \\
\end{eqnarray}
These four phase lockings are respectively shown as blue, red, green and black points in Fig.~\ref{Fig:T2g_PhaseLocking}. The gapless phase corresponds to a \emph{topological} Dirac semimetal (since nematicity preserves the Kramers degeneracy of valence and conduction bands), similar to the ones recently found in Cd$_3$As$_2$~\cite{cdas:Exp} and Na$_3$Bi~\cite{nabi:Exp}. The DoS in a Dirac semimetal vanishes as $\varrho(E) \sim |E|^2$. The Dirac points are located along the body diagonals (the $C_{3v}$ axes) of a cubic system and respectively placed at 
\begin{eqnarray}
{\bf k}= \pm \bigg\{ \underbrace{\left( 1,1,1\right)}_{1}, \underbrace{\left( 1,1,-1\right)}_{2}, \underbrace{\left( 1,-1,1\right)}_{3}, \underbrace{\left( 1,-1,-1\right)}_{4}
\bigg\} k_0, \nonumber \\
\end{eqnarray}
where $k_0=\left[ 2 m_1 \Delta_1 /3 \right]^{1/2}$, as shown in Fig.~\ref{Fig:T2g_Diracpoints}. For any other phase locking within the $T_{2g}$ sector the system becomes an \emph{insulator}. The spectral gap in the insulating phase is \emph{anisotropic} and it is energetically superior over the gapless Dirac semimetal phase.

\begin{figure}[t!]
\subfigure[]{\includegraphics[width=4cm,height=4cm]{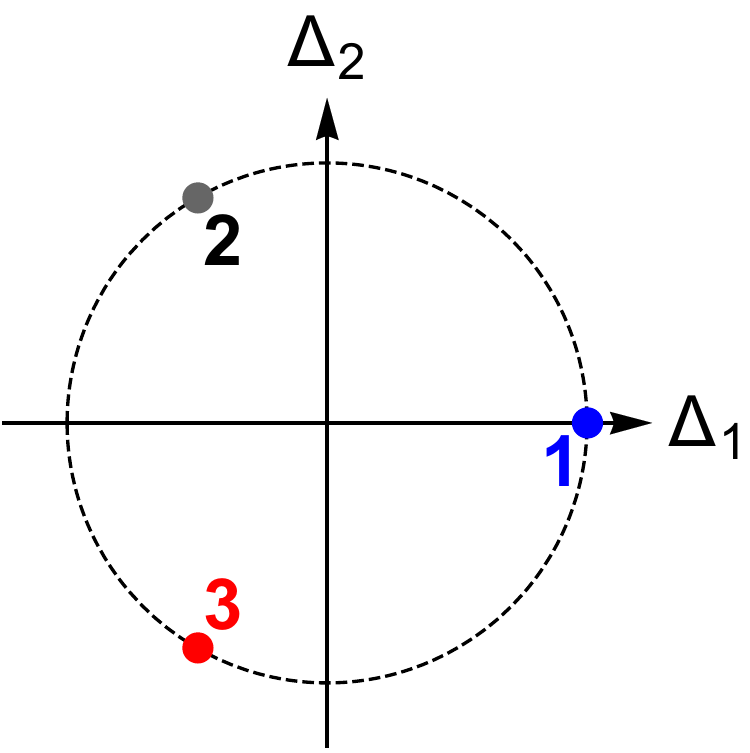}~\label{Fig:Eg_PhaseLocking}}
\subfigure[]{\includegraphics[width=4cm,height=4cm]{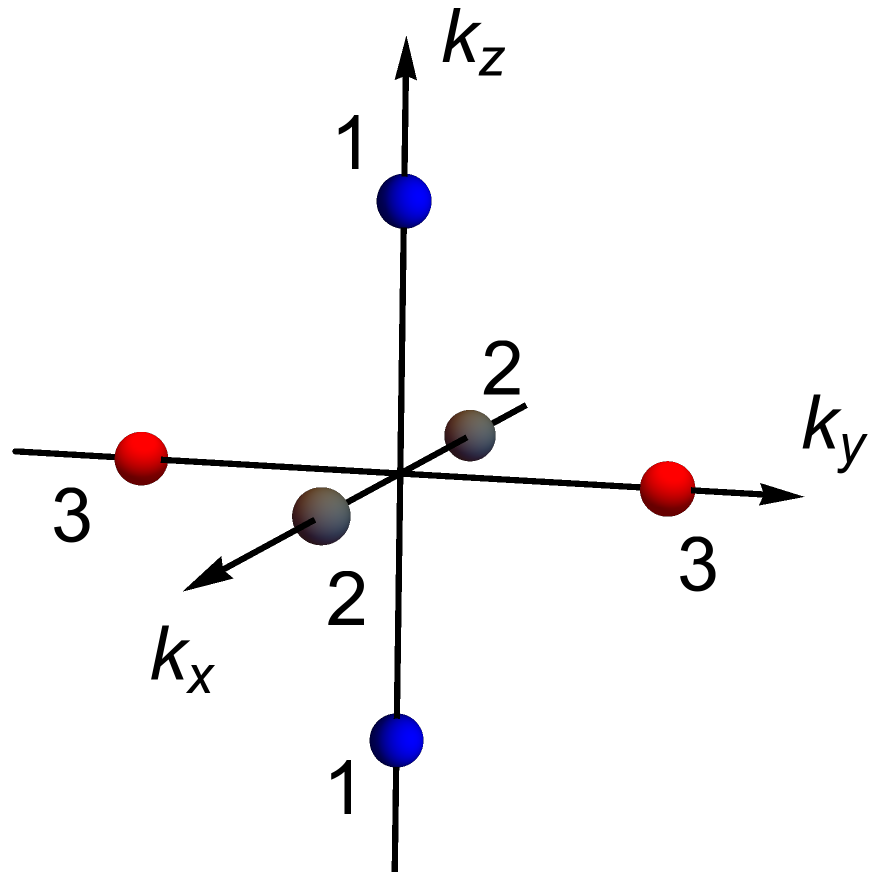}~\label{Fig:Eg_Diracpoints}}
\caption{ (a) Three possible phase lockings (shown by blue, black and red dots) of two-component $E_g$ nematic order, see Eq.~(\ref{Eq:EgPhaselocking_DSM}), that give rise to topological Dirac semimetal inside the ordered phase. The Dirac points inside the $E_g$ nematic phase are located on the $k_z$, $k_x$ and $k_y$ axes ($C_{4v}$ axes in a cubic system), as shown in panel (b), in contrast to the situation inside the $T_{2g}$ nematic phase, see Fig.~\ref{Fig:T2g_Diracpoints}. For any other phase locking the system is an insulator.     
}~\label{Fig:Eg-Topology}
\end{figure}

2. $E_{g}$ nematicity: The two component $E_g$ nematic order is most conveniently described in terms of the following parametrization
\begin{equation}
\vec{\Delta}_2= \frac{|\Delta_2|}{\sqrt{2}} \left( \sin \phi_{_{\rm E_g}}, \: \cos \phi_{_{\rm E_g}}  \right),
\end{equation}
where $\phi_{_{\rm E_g}}$ is the internal angle in the order-parameter space. Only for 
\begin{equation}~\label{Eq:EgPhaselocking_DSM}
\phi_{_{\rm E_g}}= \left\{\underbrace{0}_1, \:\: \underbrace{2 \pi/3}_{2}, \:\: \underbrace{4 \pi/3}_{3} \right\}
\end{equation}
the quasi-particle spectra are gapless, as shown in Fig.~\ref{Fig:Eg_PhaseLocking}, and the ordered phase represents a topological Dirac semimetal. Specifically, for $\phi_{_{\rm E_g}}=0, 2 \pi/3, 4 \pi/3$, the Dirac points are respectively located on $k_z$, $k_x$ and $k_y$ axes (the $C_{4v}$ axes), see Fig.~\ref{Fig:Eg_Diracpoints}, and the separation of two Dirac points is given by $2 k_0$, where $k_0=\left[ 2 m_2 \Delta_2/\sqrt{2} \right]^{1/2}$. For any other phase locking within the $E_g$ sector, the system becomes an insulator. 
Recently, it was shown that the gapless phases in both $T_{2g}$ and $E_g$ nematic phases correspond to higher-order Dirac semimetals supporting one-dimensional hinge modes, whereas the insulating phases accommodate a \emph{mixed} topology~\cite{szabo-moessner-roy}.

\begin{figure}[t!]
\subfigure[]{\includegraphics[width=4cm,height=4cm]{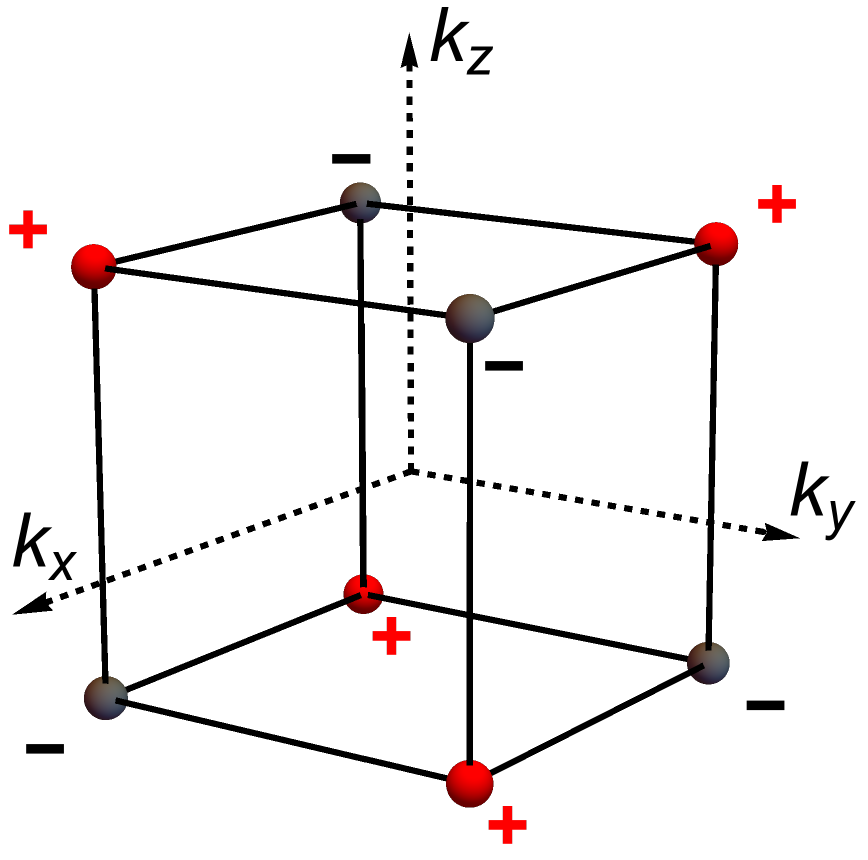}~\label{Fig:AIAO_Spectra}}
\subfigure[]{\includegraphics[width=4cm,height=4cm]{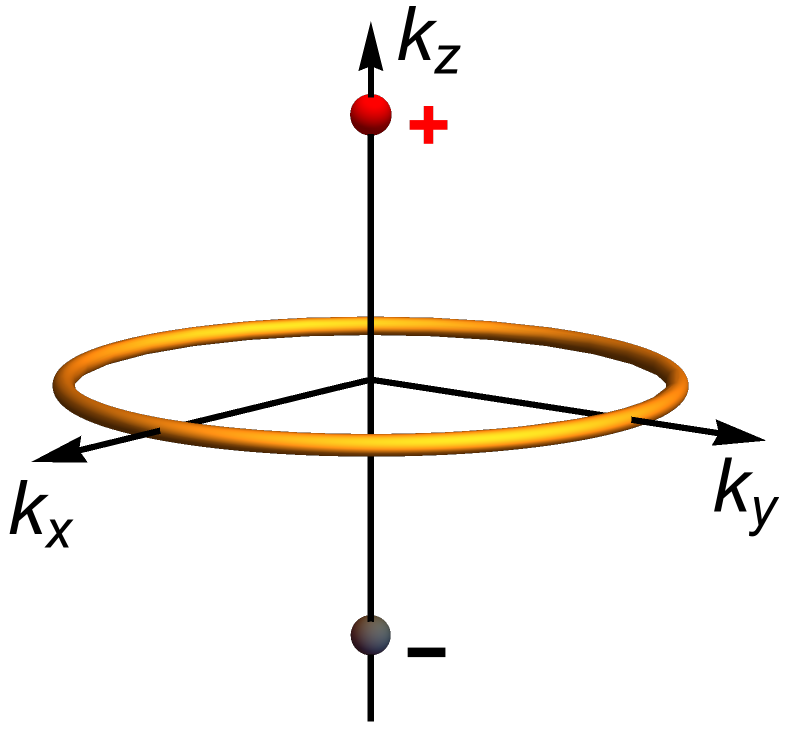}~\label{Fig:Ice_Spectra}}
\caption{(a) Location of eight Weyl nodes in the presence of $A_{2u}$ octupolar order, which in a pyrochlore lattice corresponds to the all-in all-out magnetic order. Here, red and gray dots respectively correspond to the source ($+$) and sink ($-$) of Abelian Berry curvature, with monopole charge $\pm 1$. (b) Nodal structure in the presence of a uniaxial $T_{1u}$ order (when its moment points along $\hat{z}$), supporting two Weyl nodes separated along $\hat{k}_z$ (red and gray dots) and a nodal loop (dark yellow ring) in the $k_x-k_y$ plane. When the moment of the uniaxial $T_{1u}$ order points along $\hat{x}$ and $\hat{y}$ direction, the Weyl nodes are respectively separated along the $k_x$ and $k_y$ axes, and the nodal-loops are found in the $k_y-k_z$ and $k_x-k_z$ plane. By contrast, inside a \emph{triplet} $T_{1u}$ phase, only two Weyl nodes are found along one of the body diagonals.  
}
\end{figure}

3. $A_{2u}$ magnet: In the presence of an octupolar $A_{2u}$ ordering, the two-fold degeneracy of the valence and conduction band gets lifted and a pair of Kramers non-degenerate bands touch each other at the following \emph{eight} points in the Brillouin zone [see Fig.~\ref{Fig:AIAO_Spectra}] 
\begin{equation}
{\bf k}=\left( \pm 1, \pm 1, \pm 1 \right) \; k_0,
\end{equation}
where $k_0=\sqrt{2 m_1 \Delta_3/3}$. They represent simple Weyl points, which act as source (4 of them) and sink (4 of them) of Abelian Berry curvature of unit strength. However, due to an octupolar arrangement of the Weyl nodes, the net Berry curvature through any high-symmetry plane is precisely \emph{zero} and this phase does not support any anomalous Hall effect. The DoS at low energies then scales as $\varrho(E) \sim |E|^2$~\cite{Savrasov, Balents3, goswami-roy-dassarma}.

4. $T_{1u}$ magnet: For each component of $T_{1u}$ magnetic order (represented by the matrix operator $\Gamma_{45} \Gamma_j$ with $j=1,2,3$) the ordered phase supports \emph{two} Weyl nodes along one of the $C_{4v}$ axes and a \emph{nodal-loop} in the corresponding basal plane. For example, when $\langle \Psi^\dagger \Gamma_{45} \Gamma_3 \Psi \rangle \equiv \Delta^3_4\neq 0$ the left and right chiral Weyl nodes are located at $(0,0,\pm k_0)$ where $k_0=\sqrt{2 m_2 \Delta_4}$ and a nodal-loop is found in the $k_x-k_y$ plane, as shown in Fig.~\ref{Fig:Ice_Spectra}. Similarly, for $j=1$ and $2$ the Weyl nodes are separated along the $k_x$ and $k_y$ axes, and the nodal-loops are respectively found in the $k_y-k_z$ and $k_x-k_z$ planes. Due to the presence of two Weyl nodes, each configuration of two-in two-out magnetic order supports a finite anomalous Hall effect in the plane perpendicular to the separation of the Weyl nodes. However, any \emph{triplet} magnetic order, represented by $\vec{\Delta}_4=|\Delta_4|(\pm 1, \pm 1, \pm 1)/\sqrt{3}$, gets rid of the nodal loop and supports only \emph{two} Weyl nodes along one of the body-diagonals ($C_{3v}$ axes). Hence, triplet $T_{1u}$ magnetic orders are energetically favored over their uniaxial counterparts~\cite{goswami-roy-dassarma}.~\footnote{The low energy DoS in the presence of a nodal loop and two point nodes (due to a uniaxial $T_{1u}$ order) is dominated by the former and scales as $\varrho(E) \sim |E|$, while in a triplet $T_{1u}$ state the DoS scales as $\varrho(E) \sim |E|^2$ (due to the point nodes). Hence, formation of the triplet ordering causes power-law suppression of the DoS and increases the condensation energy gain. }

5. $A_{1g}$ or $s$-wave pairing: Notice that the matrix operator representing an $s$-wave pairing fully anti-commutes with the Luttinger Hamiltonian (for any value of $\alpha$) and thus corresponds to a \emph{mass} for Luttinger fermions. The quasiparticle spectra inside the paired state is fully gapped, but the phase is topologically \emph{trivial}.

\begin{figure}[t!]
\subfigure[]{
\includegraphics[width=4cm,height=3.5cm]{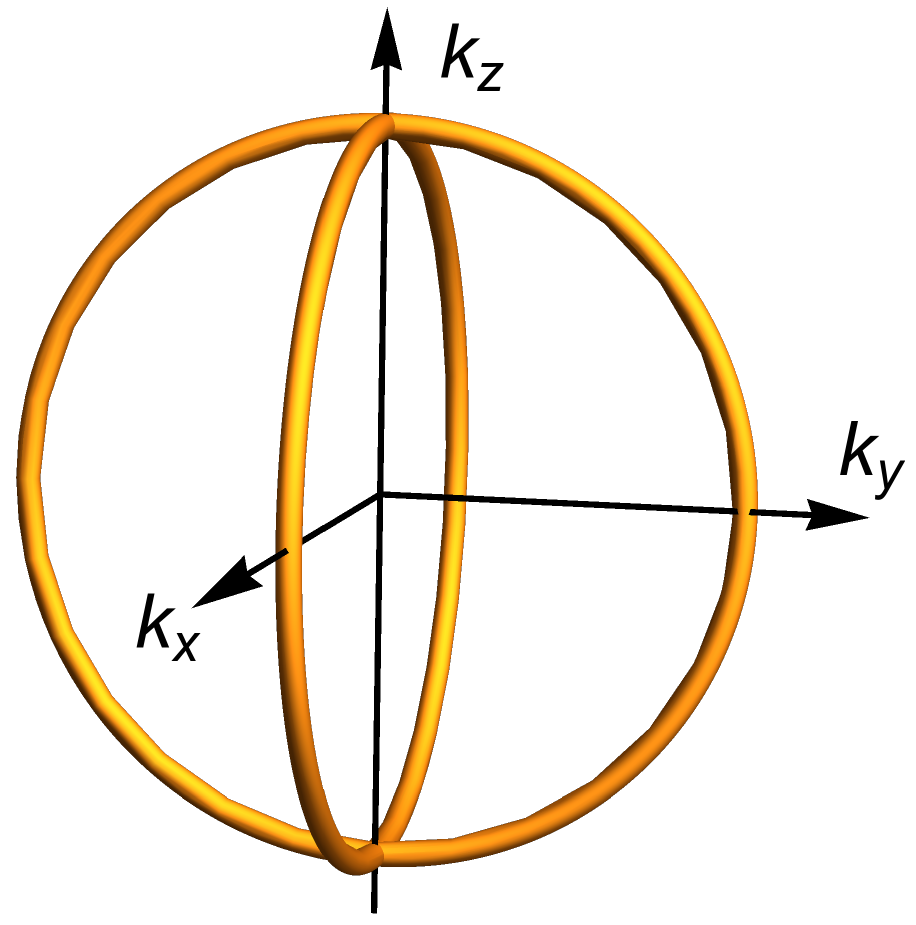}~\label{SubFig:dxypairing_loops}
}
\subfigure[]{
\includegraphics[width=4cm,height=3.5cm]{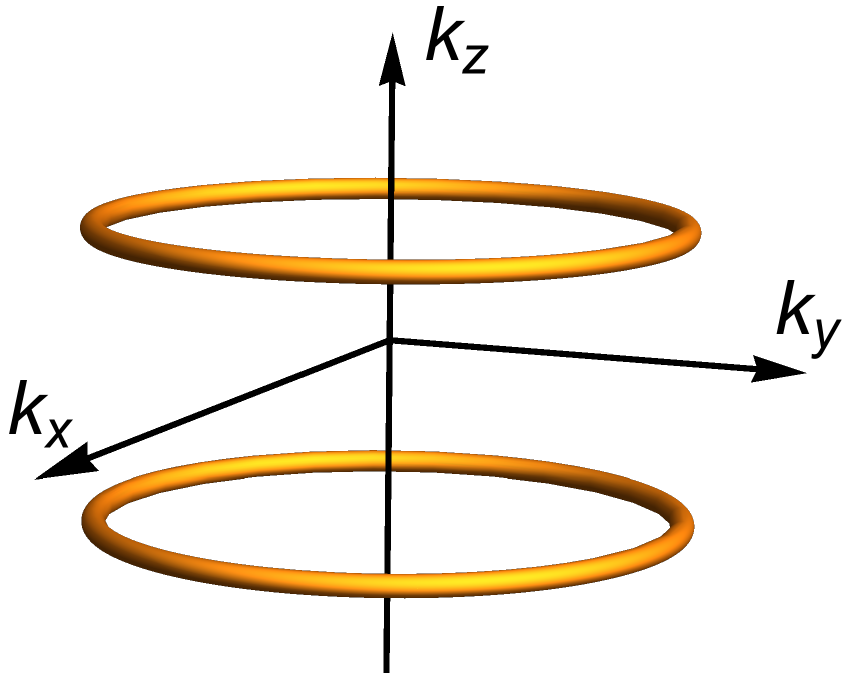}~\label{SubFig:d3zpairing_loops}
}
\caption{Structure of two nodal loops in the presence of an underlying (a) $d_{xy}$ and (b) $d_{3z^2-r^2}$ pairings. These two pairings are respectively represented by $\Gamma_3$ and $\Gamma_5$ matrices for the Luttinger fermions. Note that the nodal loops for $d_{yz} (\Gamma_1)$, $d_{xz}(\Gamma_2)$, $d_{x^2-y^2} (\Gamma_4)$ pairings can be obtained by rotating the ones shown here for $d_{xy}$ pairing about suitable momentum axes. However, the two nodal loops for the $d_{3z^2-r^2}$ pairing are disjoint from the remaining ones (note these two nodal loops do not cross each other), see Sec.~\ref{SubSec:band-topology} and Table~\ref{Tab:Spectra_dwaves}.   
}~\label{Fig:NodalLoops_dwaves}
\end{figure}

6. $T_{2g}$ pairing: Three $d$-wave pairings, proportional to $\Gamma_1$, $\Gamma_2$ and $\Gamma_3$ matrices, belong to the $T_{2g}$ representation and respectively possess the symmetry of $d_{yz}$, $d_{xz}$ and $d_{xy}$ pairings. Each component supports two nodal loops in the ordered phase, as shown in the first three rows of Table~\ref{Tab:Spectra_dwaves}~\cite{brydon-1, roy-nevidomskyy}. Two nodal loops for the $\Gamma_3$ or $d_{xy}$ pairing are shown in Fig.~\ref{SubFig:dxypairing_loops}. The two nodal loops for $\Gamma_1$ or $d_{yz}$ and $\Gamma_2$ or $d_{xz}$ pairings can respectively be obtained by rotating the ones shown for $d_{xy}$ pairing by $\frac{\pi}{2}$, with respect to the $k_y$ and $k_x$ axes.

7. $E_{g}$ pairing: $E_g$ pairings proportional to $\Gamma_4$ and $\Gamma_5$ matrices respectively possess the symmetry of $d_{x^2-y^2}$ and $d_{3z^2-r^2}$ pairings and each of them supports two nodal loops, as shown in the last two rows of Table~\ref{Tab:Spectra_dwaves}~\cite{brydon-1, roy-nevidomskyy}. Note that two nodal loops for the $d_{x^2-y^2}$ pairing can be realized by rotating the ones for the $d_{xy}$ pairing by $\frac{\pi}{4}$ about the $k_z$ axis. However, two nodal loops for the $d_{3z^2-r^2}$ pairing, shown in Fig.~\ref{SubFig:d3zpairing_loops}, cannot be rotated into the ones for $d_{x^2-y^2}$ pairing. Therefore, despite the fact that the $d_{3z^2-r^2}$ and $d_{x^2-y^2}$ pairings belong to the same $E_g$ representation, they are not energetically degenerate~\cite{roy-nevidomskyy, sigrist-RMP}. Since the radius of the nodal loops for the $d_{3z^2-r^2}$ pairing is the \emph{smallest}, this paired state is the energetically most favorable among five $d$-wave pairings.~\footnote{Even though $d+id$ type, such as $d_{x^2-y^2}+id_{3z^2-r^2}$, pairing can eliminate nodal loops from the  quasiparticle spectra in favor of point nodes around which $\varrho(E) \sim |E|^2$ in a single band Fermi liquid~\cite{sigrist-RMP}, the strong inter-band coupling causes inflation of such nodes in doped LSM and yields Fermi surface of BdG quasiparticles, leading to a constant DoS at lowest energy, followed by $\varrho(E) \sim |E|^2$ a higher energies~\cite{brydon-2}. Presently, it is not very clear between (a) individual $d$-wave pairings and (b) $d+id$ type pairings, which one is energetically more advantageous. However, based on the power-law scaling of DoS, we expect individual $d$-wave pairings to be energetically favored over $d+id$ type pairings at least when the inter-band coupling is strong, which is the case when pairing results from pure Hubbardlike repulsive interactions. This conclusion is in accordance with the organizing principle discussed in Sec.~IIB1 and the energy-entropy argument, summarized in Fig.~\ref{Fig:energyentropy}. The $|E|$-linear DoS for individual $d$-wave parings (stemming from the underlying nodal loops) results in a $T$-linear scaling of the penetration depth, as observed in YPtBi~\cite{Exp:Paglione-2}.}

\begin{table}[t!]
\renewcommand{\arraystretch}{1.4}
\begin{tabular}{|c c c c|}
\hline
Pairing & IREP. & Equations for nodal loops & Symmetry \\
\hline
$\Gamma_1$ & $T_{2g}$ & $k^2_x+k^2_y=2 m \Delta$, $k^2_x+k^2_z=2 m \Delta$ & $d_{yz}$ \\
\rowcolor{RowColor}
$\Gamma_2$ & $T_{2g}$ & $k^2_x+k^2_y=2 m \Delta$, $k^2_z+k^2_y=2 m \Delta$ & $d_{xz}$ \\
$\Gamma_3$ & $T_{2g}$ & $k^2_y+k^2_z=2 m \Delta$, $k^2_x+k^2_z=2 m \Delta$ & $d_{xy}$ \\
\rowcolor{RowColor}
$\Gamma_4$ & $E_{g}$ & $k^2_z+k^2_\perp=2 m \Delta$, $k_x=\pm k_y$ & $d_{x^2-y^2}$ \\
$\Gamma_5$ & $E_{g}$ & $k^2_\perp=4 m \Delta/3$, $k_z=\pm k_\perp/\sqrt{2}$ & $d_{3z^2-r^2}$ \\
\hline
\end{tabular}
\caption{ The structure of two nodal loops in the presence of five individual $d$-wave pairings, belonging to the $T_{2g}$ and $E_g$ representations, where $k^2_\perp =k^2_x+k^2_y$. We display the symmetry of each $d$-wave pairing in the proximity to the Fermi surface (realized on the conduction or valence band) in the last column. Note that two nodal loops for $d_{xy}$, $d_{xz}$, $d_{yz}$ and $d_{x^2-y^2}$ parings can be rotated into each other, while those in the presence of $d_{3z^2-r^2}$ pairing are disconnected from the remaining ones, see Fig.~\ref{Fig:NodalLoops_dwaves}. For the sake of simplicity we here assume $m_1=m_2=m$, for which the nodal loops are \emph{circular} in shape. For $m_1 \neq m_2$, the nodal loops become \emph{elliptic}. Here, $\Delta$ is the amplitude of $d$-wave pairings.  }~\label{Tab:Spectra_dwaves}
\end{table}

            \subsection{Energy and Entropy Inside Ordered Phases}~\label{SubSec:energy_Entropy}

From the computation of the reconstructed band structure we can gain insight into the condensation energy ($\Delta_F$) and entropy ($\Delta_S$) inside the ordered phases. While the \emph{stiffness} of the spectral gap measures the gain of condensation energy, the scaling of the DoS at low-energies (due to gapless quasiparticles) measures the entropy. Recall that the $s$-wave pairing yields fully gapped spectra (isotropic), while the nematic orders produce either an anisotropic gap or gapless quasiparticles. Hence, the former ordering is associated with higher (lower) gain in condensation energy (entropy). On the other hand, the DoS vanishes as $\varrho(E) \sim |E|$ and $|E|^2$ respectively in the presence of a nodal-loop and Dirac or Weyl points. We found that the $A_{2u}$ magnetic order gives birth to eight Weyl nodes, while only two Weyl nodes can be found inside the \emph{triplet} $T_{1u}$ magnetic order. By contrast, all five $d$-wave pairings are accompanied by two nodal loops (see Table~\ref{Tab:Spectra_dwaves}). Therefore, we can organize these ordered phases according to their contribution to (a) condensation energy and (b) entropy gain, as shown in Fig.~\ref{Fig:energyentropy}. The LSM, on the other hand, accommodates the largest amount of gapless fermionic excitations near zero energy, where the DoS vanishes as $\varrho(E) \sim \sqrt{E}$. Hence, the LSM is endowed with largest entropy.~\footnote{Such an organization of ordered phases according to their contributions to the gain of condensation energy and entropy is purely based on the power-law dependence of low-energy DoS or the stiffness (isotropic or anisotropic) of the spectral gap. This procedure, however, cannot distinguish two phases with similar scaling of the DoS, such as between $A_{2u}$ and triplet $T_{1u}$ magnetic orders (producing Weyl nodes), or the stiffness of the spectral gap, such as between $T_{2g}$ and $E_g$ nematic orders (producing anisotropic gaps). A more microscopic analysis is needed to resolve these situations. }

We now present a simple prescription to estimate (at least qualitatively) the gain of condensation energy or entropy. Let us assume that the eight-dimensional Hermitian matrix $M$, entering the definition of an order parameter as $\Delta_M \equiv \langle \Psi^\dagger_{\rm Nam} M \Psi_{\rm Nam}\rangle$ respectively anticommutes and commutes with $N^{M}_{\rm anti}$ and $N^{M}_{\rm comm}$ number of matrices appearing in the Luttinger Hamiltonian $\hat{h}^{\rm Nam}_{\rm L}({\bf k})$ (when $\mu=0$), and thus   
\begin{equation}~\label{Eq:matrix_energy_entropy}
N^{M}_{\rm anti} + N^{M}_{\rm comm}=5.
\end{equation} 
Therefore, for the $s$-wave pairing, two nematic orders, two magnetic orders and five $d$-wave pairings $N^{M}_{\rm anti}=5,4,2,1$, while $N^{M}_{\rm comm}=0,1,3,4$, respectively. Then for any ordered phase      
\begin{equation}
\Delta_F \sim N^{M}_{\rm anti}, \quad 
\Delta_S \sim N^{M}_{\rm comm}.
\end{equation}
This correspondence can be anchored from a simple example. Let us choose $s$-wave pairing, for which $N^{M}_{\rm anti}=5, N^{M}_{\rm comm}=0$ and chemical potential, for which $N^{M}_{\rm anti}=0, N^{M}_{\rm comm}=5$, as two perturbations in a LSM. While the $s$-wave pairing yields an isotropic gap, a finite chemical doping creates a Fermi surface (producing a constant DoS). Consequently, the $s$-wave pairing (chemical doping) is accompanied by larger gain of condensation energy (entropy). Therefore, the analysis of reconstructed band structure and emergent topology inside BSPs allows us to organize them according to the gain of condensation energy and entropy. The RG analysis at finite temperature (in the regime where the dimensionless temperature $t=2 m T/\Lambda^2 \ll 1$) captures such energy-entropy competition, which we discuss in Sec.~\ref{SubSec:FiniteTRG}, see also Fig.~\ref{Fig:FiniteT_Summary}.

\begin{figure}[t!]
\includegraphics[width=6.5cm,height=2.75cm]{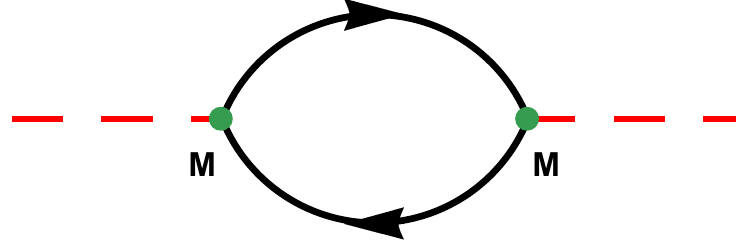}
\caption{The Feynamn diagram contributing to the leading order bare susceptibility ($\chi_{_M}$), given in Eq.~(\ref{Eq:baresusceptibility}), for zero external momentum and frequency. The \emph{red} dashed (\emph{black} solid) lines represent the order parameter (fermionic) fields. The two vertices (\emph{green} dots) are accompanied by the appropriate eight-dimensional Hermitian order-parameter matrix $M$, appearing in the corresponding fermion bilinear $\Psi^\dagger_{\rm Nam} M \Psi_{\rm Nam}$.   
}~\label{Fig:baresusceptibility}
\end{figure}

The hierarchy of the energy and entropy gains inside the ordered phases changes when the chemical potential is placed away from the band touching point (i.e., $\mu \neq 0$). Since any pairing operator \emph{anti-commutes} with the number operator ($\hat{N}=\eta_3 \Gamma_0$), superconducting orders maximally gap (either fully by the $s$-wave pairing or partially by the individual $d$-wave pairings) the Fermi surface. By contrast, any excitonic order always gives birth to a Fermi surface, according to the Luttinger theorem~\cite{Luttinger_Theorem}. Hence, at finite doping all superconductors are energetically superior over the particle-hole orders, while excitonic orders are accompanied by larger entropy (due to presence of a Fermi surface). The energy-entropy competition at finite-$\mu$ is also captured by the RG analysis, discussed in Sec.~\ref{SubSec:LuttingerMetalRG}, leading to the phase diagrams shown in Figs.~\ref{Fig:swave_IsotropicLSM}, ~\ref{Fig:dwaves_IsotropicLSM} and ~\ref{Fig:GlobalPD_Intro_FS}.

\section{Electron-electron interactions}~\label{Sec:e-e_Interaction}

Next we proceed to demonstrate the onset of various BSPs, discussed in the previous section, triggered by \emph{repulsive} (at the bare level) electron-electron interactions. As mentioned earlier we will focus only on the local or short-range part of Coulomb interaction and neglect its long-range tail. For the sake of concreteness we assume that the local interactions are density-density in nature. Any generic local density-density interaction (such as the ones appearing in an extended Hubbard model, for example) can be captured by \emph{six} quartic terms and the corresponding interacting Hamiltonian reads
\begin{eqnarray}~\label{Eq:Interaction_Hamiltonian}
H_{\rm int} &=& -\bigg[ \lambda_0 (\Psi^\dag \Psi)^2 + \lambda_1 \sum^3_{j=1} \left( \Psi^\dag \Gamma_{j} \Psi \right)^2 \nonumber \\
&+& \lambda_2 \sum^5_{j=4} \left( \Psi^\dag \Gamma_{j} \Psi \right)^2 
+ \lambda_3 (\Psi^\dag \Gamma_{45} \Psi)^2 \\
&+& \lambda_4 \sum^3_{j=1}\left( \Psi^\dag \Gamma_j \Gamma_{45} \Psi \right)^2
+ \lambda_5 \sum^3_{j=1} \sum^5_{k=4} \left( \Psi^\dag \Gamma_{jk} \Psi \right)^2  \bigg]. \nonumber
\end{eqnarray} 
In this notation $\lambda_j >0$ corresponds to repulsive interaction. However, all four-fermion interactions are \emph{not} linearly independent due to the existence of \emph{Fierz identity} among sixteen four-dimensional Hermitian matrices, closing a $U(4)$ Clifford algebra [see Appendix~\ref{Append:Fierz}]~\cite{HJR-Fierz, RoySDS-Fierz, RGJ-Fierz}. It turns out that any generic local interaction can be expressed in terms of only \emph{three} quartic terms, and we conveniently (without any loss of generality) choose them to be $\lambda_0$, $\lambda_1$ and $\lambda_2$. Following the Fierz relations we can express local quartic terms proportional to $\lambda_3$, $\lambda_4$, $\lambda_5$ as linear combinations of above three, see Eq.~(\ref{EqAppend:FierzRelations}). Whenever we generate four-fermion interactions proportional to $\lambda_{3,4,5}$ during the coarse-graining (discussed in Sec.~\ref{Sec:RG}), they can immediately be expressed in terms of $\lambda_{0,1,2}$, and the interacting model defined in terms of $\lambda_{0,1,2}$ [see Eq.~(\ref{Eq:Action_Int}) below] always remains closed under the RG procedure to any order in the perturbation theory.

The imaginary time Euclidean action for the interacting system is given by 
\begin{eqnarray}~\label{Eq:Action_Int}
S_{\rm int} &=& S_0 -\int d\tau d^d{\bf r} \bigg[ \lambda_0 (\Psi^\dag \Psi)^2 + \lambda_1 \sum^3_{j=1} \left( \Psi^\dag \Gamma_{j} \Psi \right)^2 \nonumber \\
&+& \lambda_2 \sum^5_{j=4} \left( \Psi^\dag \Gamma_{j} \Psi \right)^2  \bigg],
\end{eqnarray}
where $\Psi \equiv \Psi (\tau, {\bf r})$ and $\Psi^\dagger \equiv \Psi^\dagger (\tau, {\bf r})$. Under the rescaling of space-time(imaginary) coordinates and the fermionic fields, see Eq.~(\ref{Eq:rescaling}), the local four-fermion interaction scales as
\begin{equation}
\lambda_j \to e^{(d-z)\ell} \lambda_j. 
\end{equation} 
Hence the scaling dimension of quartic couplings is $[\lambda_j]=z-d$. For $z=2$ and $d=3$, $[\lambda_j]=-1$, and any weak local interaction is an \emph{irrelevant} perturbation and leaves the Luttinger fermions unaffected. Therefore, any ordering sets in at an intermediate strength of coupling through QPT. In Sec.~\ref{Sec:RG} we demonstrate appearances of various BSPs using a RG analysis, controlled via an $\epsilon$-expansion, where $\epsilon=d-2$, about the \emph{lower-critical} two spatial dimension of this theory. Within the framework of the $\epsilon$-expansion, such QPTs take place at a critical interaction strength $\lambda^\ast_j \sim \epsilon$, and in three spatial dimensions ($d=3$) $\epsilon=1$. Before proceeding to the RG analysis, we seek to gain some insight into the propensity toward various orderings by computing the corresponding mean-field susceptibility for a wide range of the mass anisotropy parameter ($\alpha$). Readers interested in the RG analysis may skip the following discussion and directly go to Sec.~\ref{Sec:RG}.     

\begin{figure}[t!]
\includegraphics[width=8.5cm,height=6cm]{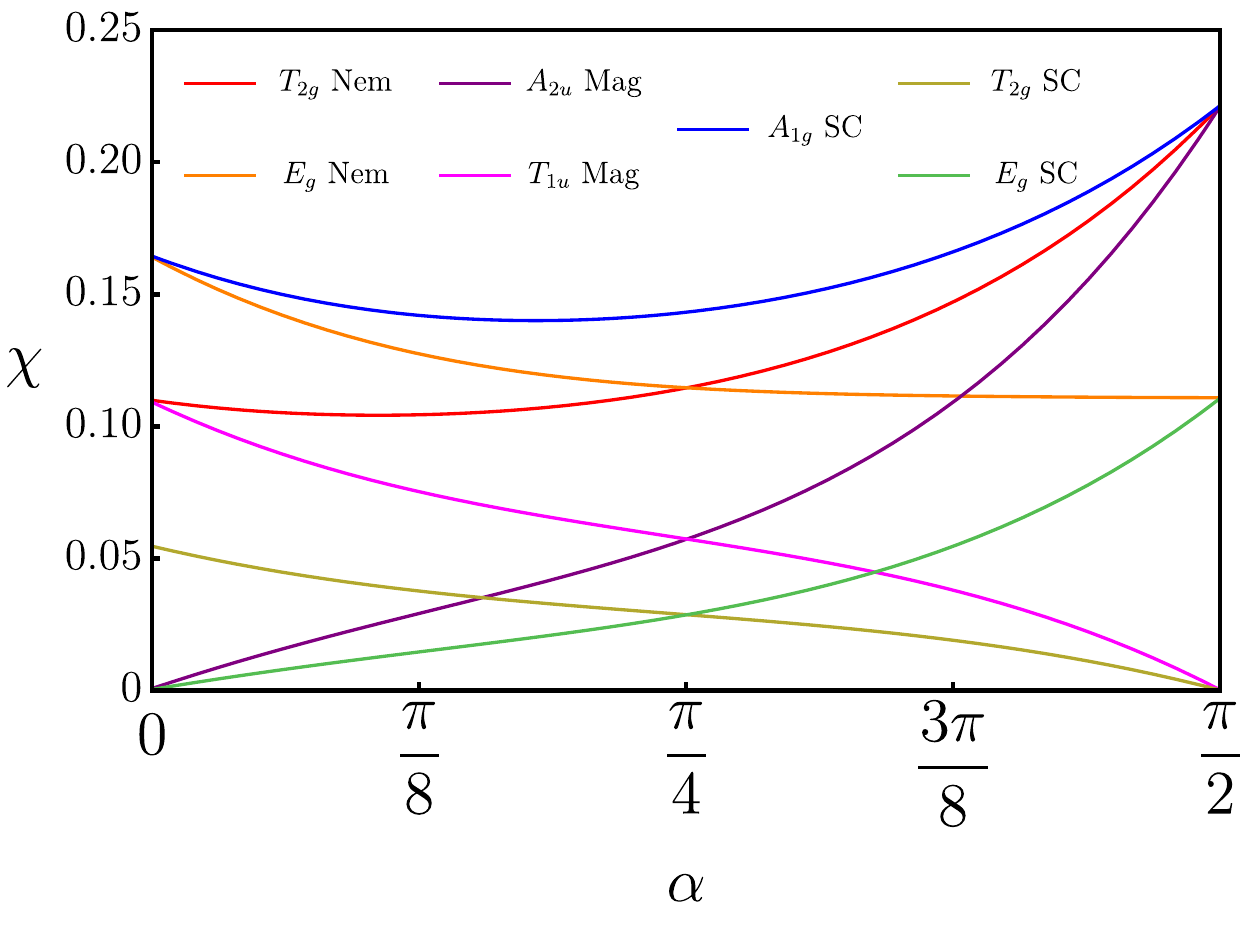}
\caption{ Bare mean-field susceptibility [see Eq.~(\ref{Eq:baresusceptibility})] for zero external momentum and frequency [see Fig.~\ref{Fig:baresusceptibility} for the relevant Feynman diagram] for various orderings at zero temperature and chemical doping, as a function of $\alpha$ [parametrizing the anisotropy between the mass parameters in the $T_{2g}$ and $E_g$ orbitals]. Here, $\chi$ is measured in units of $m \Lambda$. For $\alpha=\frac{\pi}{4}$ two nematic orders and two $d$-wave pairings (belonging to the $T_{2g}$ and $E_g$ representations) possess equal susceptibilities, and so do two magnetic orders (within the $A_{2u}$ and $T_{1u}$ representations). The $A_{1g}$ $s$-wave pairing always possesses the largest susceptibility (for any $\alpha$) as it represents a \emph{mass} for spin-3/2 fermions. 
Susceptibilities for the $s$-wave pairing, $T_{2g}$ nematic and $A_{2u}$ magnetic orders display exact degeneracy as $\alpha \to \frac{\pi}{2}$, when all of them become \emph{mass} [see Sec.~\ref{SubSubSec:Susceptibility_piover2}]. On the other hand, as $\alpha \to 0$ the $s$-wave pairing and $E_g$ nematicity become \emph{mass} and their bare susceptibilities are degenerate and largest [see Sec.~\ref{SubSubSec:Susceptibility_0}]. For detailed discussion consult Sec.~\ref{SubSec:susceptibility}.  
}~\label{Fig:susceptibility_NoFS}
\end{figure}

                 \subsection{Mean-field susceptibility}~\label{SubSec:susceptibility}

To gain insights into the propensity toward the formation of various orderings, we first compute the bare mean-field susceptibility ($\chi_{_M}$) of all possible symmetry allowed fermionic bilinears $\Psi_{\rm Nam}^\dagger M \Psi_{\rm Nam}$, where $M$ is an eight dimensional Hermitian matrix (see Sec.~\ref{Sec:BSPs}). For simplicity we set $\mu=0$. For zero external momentum and frequency this quantity is given by 
\begin{equation}~\label{Eq:baresusceptibility}
\chi_{_M} =- \frac{1}{2}\int \frac{d^3 {\bf k}}{(2 \pi)^3} \sum_{i\omega_n} {\bf Tr} \left[ M G_{\bf k} (i \omega_n) M G_{\bf k} (i \omega_n)\right]. 
\end{equation} 
The relevant Feynman diagram is shown in Fig.~\ref{Fig:baresusceptibility} and the ``$-$" sign arises from the fermion bubble. Here $G_{\bf k}(i \omega_n)$ is the fermionic Green's function in the Nambu doubled basis. The factor of $1/2$ takes care of the artificial Nambu doubling. Results are displayed in Fig.~\ref{Fig:susceptibility_NoFS}. Next we discuss the scaling of $\chi_{_M}$ in different channels for a few specific values of the mass anisotropy parameter.

\subsubsection{ \bf Isotropic Luttinger Semimetal ($\alpha=\frac{\pi}{4}$)}~\label{SubSubSec:Susceptibility_isotropic}

For $\alpha=\frac{\pi}{4}$, the effective masses for the $T_{2g}$ and $E_g$ orbitals are equal (i.e. $m_1=m_2$) and the system enjoys an enlarged spherical symmetry. Since each one of the five $\Gamma$-matrices (representing two nematic orders) anti-commutes with four matrices and commutes with one matrix appearing in the Luttinger Hamiltonian, two nematic orders belonging to the $T_{2g}$ (red curve) and $E_g$ (orange curve) representations possess \emph{equal} susceptibility. On the other hand, all ten commutators (representing various magnetic orders) anti-commute with three and commute with two matrices appearing in this model. Hence, magnetic orders in the $A_{2u}$ (purple curve) and $T_{1u}$ (magenta curve) channels also possess equal susceptibility. Two copies of the $d$-wave pairing, transforming under the $T_{2g}$ (dark green curve) and $E_g$ (dark yellow curve) representations, have degenerate susceptibilities, as all five $d$-wave pairing matrices commute with four matrices and anti-commute with only one matrix appearing in the Luttinger model. As the $s$-wave pairing fully anti-commutes with the Luttinger Hamiltonian, it always possesses the largest susceptibility (blue curve) for any $\alpha$.

Susceptibilities for different BSPs (characterized by the fermion bilinear $\Psi_{\rm Nam}^\dagger M \Psi_{\rm Nam}$) $\sim N^M_{\rm anti}$, the number of matrices in $\hat{h}^{\rm Nam}_{\mathrm L} ({\bf k})$ anti-commuting with $M$. This simple correspondence is operative irrespective of the choice of $\alpha$. Note that in an isotropic system $N_{\rm anti}^M=5, 4, 3$ and $1$ for the $s$-wave pairing, nematicity, magnetic orders and $d$-wave pairings, respectively, hence 
\begin{equation}
\chi_{_M} \left( {\alpha=\frac{\pi}{4}} \right): {s{\rm-wave}} > {\rm nematic} > {\rm magnetic} > {d{\rm-wave}}. \nonumber
\end{equation} 
Computation of the bare susceptibility suggests a strong propensity toward the formation of $s$-wave pairing and two nematic orders in the world of interacting spin-3/2 fermions with isotropic dispersion. The magnetic orders and $d$-wave pairings are expected to be suppressed near $\alpha=\frac{\pi}{4}$, at least at zero temperature [see Figs.~\ref{Fig:FiniteT_Summary} and ~\ref{Fig:GlobalPD_Intro_noFS}]. We also note that the gain in free-energy [see Sec.~\ref{SubSec:energy_Entropy}] and the mean-field susceptibility follow the same hierarchy $\Delta_F, \chi_{_M} \sim N^M_{\rm anti}$.

\subsubsection{\bf Anisotropic Luttinger Semimetal near $\alpha = \frac{\pi}{2}$}~\label{SubSubSec:Susceptibility_piover2}

When the effective mass in the $T_{2g}$ orbital becomes sufficiently large, the Luttinger model simplifies to 
\begin{equation}
\lim_{{\alpha \to \frac{\pi}{2}}}\hat{h}^{\rm Nam}_{\rm L} ({\bf k})= -\eta_3 \: \frac{k^2}{2 m_2} \sum^5_{j=4}  \Gamma_j \hat{d}_j (\hat{{\bf k}}).
\end{equation}
This Hamiltonian possesses an emergent $SU(2) \otimes U(1)$ chiral symmetry, where $\{ \Gamma_{45} \Gamma_j\}$ with $j=1,2,3$ are the three generators of an $SU(2)$ rotation, whereas a $U(1)$ rotation is generated by $\hat{N}= \eta_3 \Gamma_0$, the number operator. In this limit the Hamiltonian is similar to the one for spinless fermions in Bernal-stacked bilayer graphene, which altogether supports \emph{six} masses, given by~\footnote{When an order parameter matrix fully anticommutes with the noninteracting Hamiltonian, we coin it as mass order.}
\begin{equation}~\label{Eq:masspiover2}
{\bf M}_\frac{\pi}{2}=\left\{ \overbrace{ \underbrace{\eta_3 \left(\Gamma_1, \Gamma_2, \Gamma_3 \right) }_{\mbox{$T_{2g}$ nematic}},
\underbrace{\left( \eta_1,\eta_2 \right) \Gamma_0}_{\mbox{$s$-wave}}  }^{\mbox{SO(5) vector}},
\underbrace{\eta_0 \Gamma_{45}}_{\mbox{$A_{2u}$ magnet}}    \right\}. 
\end{equation}
Note that three components of the $T_{2g}$ nematicity break the continuous $SU(2)$ chiral symmetry, while the $s$-wave pairing breaks the global $U(1)$ symmetry. On the other hand, the $A_{2u}$ magnet transforms as a scalar under the chiral rotation and breaks only time-reversal-symmetry. These three mass orders possess the largest and equal susceptibilities as $\alpha \to \frac{\pi}{2}$, see Fig.~\ref{Fig:susceptibility_NoFS}. Therefore, repulsive interactions favor two excitonic masses for zero [see Figs.~\ref{Subfig:T2gswave} and ~\ref{Subfig:A2uEg}] and $s$-wave pairing [see Fig.~\ref{Subfig:T2gswaveMU}] for finite chemical doping.

Also note that each member of the following vector 
\begin{equation}~\label{Eq:semimasspiover2}
{\bf M}^{\prime}_\frac{\pi}{2}=\left\{ \overbrace{  \underbrace{\eta_3 \left(\Gamma_4, \Gamma_5\right) }_{\mbox{$E_g$ nematic}},
\underbrace{\left( \eta_1,\eta_2 \right) (\Gamma_4,\Gamma_5)}_{\mbox{$E_g$ d-wave}} }^{\mbox{multiplet of SO(3) vectors} }
\right\}. 
\end{equation}
anti-commutes and commutes with one matrix appearing in $\lim_{{\alpha \to \frac{\pi}{2}}}\hat{h}^{\rm Nam}_{\rm L} ({\bf k})$. Hence, $E_g$ nematicity and $d$-wave pairing have identical susceptibilities as $\alpha \to \frac{\pi}{2}$, but 
\begin{equation}
\chi_{_{{\bf M}^{\prime}_\frac{\pi}{2}}} \: < \: \chi_{_{{\bf M}_\frac{\pi}{2}}}. \nonumber 
\end{equation} 
As a result such an anisotropic system can accommodate an $E_g$ $d$-wave pairing at finite chemical doping, see Fig.~\ref{Subfig:A2udwaveMU}. However, $T_{1u}$ magnet and $T_{2g}$ $d$-wave superconductor have exactly \emph{zero} susceptibility as they \emph{fully commute} with $\lim_{{\alpha \to \frac{\pi}{2}}}\hat{h}^{\rm Nam}_{\rm L} ({\bf k})$. Hence, onset of these two orders is unlikely when $\alpha \approx \frac{\pi}{2}$.

\begin{figure}[t!]
\includegraphics[width=8.25cm,height=5cm]{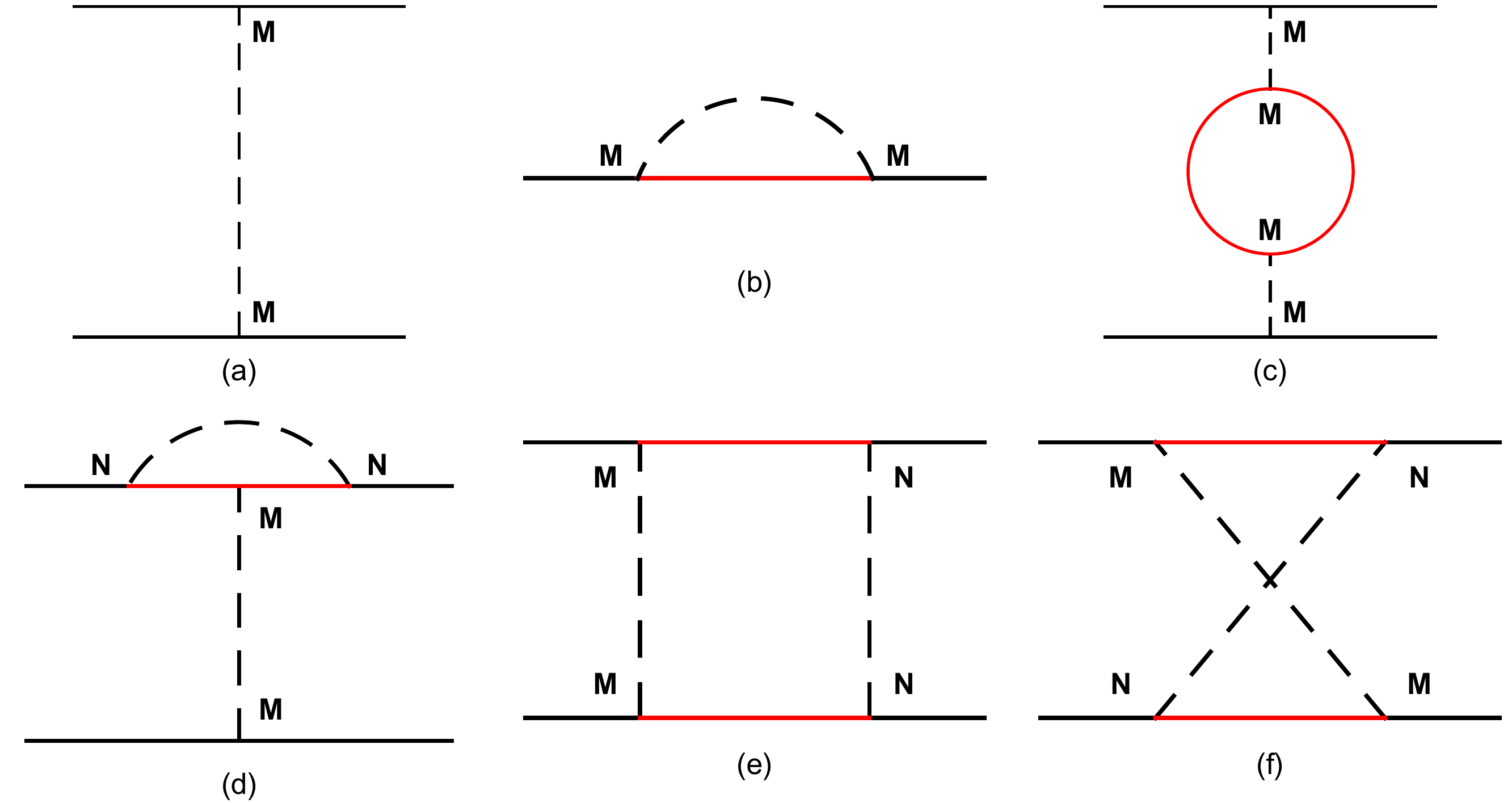}
\caption{ (a) Bare four-fermion interaction vertex $\left( \Psi^\dagger M \Psi \right)^2$, and (b) self-energy correction due to four-fermion interaction. Note contribution from Feymann diagram (b) is finite only when the chemical potential ($\mu$) is finite, and renormalizes $\mu$ (see Sec.~\ref{Sec:RG}). Feynman diagrams (c)-(f) yield corrections to the bare interaction vertex to the leading order in the $\epsilon$-expansion, where $\epsilon=d-2$. Here, solid lines represent fermions, and $M$ and $N$ are four-dimensional Hermitian matrices. While the red lines in (b)-(f) correspond to the fast modes (living within a thin Wilsonian shell $\Lambda \; e^{-\ell} <|{\bf k}|<\Lambda$, where $\Lambda$ is the ultraviolet momentum cut-off), the black lines are the slow modes with $|{\bf k}|<\Lambda \; e^{-\ell}$. Recall $\ell$ is the logarithm of the renormalization group scale.  
}~\label{Fig:FeynDiag_Interaction}
\end{figure}

\subsubsection{\bf Anisotropic Luttinger Semimetal near $\alpha = 0$}~\label{SubSubSec:Susceptibility_0}

Finally, we compare the susceptibility for various orders when the mass of the $E_{g}$ orbital becomes sufficiently large. The Luttinger Hamiltonian then takes the form~\footnote{This Hamiltonian is quite similar to the one for three-dimensional massless Dirac fermions, with the crucial difference that for the Dirac Hamiltonian $\hat{d}_j(\hat{k}) \sim k_j$, while for the Luttinger Hamiltonian $\hat{d}_{j}(\hat{k}) \sim |\epsilon_{jlm}| \; \hat{k}_l \; \hat{k}_m$, where $j, l, m=1,2,3$.} 
\begin{equation}~\label{Eq:LuttingerHamil_alpha0}
\lim_{\alpha \to 0} \hat{h}^{\rm Nam}_{\rm L} ({\bf k})= -\eta_3 \: \frac{k^2}{2 m_1} \sum^{3}_{j=1} \Gamma_j \hat{d}_j (\hat{{\bf k}}),
\end{equation}
which possesses an emergent $U(1) \otimes U(1)$ symmetry, generated by $\eta_0 \Gamma_{45}$ and $\eta_3 \Gamma_0$. The mass orders in this limiting scenario constitute the following vector  
\begin{equation}~\label{Eq:mass0}
{\bf M}_0 =\left\{  \overbrace{ \underbrace{\eta_3 \left( \Gamma_4, \Gamma_5 \right)}_{\mbox{$E_g$ nematic}},
\underbrace{\left( \eta_1,\eta_2 \right) \Gamma_0}_{\mbox{$s$-wave}} }^{\mbox{SO(4) vector}}
\right\}. 
\end{equation} 
and consequently the $E_g$ nematicity and $s$-wave pairing acquire an identical and the largest susceptibility as $\alpha \to 0$, see Fig.~\ref{Fig:susceptibility_NoFS}. Hence, a competition between these two ordered phases can be anticipated near $\alpha=0$ [see Figs.~\ref{Subfig:EgswaveMU} and ~\ref{Subfig:Egswave}]. Any order parameter from the following vector anti-commutes with two matrices and commutes with one matrix appearing in Eq.~(\ref{Eq:LuttingerHamil_alpha0})
\begin{equation}~\label{Eq:semimass0} 
{\bf M}^\prime_0=\left\{ \overbrace{ \underbrace{\eta_3 \left(\Gamma_1, \Gamma_2, \Gamma_3 \right) }_{\mbox{$T_{2g}$ nematic}}, \:\:
\underbrace{\eta_0 \Gamma_{45}\left(\Gamma_1, \Gamma_2, \Gamma_3 \right) }_{\mbox{$T_{1u}$ magnet}} }^{\mbox{multiplet of SO(3) vectors}}
\right\},
\end{equation} 
and they also possess degenerate susceptibilities, but 
$$\chi_{_{{\bf M}^{\prime}_0}} \: < \: \chi_{_{{\bf M}_0}}.$$ 
As a result, a competition between $T_{1u}$ magnet and $T_{2g}$ nematicity can also be observed around $\alpha=0$, see Fig.~\ref{Subfig:T1uT2g}. On the other hand, the $T_{2g}$ $d$-wave pairing matrices anti-commute with one matrix and commute with two matrices appearing in Eq.~(\ref{Eq:LuttingerHamil_alpha0}) and its susceptibility is smaller than the orders appearing in ${\bf M}_0$ and ${\bf M}^\prime_0$. Nonetheless, when assisted by finite chemical doping, the $T_{2g}$ $d$-wave pairing can be realized even for repulsive magnetic interaction in the $T_{1u}$ channel, as shown in Fig.~\ref{Subfig:T1udwaveMU}. Finally, we note that $A_{2u}$ magnet and $E_g$ $d$-wave pairing fully commute with $\lim_{\alpha \to 0} \hat{h} ({\bf k})$ and possess \emph{zero} susceptibility. Hence, onset of these two orders around $\alpha=0$ is unlikely.

\begin{figure}[t!]
\includegraphics[width=8cm,height=2.5cm]{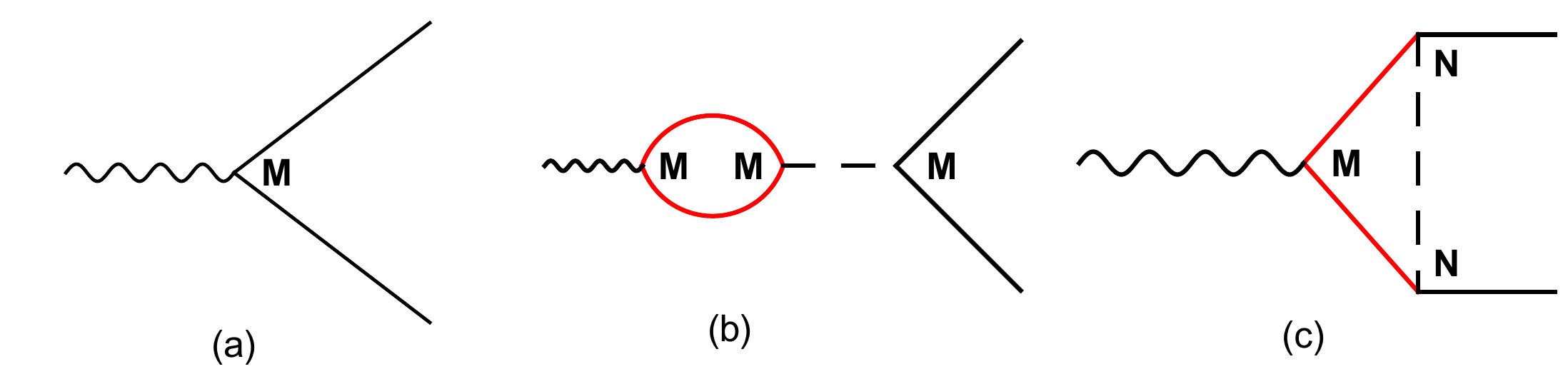}
\caption{ (a) The bare vertex associated with the source term $\Psi^\dagger_{\rm Nam} M \Psi_{\rm Nam}$. The leading order renormalization of such vertices arises from Feynman diagrams (b) and (c), yielding the RG flow of the source terms, displayed in Eq.~(\ref{Eq:sourceRG}). Here, wavy lines stand for the source field, while solid lines for fermions, and the dashed lines for the interaction vertex. The black (red) solid lines represent slow (fast) modes. 
}~\label{Fig:FeynDiag_Susceptibility}
\end{figure}

Note that various matrices appearing in ${\bf M}_\frac{\pi}{2}$, ${\bf M}^\prime_\frac{\pi}{2}$, ${\bf M}_0$, ${\bf M}^{\prime}_0$ can form composite order-parameters and the enlarged symmetries among distinct orderings are displayed in Eqs.~(\ref{Eq:masspiover2}), (\ref{Eq:semimasspiover2}), (\ref{Eq:mass0}), (\ref{Eq:semimass0}). Such enlargement of order-parameter vectors plays an important role in determining the confluence of competing orders, which we discuss in Sec.~\ref{SubSec:SelectionRule}.   

\section{Renormalization group analysis}~\label{Sec:RG}

After gaining insight into the propensity toward various orderings in the Luttinger system, next we seek to investigate the onset of different BSPs and the competition among them within the framework of an unbiased RG analysis. This will allow us to go beyond the mean-field analysis, presented in the last section, and systematically incorporate fluctuations. In what follows we here restrict ourselves to the leading order in the $\epsilon$ expansion, where $\epsilon=d-2$, and account for corrections to the bare interaction vertices ($\lambda_j$s) to quadratic order in the coupling constants. The relevant Feynman diagrams are shown in Fig~\ref{Fig:FeynDiag_Interaction}. After performing summation over fermionic Matsubara frequencies $\omega_n=(2 n+1) \pi T$ with $-\infty \leq n \leq \infty$, we integrate out a thin Wilsonian shell $\Lambda e^{-\ell} <|{\bf k}| <\Lambda$ to arrive at the following RG flow equations
\begin{equation}~\label{Eq:RGgeneral}
\beta_{g_{_i}}= -\epsilon g_{_i} + \sum^2_{j=0} g^2_{_j} H^i_{jj}(\alpha,t,\mu) 
+ {\sum^{2}_{j, k=0}}^\prime  g_{_j} g_{_k} H^i_{jk} (\alpha,t,\mu),
\end{equation}
for $i=0,1,2$, where $\beta_X \equiv dX/d\ell$, in terms of dimensionless quantities defined in Eq.~(\ref{eq:dimensionlessparameters}). The prime symbol in the summation indicates that $j>k$. For notational clarity we take $\tilde{\mu} \to \mu$, and the functions $H^{i}_{jk}(\alpha, t, \mu)$ for $i,j,k=0,1,2$ are shown in Appendix~\ref{Append:RGdetails}. Due to their lengthy expressions (which are not very instructive in particular) we here only display the schematic form of the flow equations. Both temperature and chemical potential also flow under coarse graining [see first line of Eq.~(\ref{Eq:RG_Introduction})] as relevant perturbations with bare scaling dimensions $[t]=[\mu]=z$, where $z=2$ for the Luttinger system. These two flow equations can then be solved, respectively yielding   
\begin{equation}
t(\ell) =t(0) \; e^{z \ell}, \quad \mu(\ell)=\mu(0) \; e^{z \ell},
\end{equation}
where $t(0)$ and $\mu(0)$ are the bare values. We supply these solutions to Eq.~(\ref{Eq:RGgeneral}) to find the phase diagram of interacting Luttinger fermions using the following prescription.~\footnote{We should note that Feynman diagram (b) in Fig.~\ref{Fig:FeynDiag_Interaction} provides interaction driven corrections (linear in $g_{_i}$) to $\mu$. However, to maintain the order by order correction to quartic interactions, we neglect such corrections in Eq.~(\ref{Eq:RGgeneral}) within the framework of the leading order $\epsilon$-expansion.}

\begin{table}[t!]
\renewcommand{\arraystretch}{1.4}
\begin{tabular}{|c|c|c c c|c c c|}
\hline
\multirow{2}{*}{Coup.} &  $\alpha=\frac{\pi}{4}$ &  \multicolumn{3}{c|}{$\alpha=1.5$}  & \multicolumn{3}{c|}{$\alpha=0.04$} \\ 
  &QCP$^1_{\frac{\pi}{4}}$ &  QCP$^1_\frac{\pi}{2}$ &  QCP$^2_\frac{\pi}{2}$ & BCP$_\frac{\pi}{2}$ & QCP$^1_0$ & QCP$^2_0$ &  BCP$_0$     \\ \hline 
$g^\ast_{_0} \times 10^{3}$ &	$2.24$ &	 $1.49$ &	$-2.93$ &	$-1.58$ &  $1.95$ & $-5.98$ & $-5.60$       \\ 
\rowcolor{RowColor}$g^\ast_{_1} \times 10^{3}$ &	$2.03$ &	 $1.47$ &  $-2.84$ &  $-1.33$ &  $1.71$ &  $4.80$ &  $4.97$       \\ 
$g^\ast_{_2} \times 10^{3}$ &	$2.03$ &	 $1.26$ &  $3.76$  &  $5.54$  &  $1.94$ & $-5.87$ & $-5.48$    \\ 
\hline 
\end{tabular}
\caption{Locations of quantum critical points (QCPs) possessing one unstable direction and bi-critical points (BCPs) possessing two unstable directions for three specific choices of the mass anisotropy parameter $\alpha$ [see Eq.~(\ref{Eq:anisotropicparameter})]. Note for $\alpha=\frac{\pi}{4}$ the Luttinger model possesses an emergent spherical symmetry, while for $\alpha \to 0 \; (\frac{\pi}{2})$ the band dispersion in the $E_g(T_{2g})$ orbitals becomes almost flat. For $\alpha=\frac{\pi}{4}$ the coupled RG flow equations support only one QCP, denoted by QCP$^1_{\frac{\pi}{4}}$, while for $\alpha \to \frac{\pi}{2}$ and $\alpha \to 0$ we find two QCPs, respectively denoted by QCP$^j_\frac{\pi}{2}$ and QCP$^j_0$ for $j=1,2$, see also Ref.~\cite{Herbut-2}. For these two limiting cases the flow equations also support one BCP, respectively identified as BCP$_\frac{\pi}{2}$ and BCP$_0$. Existence of such BCP in the presence of two QCPs is necessary to ensure the continuity of the RG flow trajectories and separate the basins of attraction of two QCPs. All coupling constants at the fixed points are measured in units of $\epsilon$, where $\epsilon=d-2$ measures the deviation from the lower critical two spatial dimensions, where all local quartic interactions are \emph{marginal}. Note that QCP$^1_{\frac{\pi}{4}}$, QCP$^1_{\frac{\pi}{2}}$, QCP$^1_{0}$ represent the same QCP, only its location shifts as we tune $\alpha$. By contrast, QCP$^2_{\frac{\pi}{2}}$ and QCP$^2_{0}$ are solely introduced by mass anisotropy and bear no analog to the fixed points found around $\alpha=\pi/4$. Besides the above QCPs and the BCPs, there always exists a trivial Gaussian fixed point, representing the non-interacting Luttinger semimetal, at $(g^\ast_{_0}, g^\ast_{_1}, g^\ast_{_2})=(0,0,0)$, endowed with three stable directions. See Appendix~\ref{Append:Stabilitymatrix} for details. 
}~\label{Table:CriticalPoints}
\end{table}

While the \emph{divergence} of at least one of the quartic couplings (i.e., $g_{_i} \to + \infty$) under coarse graining indicates the onset of a BSP, to unambiguously determine the pattern of symmetry breaking we also account for the leading order RG flow of all source terms, associated with different BSPs, discussed in Sec.~\ref{Sec:BSPs}. The effective action in the presence of all symmetry allowed fermionic bilinears [see Eqs.~(\ref{Eq:excitonicOP_Hamil})-(\ref{Eq:superconductingOP})] is given by 
\begin{eqnarray}~\label{Eq:Action_SourceTerm}
S_{s}=\int d\tau \; d^3{\bf r} \; \Psi^\dagger_{\rm Nam} \; \left( \hat{h}^{\rm exc}_{\rm local} + \hat{h}^{\rm pair}_{\rm local} \right)\; \Psi_{\rm Nam},
\end{eqnarray}
with $\Psi^\dagger_{\rm Nam} \equiv \Psi^\dagger_{\rm Nam}(\tau,{\bf x})$ and $\Psi_{\rm Nam} \equiv \Psi_{\rm Nam}(\tau,{\bf x})$ as two independent Grassmann variables. Relevant Feynman diagrams are shown in Fig.~\ref{Fig:FeynDiag_Susceptibility}. The resulting RG flow equations take the following schematic form
\begin{equation}~\label{Eq:sourceRG}
\frac{d \ln \Delta_i}{d \ell}-2= \sum^{2}_{j=0} F^{j}_i \left( \alpha, t, \mu \right) \; g_j. 
\end{equation}    
See Appendix~\ref{Append:RGdetails} for explicit form of these flow equations. The quantities appearing on the right hand side of each equation, represent the \emph{scaling dimension} of the corresponding order-parameter.

\begin{table}[t!]
\renewcommand{\arraystretch}{1.4}
\begin{tabular}{|c|c|c c c|c c c|}
\hline
\multirow{2}{*}{Sr} & $\alpha=\frac{\pi}{4}$ &  \multicolumn{3}{c|}{$\alpha=1.5$}  & \multicolumn{3}{c|}{$\alpha=0.04$} \\ 
   & QCP$^1_{\frac{\pi}{4}}$ &  QCP$^1_\frac{\pi}{2}$ &  QCP$^2_\frac{\pi}{2}$ & BCP$_\frac{\pi}{2}$ & QCP$^1_0$ & QCP$^2_0$ &  BCP$_0$     \\ \hline 
$\Delta_0$ &	0 & 0 & 0 & 0 & 0 & 0 & 0 \\ \rowcolor{RowColor}
$\Delta_1$ &	\emph{0.426} & \emph{0.531} & -0.238 & 0.378 & 0.353 & \emph{0.614} &  0.656 \\ 
$\Delta_2$ &	\emph{0.426} & 0.275 & \emph{0.325} & 0.678 & \emph{0.539} & -0.153 & -0.070  \\ \rowcolor{RowColor}
$\Delta_3$ &	-0.076 & -0.204 & {\bf 1.147} & 1.006 & -0.004 &  -0.024 & -0.024  \\ 
$\Delta_4$ &	-0.076 &  -0.007 &	-0.021 & -0.030 & -0.134 & {\bf 0.731} & 0.699  \\ \rowcolor{RowColor}
$\Delta_5$ &	-0.076 &  -0.094 & 0.182 & 0.092 & -0.062 & 0.020 & 0.011  \\ 
\hline 
$\Delta^{\rm p}_{A_{1g}}$ &	{\bf 0.551} &	{\bf 0.545} & -0.254 & 0.355 & {\bf 0.547} & -0.166 & -0.083 \\ \rowcolor{RowColor}
$\Delta^{\rm p}_{T_{2g}}$ &	-0.034 & -0.004 & -0.011 & -0.016 & -0.059 & {\color{blue}0.016} & 0.006  \\ 
$\Delta^{\rm p}_{E_g}$ &	-0.034 &  -0.088 & {\color{blue}0.169} & 0.073 &  -0.002 &-0.012 & -0.012  \\ \hline 
\end{tabular}
\caption{ Scaling dimensions (in units of $\epsilon$) of various source (Sr) terms or fermion bilinears at different fixed points (reported in Table~\ref{Table:CriticalPoints}), obtained by substituting fixed point values of the coupling constants $g^\ast_{_0}$, $g^\ast_{_1}$ and $g^\ast_{_2}$ on the right-hand side of the corresponding flow equation [see Eq.~(\ref{Eq:sourceRG})] at $t=0$ and $\mu=0$. At each QCP the largest scaling dimension is shown in bold, while the second largest ones are shown in italic. At the two magnetic QCPs (QCP$^2_\frac{\pi}{2}$ and QCP$^2_0$), the largest scaling dimensions for the superconducting channel (namely $d$-wave pairings) are shown in blue.    
}~\label{Table:scalingdimensions}
\end{table}

We simultaneously run the flow of the quartic couplings ($g_{_j}$s) and the source terms ($\Delta_j$s). When at least one of the quartic couplings \emph{diverges} and flows toward $\to + \infty$ (thus indicating onset of a BSP), we identify the source term (say $\Delta_j$) that diverges toward $\to + \infty$ \emph{fastest} (assuming a possible scenario when more than one source term diverge toward $+\infty$). The BSP is then characterized by the order-parameter $\Delta_j \neq 0$. We use this strategy to determine various cuts of the global phase diagram of spin-3/2 Luttinger fermions, displayed in Fig.~\ref{Fig:GlobalPD_Intro_noFS} (for $t=\mu=0$), Fig.~\ref{Fig:FiniteT_Summary} (for $\mu=0$, but finite-$t$) and Figs.~\ref{Fig:swave_IsotropicLSM}, ~\ref{Fig:dwaves_IsotropicLSM} and ~\ref{Fig:GlobalPD_Intro_FS} (for finite-$t$ and finite-$\mu$). Next we discuss these cases in three subsequent sections.

Unless the coupling constants are fine tuned, only one of them diverges fastest toward $+\infty$. The selection rule among the competing orders, discussed in Sec.~IIB3 (see also Sec.~\ref{SubSec:SelectionRule} for details), is then determined in terms of the fastest diverging coupling constant. Appearance of various ordered phases in all the cuts of the global phase diagram are then consistent with the selection rule, even in the presence of multiple running coupling constants.

         \subsection{Quantum criticality in Luttinger semimetal}~\label{SubSec:QPTinLSM}

We first discuss the effects of electronic interactions on a LSM (i.e. when the chemical potential is pinned at the band touching point) at zero temperature. The RG flow equations for $\mu=0$ and $t=0$ can be derived by taking the limit $\mu \to 0$ and then $t \to 0$ in Eq.~(\ref{Eq:RGgeneral}), suggesting that weak interactions are irrelevant perturbations and any ordering takes place at finite coupling $g_{_i}\sim \epsilon$ through a QPT. Next we discuss the following three cases separately (i) isotropic Luttinger system ($\alpha=\frac{\pi}{4}$), large mass for (ii) the $T_{2g}$ orbital ($\alpha \to \frac{\pi}{2}$) and (iii) the $E_g$ orbital ($\alpha \to 0$). Such systematic analysis will allow us to anchor our anticipations regarding the nature of BSPs from the mean-field susceptibility, discussed in Sec.~\ref{SubSec:susceptibility}. Note at $t=\mu=0$, the system is devoid of any natural infrared cutoff as $\ell^t_\ast,\ell^\mu_\ast \to \infty$. Hence, we run the flows of quartic couplings up to an RG time $\ell_\ast \to \infty$ to determine the stability of LSM and the flows of the source terms to pin the pattern of symmetry breaking.

\subsubsection{ Isotropic Luttinger semimetal ($\alpha=\frac{\pi}{4}$)}

For $\alpha=\frac{\pi}{4}$ the coupled RG flow equations support only one QCP, reported in Table~\ref{Table:CriticalPoints} and identified as QCP$^1_{\frac{\pi}{4}}$, besides the trivial (and fully stable) Gaussian fixed point at $g^\ast_{_0}=g^\ast_{_1}=g^\ast_{_2}=0$ (representing the stable LSM). This QCP controls QPTs from LSM to various BSPs (depending on the interaction channel). To gain insight into the nature of the candidate competing BSPs, we compute the scaling dimensions for all fermion bilinears at this QCP. The results are summarized in Table~\ref{Table:scalingdimensions}. Note that the $s$-wave pairing has the largest scaling dimension at this QCP, while two nematic orders possess degenerate but second largest (and positive) scaling dimensions. But, the rest of the fermion bilinears possess \emph{negative} scaling dimensions. Notice that the scaling dimensions for different orders at this QCP follow the same hierarchy as the mean-field susceptibilities, discussed in Sec.~\ref{SubSubSec:Susceptibility_isotropic}.

The phase diagrams in an interacting LSM are displayed in Fig.~\ref{Fig:FiniteT_Summary} for various interaction channels. Around $\alpha=\frac{\pi}{4}$ strong repulsive nematic interactions ($g_{_1}$ and $g_{_2}$) favor $s$-wave pairing even in the absence of a Fermi surface (since $\mu=0$), see Figs.~\ref{Subfig:FiniteT_g1} and ~\ref{Subfig:FiniteT_g2}. On the other hand, strong magnetic interactions in the $A_{2u}$ ($g_{_3}$) and $T_{1u}$ ($g_{_4}$) channels respectively support $T_{2g}$ [see Fig.~\ref{Subfig:FiniteT_g4}] and $E_g$ [see Fig.~\ref{Subfig:FiniteT_g3}] nematicities. However, we could not find any magnetic ordering or $d$-wave pairing in the very close vicinity to $\alpha=\frac{\pi}{4}$, at least when $t=0$, see Fig.~\ref{Fig:GlobalPD_Intro_noFS}. Therefore, the computation of mean-field susceptibilities and scaling dimensions of fermion bilinears, in corroboration with our unbiased RG calculation, show that the strongly interacting isotropic LSM becomes unstable toward the formation of two nematic orders and $s$-wave pairing at the lowest temperature.

\subsubsection{Anisotropic Luttinger semimetal: $\alpha \to \frac{\pi}{2}$}

Next we turn our focus to the vicinity of $\alpha=\frac{\pi}{2}$. The coupled flow equations then support \emph{two} QCPs [denoted by QCP$^1_{\frac{\pi}{2}}$ and QCP$^2_{\frac{\pi}{2}}$] and \emph{one} bi-critical point [denoted by BCP$_{\frac{\pi}{2}}$], see Table~\ref{Table:CriticalPoints}. The BCP possesses \emph{two} unstable directions. Notice QCP$^1_{\frac{\pi}{2}}$ is the same as QCP$^1_{\frac{\pi}{4}}$, only shifted toward weaker coupling, which can be verified from the fact that the signs of the coupling constants and scaling dimensions for all fermion bilinears are \emph{identical} at these two QCPs [see Tables~\ref{Table:CriticalPoints} and ~\ref{Table:scalingdimensions}].

On the other hand, QCP$^2_{\frac{\pi}{2}}$ is new and bears no resemblance to any fixed points we found for $\alpha=\frac{\pi}{4}$. This QCP is induced by the mass anisotropy of Luttinger fermions. At this QCP the $A_{2u}$ magnetic ($E_g$ nematic) order possesses the largest (second largest) scaling dimension, see Table~\ref{Table:scalingdimensions}. Therefore, when repulsive interaction in the $A_{2u}$ channel dominates among various finite range components of the Coulomb interaction, the Luttinger semimetal can display a competition between these two orders as we approach $\alpha=\frac{\pi}{2}$ from an isotropic system, see Fig.~\ref{Subfig:A2uEg} [consult also Sec.~\ref{SubSec:SelectionRule}].

Among three possible local pairings the $E_g$ $d$-wave superconductor possesses the \emph{largest} (and positive) scaling dimension at this QCP. Hence, the emergence of this paired state can be anticipated when the LSM is doped away from the charge-neutrality point around $\alpha=\frac{\pi}{2}$, see Fig.~\ref{Subfig:A2udwaveMU} [consult Sec.~\ref{SubSec:LuttingerMetalRG} for details].

\subsubsection{Anisotropic Luttinger semimetal: $\alpha \to 0$} 

\begin{figure*}[t!]
\subfigure[]{\includegraphics[width=0.40\linewidth]{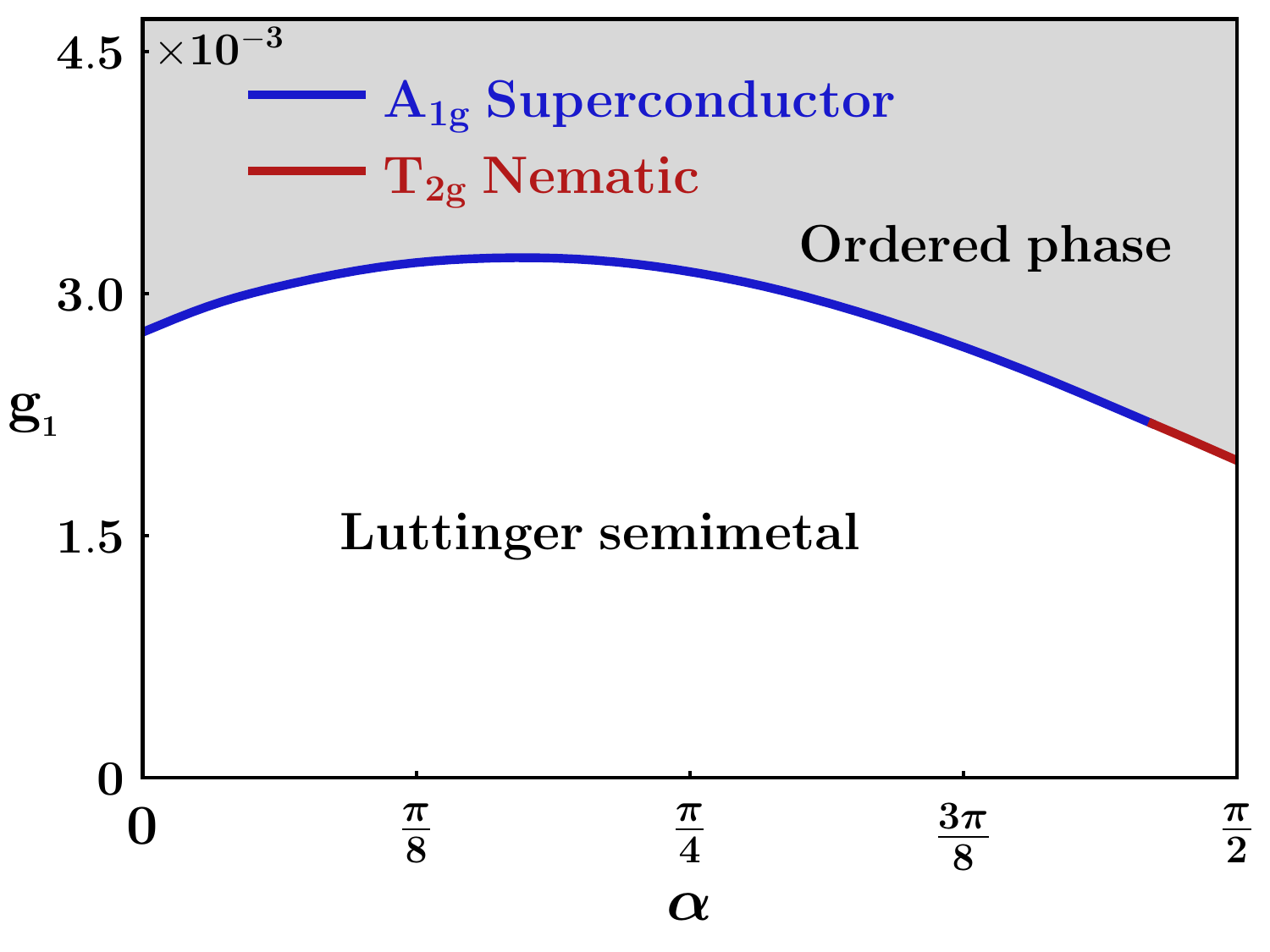}~\label{Subfig:T2gswave}}\hspace{0.5cm}
\subfigure[]{\includegraphics[width=0.40\linewidth]{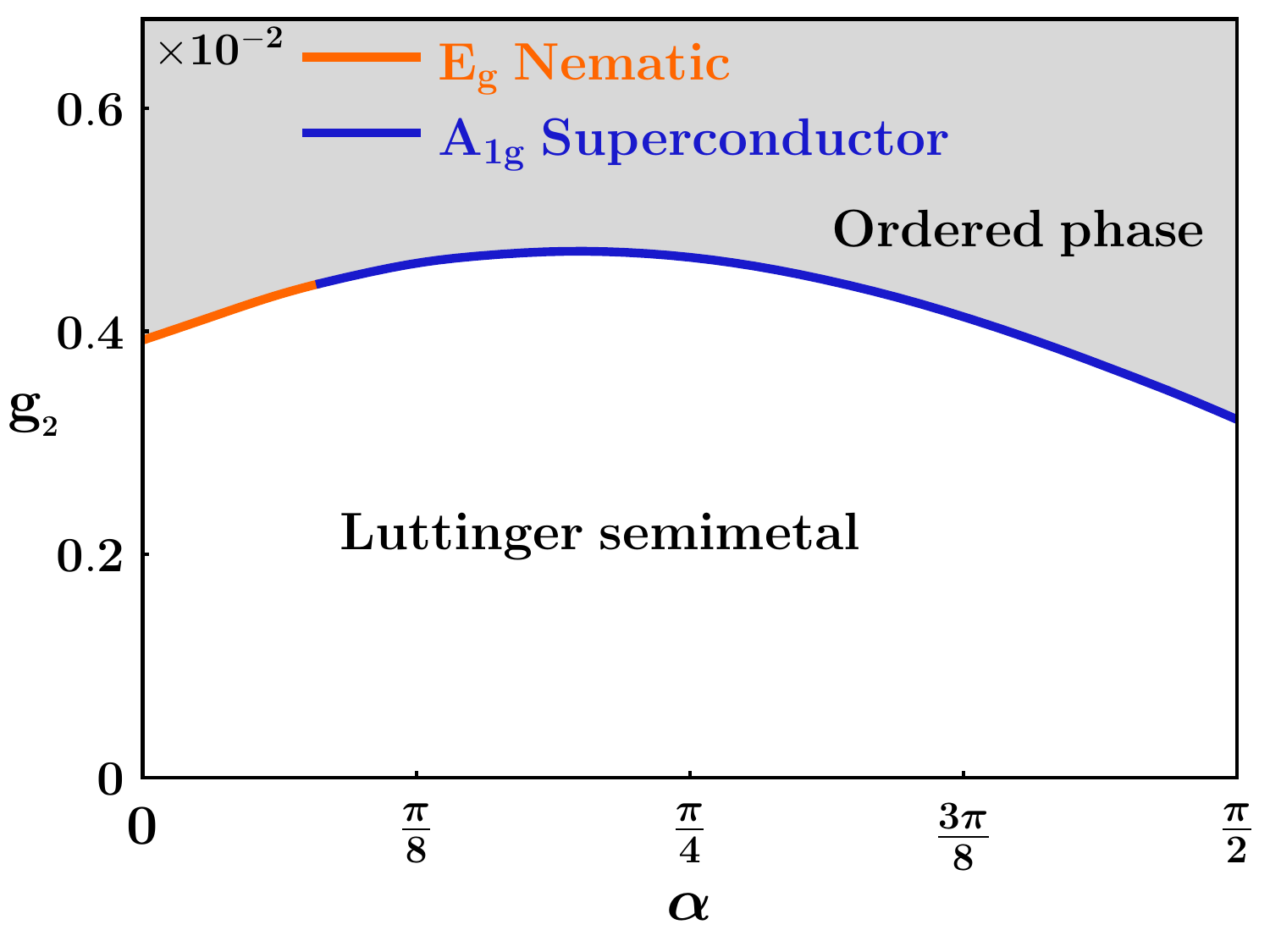}~\label{Subfig:Egswave}}
\subfigure[]{\includegraphics[width=0.40\linewidth]{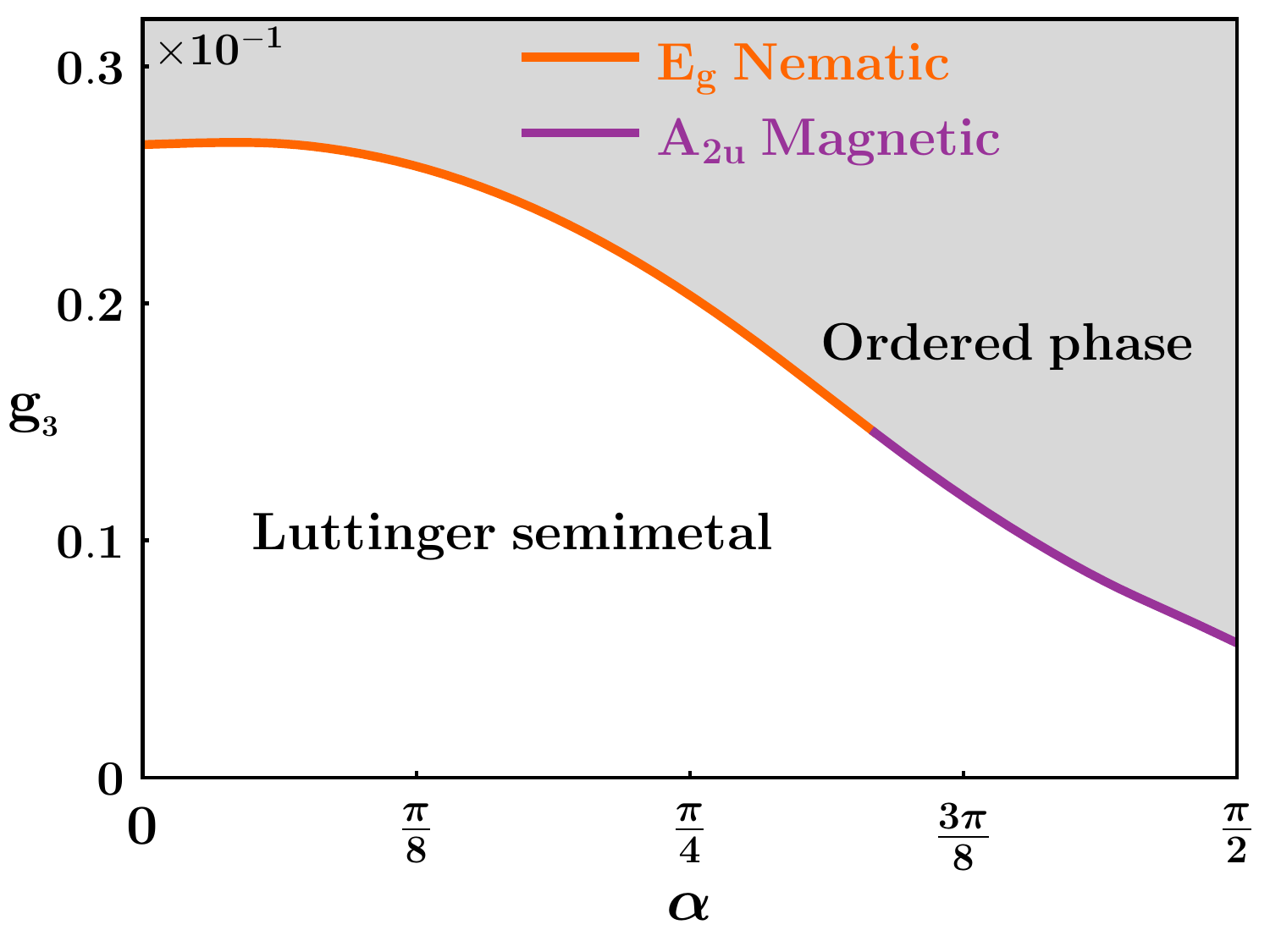}~\label{Subfig:A2uEg}}\hspace{0.5cm}
\subfigure[]{\includegraphics[width=0.40\linewidth]{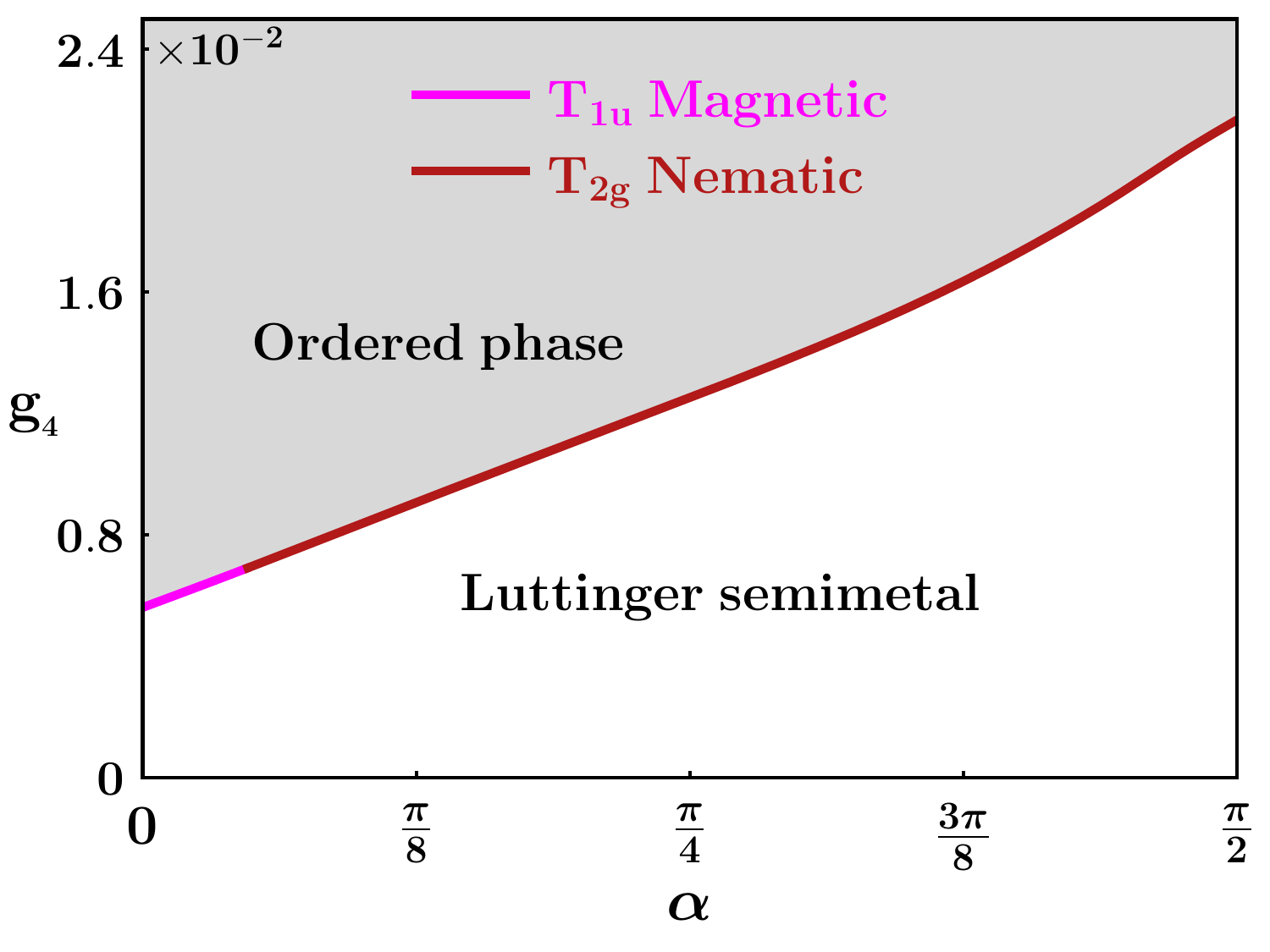}~\label{Subfig:T1uT2g}}
\caption{Specific cuts of the global phase diagram of an interacting Luttinger semimetal at zero temperature, obtained from an RG analysis [see Sec.~\ref{SubSec:QPTinLSM}]. Here $\alpha$ is the mass anisotropy parameter. For $\alpha=\frac{\pi}{4}$ the system possesses an enlarged spherical symmetry, whereas for $\alpha=0$ ($\frac{\pi}{2}$) the $E_{g}$ ($T_{2g}$) orbital possess \emph{infinite} mass [see Sec.~\ref{Sec:Luttinger_Model}]. Here $g_{_1}$ ($g_{_2}$) captures the strength of repulsive interactions in the $T_{2g}$ ($E_g$) nematic channels. Magnetic interactions in the $A_{2u}$ ($T_{1u}$) channel is denoted by $g_{_3}$ ($g_{_4}$). Notice that even in the absence of a Fermi surface, pure repulsive nematic interactions are conducive to $s$-wave pairing among spin-3/2 fermions at zero temperature [see panels (a) and (b)]. In all the panels the coupling constants are measured in units of $\epsilon$. The ordered states correspond to the gray shaded regions (see footnote~2), and its boundaries with the Luttinger semimetal (white regions), occupied by distinct broken symmetry phases are identified by different colors.
}~\label{Fig:GlobalPD_Intro_noFS}
\end{figure*}

Finally, we approach the opposite limit, when the mass in the $E_g$ orbital becomes sufficiently large, i.e. $m_2 \gg m_1$ or equivalently $\alpha \to 0$. In this regime the RG flow equations support two QCPs [denoted by QCP$^1_0$ and QCP$^2_0$] and a BCP [denoted by BCP$_0$], see Table~\ref{Table:CriticalPoints}. Note that QCP$^1_0$ is similar to QCP$^1_{\frac{\pi}{4}}$, only shifted toward weaker coupling. By contrast, QCP$^2_0$ is induced by the mass anisotropy. At this QCP, the $T_{1u}$ magnetic ($T_{2g}$ nematic) order possesses the largest (second largest) scaling dimension, see Table~\ref{Table:scalingdimensions}. Therefore, when repulsive interaction in the $T_{1u}$ channel dominates we expect a strong competition between these two orderings, as one tunes toward $\alpha \to 0$ starting from an isotropic system, see Fig.~\ref{Subfig:T1uT2g}.

Also note that among three local pairings, the $d$-wave one transforming under the $T_{2g}$ representation possesses the largest (and positive) scaling dimension. Hence, around $\alpha=0$ we expect onset of this paired state when the chemical potential is placed away from the band touching point, see Fig.~\ref{Subfig:T1udwaveMU} and discussion in Sec.~\ref{SubSec:LuttingerMetalRG}.

Finally, we comment on the role of the BCPs possessing two unstable directions, see Table~\ref{Table:CriticalPoints}. Note that a BCP can only be found when there exists two QCPs in the three dimensional coupling constant space $\left( g_{_0}, g_{_1}, g_{_2} \right)$. The existence of a BCP separates the basin of attraction of two QCPs and ensures continuity of the RG flow trajectories in coupling constant space.

              \subsection{Universality class and critical exponents}~\label{SubSec:exponents}

All QCPs, listed in Table~\ref{Table:CriticalPoints}, are characterized by only \emph{one} unstable direction or \emph{positive} eigenvalue of the corresponding \emph{stability matrix} [see Appendix~\ref{Append:Stabilitymatrix}], defined as 
\begin{equation}
M_{ij}\left( g_{_0}, g_{_1}, g_{_2} \right)= \frac{d}{d g_{_j}} \beta_{g_{_i}}.
\end{equation} 
To the leading order in the $\epsilon$-expansion the positive eigenvalue at all QCPs is exactly $\epsilon$, which in turn determines the \emph{correlation length exponent} ($\nu$) according to
\begin{equation}
\nu^{-1}=\epsilon + {\mathcal O} \left( \epsilon^2 \right).
\end{equation}   
For the physically relevant situation $\epsilon=1$ and we obtain $\nu=1$. The fact that $\nu$ is the same at all QCPs is, however, only an artifact of the leading order $\epsilon$-expansion. Generically $\nu$ is expected to be distinct at different QCPs, once we account for higher order corrections in $\epsilon$.

Since to the leading order in the $\epsilon$-expansion, local interactions do not yield any correction to fermion self-energy, the \emph{dynamic scaling exponent} ($z$) at all interacting QCPs is
\begin{equation}
z= 2 + {\mathcal O} \left( \epsilon \right).
\end{equation} 
Together the correlation length and dynamic scaling exponents determine the universality class of all continuous QPTs from a LSM to various BSPs. In the next section, we will discuss the imprint of these two exponents on the scaling of the transition temperature ($t_c$) associated with the finite temperature order-disorder transitions.  
\\
\vspace{2cm}


   \subsection{Interacting Luttinger Semimetal at Finite Temperature}~\label{SubSec:FiniteTRG}

\begin{figure*}[t!]
\includegraphics[width=.24\linewidth]{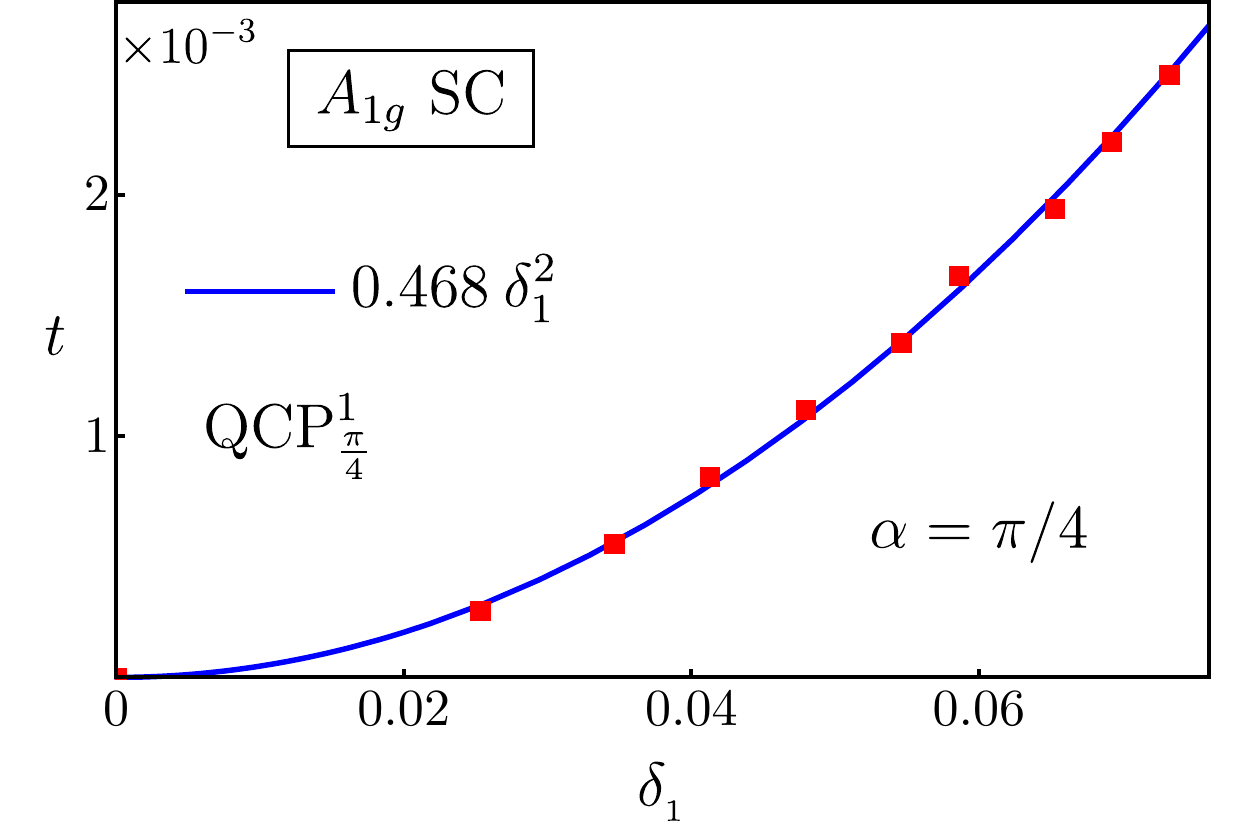}
\includegraphics[width=.24\linewidth]{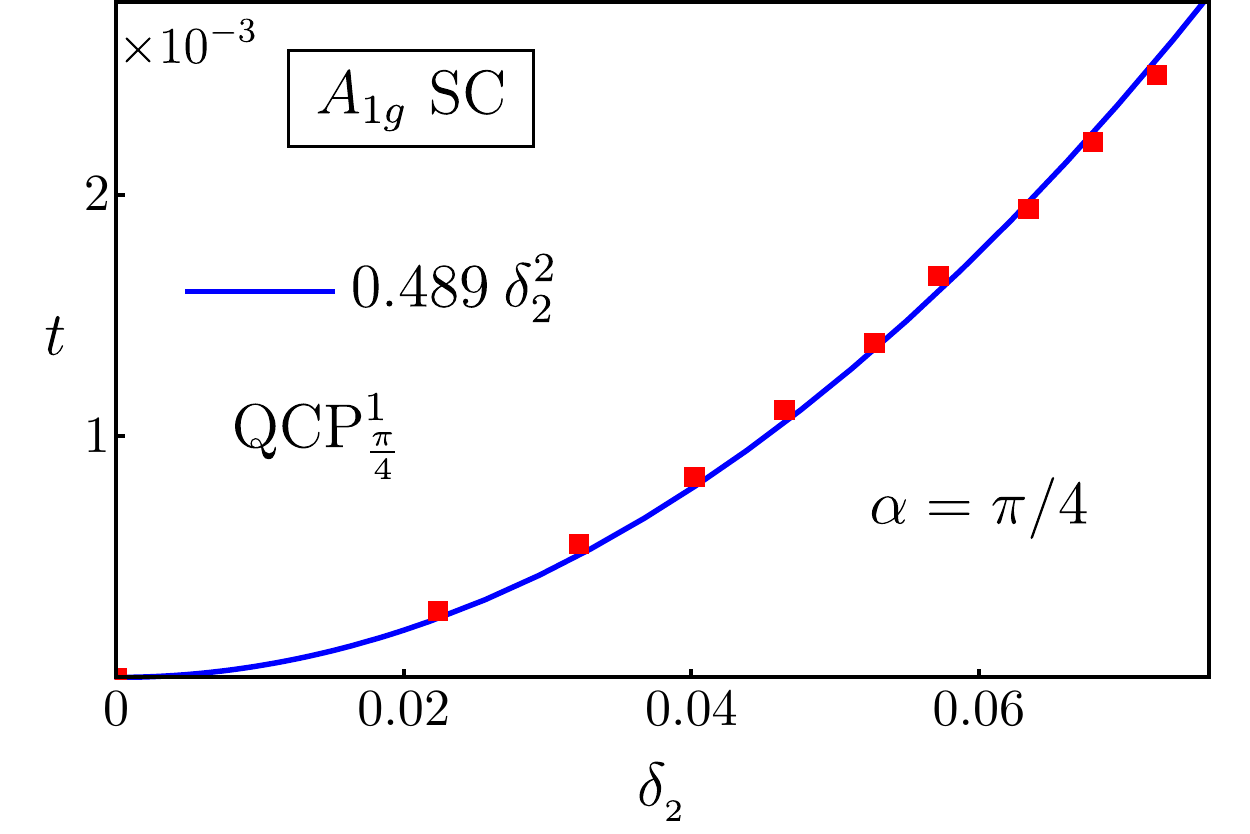}
\includegraphics[width=.24\linewidth]{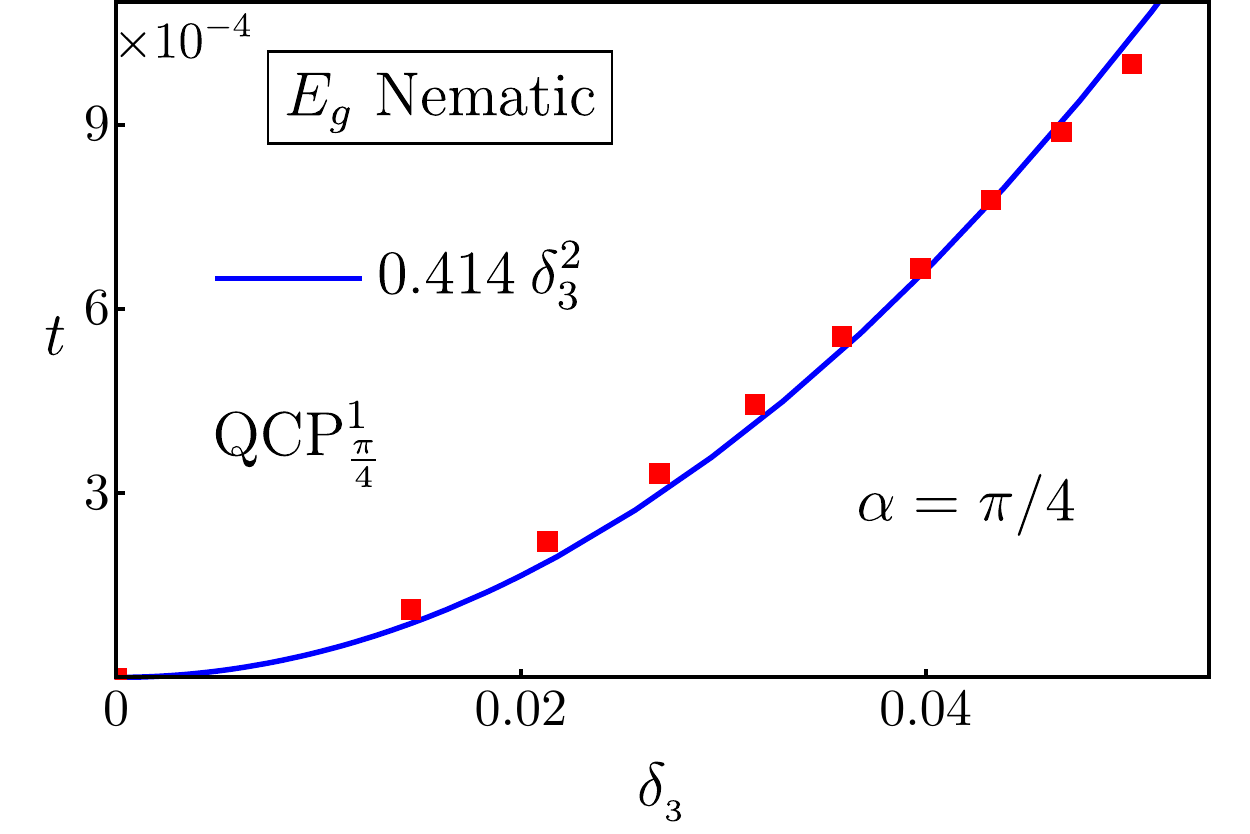}
\includegraphics[width=.24\linewidth]{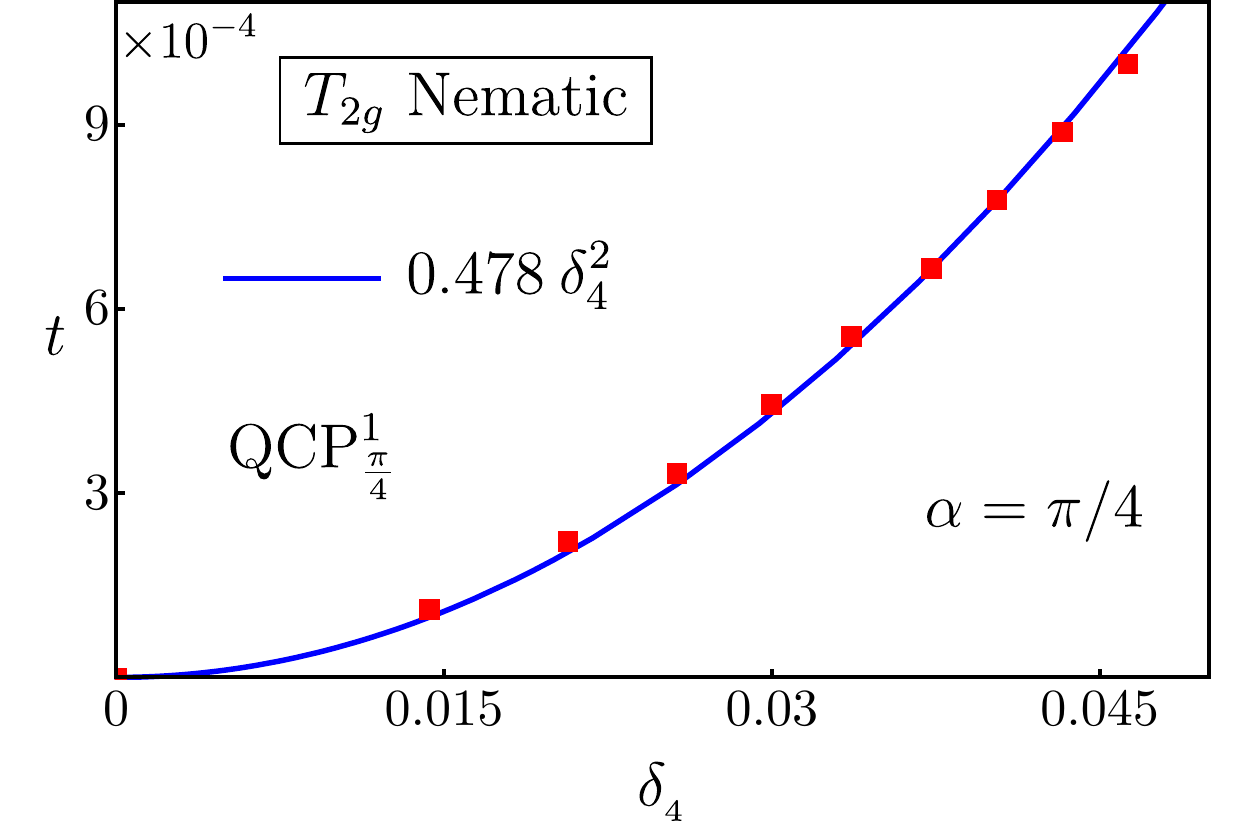}
\includegraphics[width=.24\linewidth]{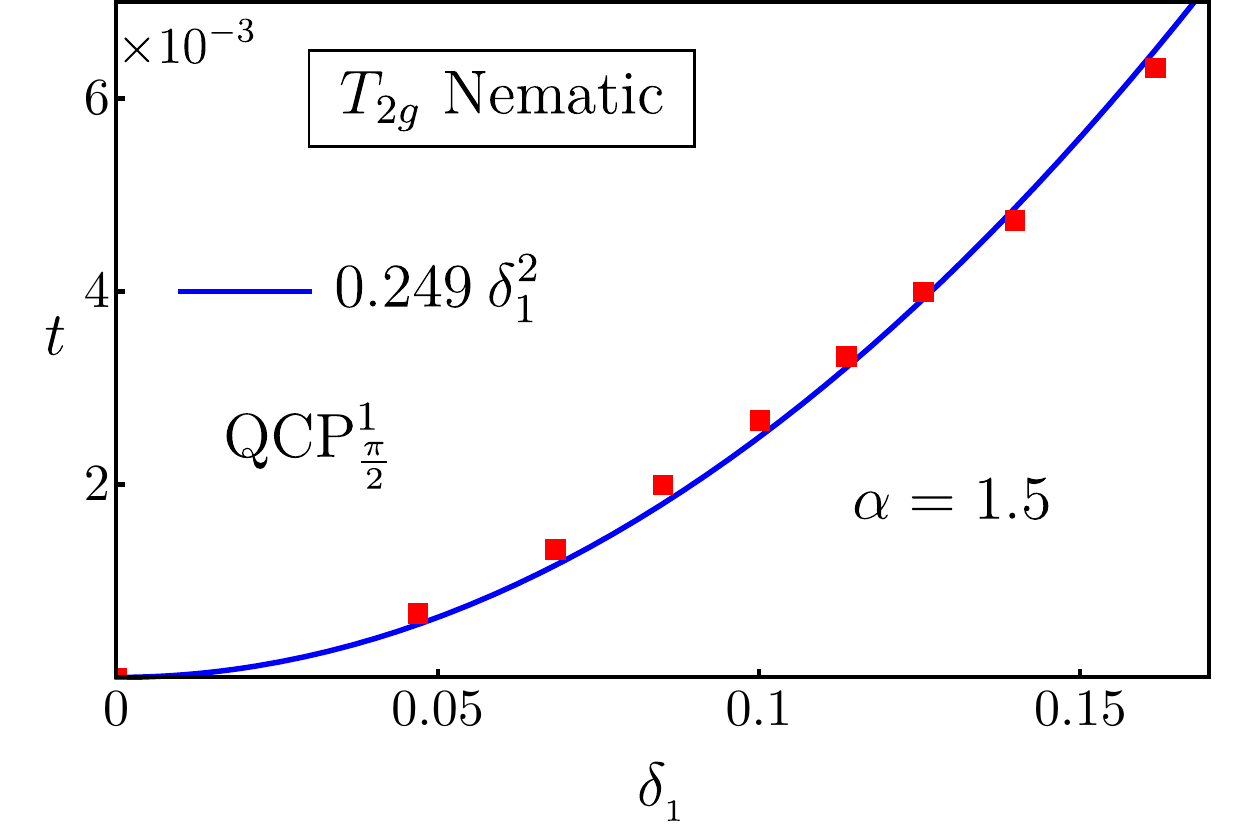}
\includegraphics[width=.24\linewidth]{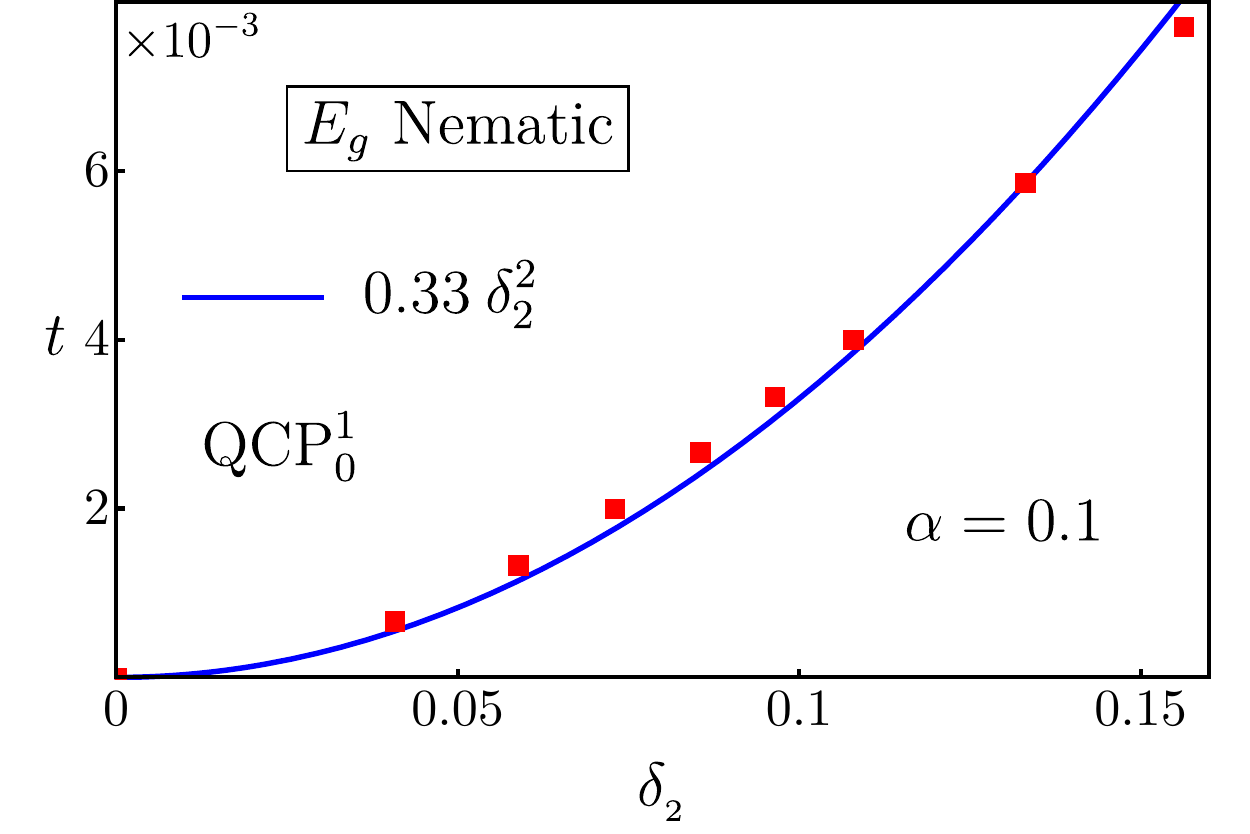}
\includegraphics[width=.24\linewidth]{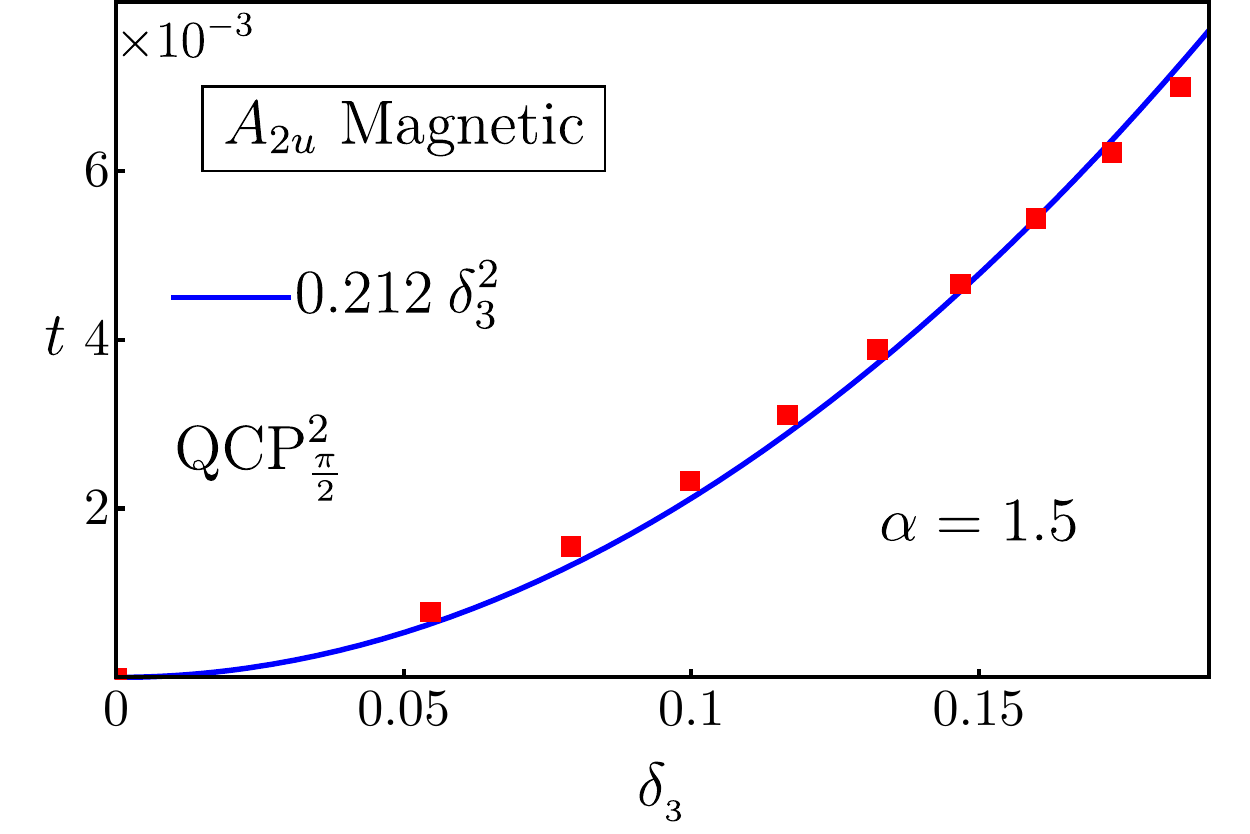}
\includegraphics[width=.24\linewidth]{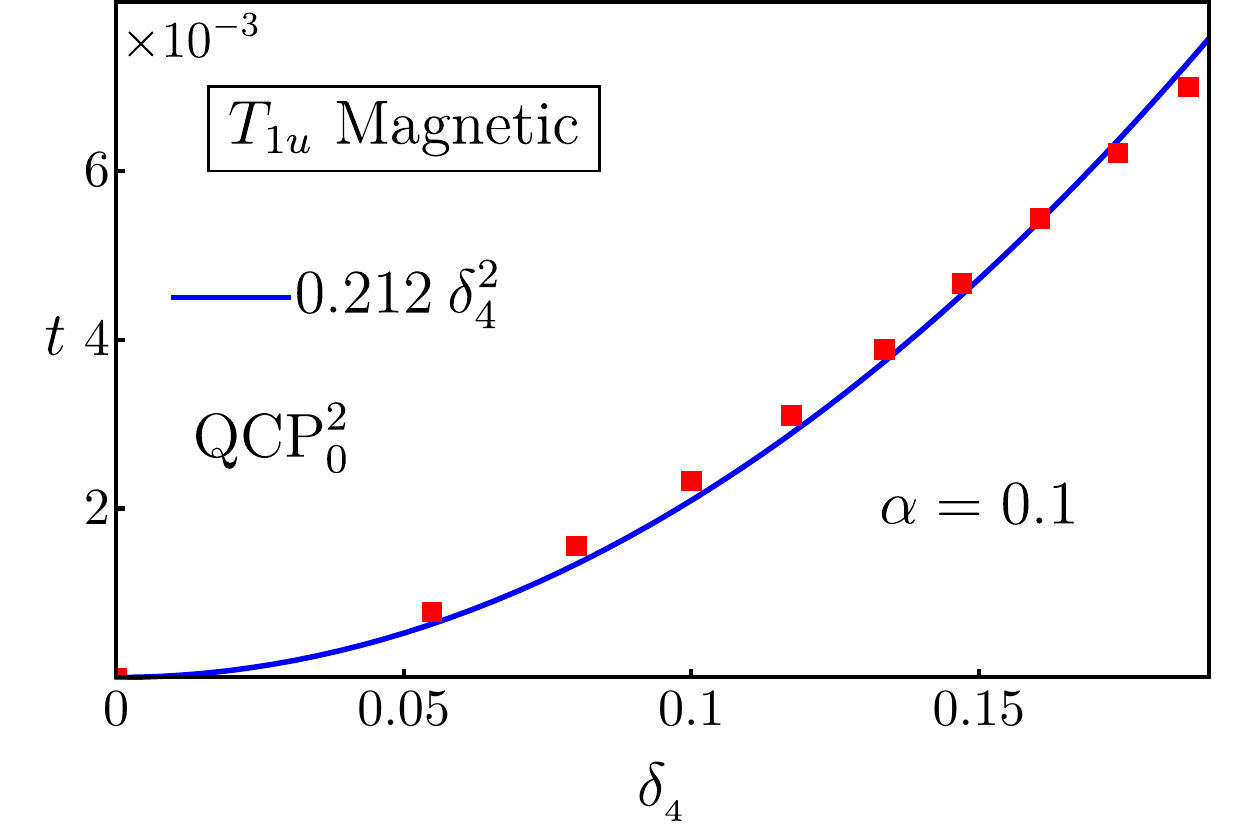}
\caption{Scaling of dimensionless transition temperature ($t_c$) with the reduced distance ($\delta_j=( g_{_j}-g^\ast_{_j} )/g^\ast_{_j}$) from the quantum critical point, located at $g_{_j}=g^\ast_{_j}$ for $j=1,2,3,4$, see Sec.~\ref{SubSec:FiniteTRG}. The mass anisotropy parameter $\alpha$ and nature of the ordered phase for $\delta_j>0$ in each panel are quoted explicitly. Note that $t_c \sim \delta^2_j$ for $\delta \ll 1$, in agreement with field theoretic prediction. The QCPs controlling the Luttinger semimetal to ordered states quantum phase transitions at $t=0$ are quoted in each panel. Red dots represent numerically obtained transition temperature and the blue lines correspond to least-square fit of $t_c \sim \delta^2_j$.  
}~\label{Fig:ScalingTc}
\end{figure*}

Even though our RG analysis in the previous section was performed at zero temperature, one can still find the imprint of various interacting QCPs at finite temperatures. The RG flow equations at finite temperature can be derived from Eq.~(\ref{Eq:RGgeneral}) by taking the $\mu \to 0$ limit in $H^{l}_{jk} (\alpha, t, \mu)$. Recall that temperature introduces a natural \emph{infrared} cutoff for the flow equations $\ell^t_\ast$ [see Eq.~(\ref{Eq:RG_Introduction})]. Physically such an infrared cutoff corresponds to a scenario when the renormalized temperature $t(\ell)$ becomes comparable to the ultraviolet energy $E_\Lambda=\Lambda^2/(2m)$ [see Fig.~\ref{Fig:CriticalFan_Noninteracting}], beyond which the notion of quadratically dispersing fermions becomes moot and the flow equations from Eq.~(\ref{Eq:RGgeneral}) lose their jurisdiction.

To capture the effects of electronic interactions in a LSM at finite-$t$, we run three quartic coupling constants up to a scale $\ell \leq \ell^t_\ast$. Now depending on the bare strength of interactions, two situations arise 
\begin{equation}
\text{(a)} \:\:\: g\left(\ell^t_\ast \right)<1, \quad {\rm or} \quad  
\text{(b)} \:\:\: g\left(\ell^t_\ast \right)>1, \nonumber 
\end{equation}    
respectively representing a disordered LSM (without any long-range ordering) or onset of a BSP at finite temperature. Hence, for a given strength of interaction $g>g_\ast$, where $g_\ast$ is the requisite critical strength of interaction for a BSP at $t=0$, we always find a temperature $t_c$ above (below) which the BSP disappears (appears). We identify $t_c$ as the \emph{critical or transition temperature}. None of the coupling constants diverge for $t>t_c$. All ordered phases can display true long-range order at finite $t$ in three dimensions and $t_c$ corresponds to a genuine transition temperature.

General scaling theory suggests that the transition temperature scales as~\cite{sondhi-RMP, sachdev-book} 
\begin{equation}
t_c \sim \delta^{\nu z},
\end{equation} 
for $\delta \ll 1$, where $\delta=\left( g-g_\ast \right)/g_\ast$ is the \emph{reduced distance} from a QCP, located at $g=g_\ast$ (say). Hence, for interacting LSM, $t_c \sim \delta^{2}$ for $\epsilon=1$ (i.e., the prediction from the leading order $\epsilon$-expansion). The scaling of critical or transition temperature for various choices of coupling constants and resulting BSPs, and different choices of the mass anisotropy parameter $\alpha$, are shown in Fig.~\ref{Fig:ScalingTc}, indicating a fairly good agreement with the field theoretic prediction $t_c \sim \delta^2$ around all QCPs, reported in Table~\ref{Table:CriticalPoints}.

     \subsubsection{Phase diagrams at finite temperature}~\label{subsubSec:energy_entropy_finiteT}

Besides the scaling of the transition temperature, we also investigate the phase diagram of an interacting LSM at finite temperature,  allowing us to demonstrate the competition between condensation energy gain and entropy. For concreteness we focus on the isotropic system ($\alpha=\frac{\pi}{4}$), where this competition is most pronounced. As argued in Sec.~\ref{SubSec:energy_Entropy}, the onset of $s$-wave pairing leads to the maximal gain in condensation energy, while the two nematic orders produce higher entropy in comparison to the former. Two specific cuts of the global phase diagram, Figs.~\ref{Subfig:FiniteT_g1} and \ref{Subfig:FiniteT_g2}, show that while $s$-wave pairing is realized at low temperature, nematicities set in at higher temperature as we increase the strength of nematic interactions ($g_{_1}$ and $g_{_2}$) in the system.

By contrast, when we tune the magnetic interactions (namely $g_{_3}$, $g_{_4}$), an isotropic LSM becomes unstable in favor of two nematic orders at $t=0$. Such an outcome can be substantiated from the simple picture of condensation energy gain, as $A_{2u}$ and $T_{1u}$ magnetic orders accommodate Weyl nodes (yielding more entropy), while nematicities produce anisotropic spectral gap (leading to higher gain of condensation energy), see Sec.~\ref{SubSec:energy_Entropy}. As we tune the strength of the $A_{2u}$ ($T_{1u}$) magnetic interaction, $E_g$ ($T_{2g}$) nematic order sets in at lower and $A_{2u}$ ($T_{1u}$) magnet at higher temperature, see Figs.~\ref{Subfig:FiniteT_g3} and \ref{Subfig:FiniteT_g4}.

The LSM is endowed with the largest entropy in the global phase diagram of interacting spin-3/2 fermions, since $\varrho(E) \sim \sqrt{E}$, in comparison to any BSP. Consequently, the requisite strength of interactions for any ordering increases with increasing temperature, irrespective of the nature of the BSP, see Fig.~\ref{Fig:FiniteT_Summary}. Therefore, our RG analysis for an interacting LSM at finite temperature corroborates the energy-entropy competition picture and substantiates the following outcome: \emph{ordered phases providing larger condensation energy gain are found at lower temperature, while at higher temperature, phases with larger entropy are favored}. This observation is also consistent with the notion of reconstructed band structure and emergent topology, discussed in Sec.~\ref{SubSec:band-topology}. Therefore, the phase diagram of an interacting LSM at finite temperature is guided by topological structure (gapped or nodal) of competing BSPs.

We close this section by answering the following question: Why do we find two magnetic orders in an isotropic LSM at finite temperature, since this system supports only one QCP, see Table~\ref{Table:CriticalPoints}, where all the magnetic orders bear negative scaling dimension (see Table~\ref{Table:scalingdimensions})? Note that any QCP can only be accessed at $t=0$, whereas finite-$t$ introduces an infrared cutoff ($\ell^t_\ast$) for the RG flow of the quartic couplings, and thus prohibits a direct access to any QCP. Hence, at finite-$t$, when magnetic interactions are sufficiently strong, the system can bypass the basin of attraction of QCP$^1_\frac{\pi}{4}$ and nucleate magnetic phases at moderately high temperature.


    \subsection{RG analysis in Luttinger metal}~\label{SubSec:LuttingerMetalRG}

Finally we proceed to the RG analysis when the chemical potential ($\mu$) is placed away from the band touching point. The chemical potential introduces yet another infrared cutoff $\ell^\mu_\ast$ [see Eq.~(\ref{Eq:Infraredcutoffs_Intro})], suggesting that the RG flow equations of the three quartic coupling constants should be stopped when the renoramlized chemical potential $\mu(\ell)$ reaches the scale of the band-width $E_\Lambda=\Lambda^2/(2m)$. At finite temperature and chemical doping, two infrared scales compete and the \emph{smaller} one $\ell_\ast$ (say), given by   
\begin{equation}
\ell_\ast= {\rm min.} \; \left( \ell^\mu_\ast, \ell^t_\ast \right)
\end{equation}
determines the ultimate infrared cutoff for the RG flow of $ g_{_j}$. Now depending on the bare strength of the coupling constants one of the following two situations arises 
\begin{equation}
\text{(a)}\:\:\: g_j (\ell_\ast) >1 \quad {\rm or} \quad 
\text{(b)}\:\:\: g_j (\ell_\ast) <1. \nonumber 
\end{equation}
While (a) indicates onset of a BSP, (b) represents a stable Luttinger metal. The resulting phase diagrams for various choices of chemical potential, temperature, coupling constants and the mass anisotropy parameter ($\alpha$) are shown in Figs.~\ref{Fig:swave_IsotropicLSM}, ~\ref{Fig:dwaves_IsotropicLSM} and ~\ref{Fig:GlobalPD_Intro_FS}.

We note that in an isotropic system and for strong enough nematic interactions an $s$-wave pairing can be realized even for zero chemical doping [see Figs.~\ref{Subfig:FiniteT_g1} and \ref{Subfig:FiniteT_g2}]. With increasing doping the $s$-wave pairing occupies larger portion of the phase diagram, while two nematic phases get pushed toward stronger coupling, see Fig.~\ref{Fig:swave_IsotropicLSM}. However, the $d$-wave pairings do not set in for zero chemical doping. Nonetheless, when the magnetic interactions in the $A_{2u}$ and $T_{1u}$ channels are strong, the presence of a Fermi surface is conducive to the nucleation of $d$-wave pairings, belonging to the $E_g$ and $T_{2g}$ representations, respectively, see Fig.~\ref{Fig:dwaves_IsotropicLSM}.

Now we focus on the anisotropic system. We chose the mass anisotropy parameter $\alpha$ such that at zero chemical doping the repulsive electronic interactions accommodate either nematic or magnetic orders, see Fig.~\ref{Fig:GlobalPD_Intro_noFS}. Specifically for $\alpha=1.5$ (close to $\frac{\pi}{2}$) the system enters into the $T_{2g}$ nematic [see Fig.~\ref{Subfig:T2gswaveMU}] or $A_{2u}$ magnetic [see Fig.~\ref{Subfig:A2udwaveMU}] phase, while for $\alpha=0.1$ (close to $0$) we find $E_g$ nematic [see Fig.~\ref{Subfig:EgswaveMU}] or $T_{1u}$ magnetic [see Fig.~\ref{Fig:T1uMagnetg4smallalpha}] order. For such specific choices of $\alpha$, the QPTs into $T_{2g}$ nematic, $E_g$ nematic, $A_{2u}$ magnetic and $T_{1u}$ magnetic orders are respectively controlled by QCP$^1_{0}$, QCP$^1_{\frac{\pi}{2}}$, QCP$^2_{\frac{\pi}{2}}$ and QCP$^2_0$ [see Sec.~\ref{SubSec:QPTinLSM} and Table~\ref{Table:CriticalPoints}]. At QPT$^1_0$ and QPT$^1_{\frac{\pi}{2}}$ the $s$-wave pairing possesses the largest scaling dimension among all possible local pairings [see Table~\ref{Table:scalingdimensions}]. Hence, in the presence of finite chemical doping, repulsive interactions in the nematic channels become conducive to $s$-wave pairing, as shown in Figs.~\ref{Subfig:T2gswaveMU} and ~\ref{Subfig:EgswaveMU}. On the other hand, at QPT$^2_{\frac{\pi}{2}}$ (QCP$^2_0$), the $d$-wave pairing belonging to the $E_g$ ($T_{2g}$) representation possesses the largest scaling dimension. Therefore, repulsive interactions in the $A_{2u}$ ($g_{_3}$) and $T_{1u}$ ($g_{_4}$) magnetic channels become conducive to the nucleation of $E_g$ [for $\alpha=1.5$] and $T_{2g}$ [for $\alpha=0.1$] $d$-wave pairings, respectively, as shown in Figs.~\ref{Subfig:A2udwaveMU} and ~\ref{Subfig:T1udwaveMU}. We conclude that \emph{nematic and magnetic interactions among spin-3/2 Luttinger fermions are respectively conducive to the $s$-wave and $d$-wave pairings}. Otherwise, at finite chemical doping the excitonic orderings set in only for stronger couplings. Such a generic feature is also consistent with the energy-entropy competition picture as the superconducting phases maximally gap the Fermi surface (yielding optimal gain of condensation energy), while exitonic orders are accompanied by a Fermi surface with constant DoS (producing more entropy).


\begin{figure}[t!]
\includegraphics[width=.95\linewidth]{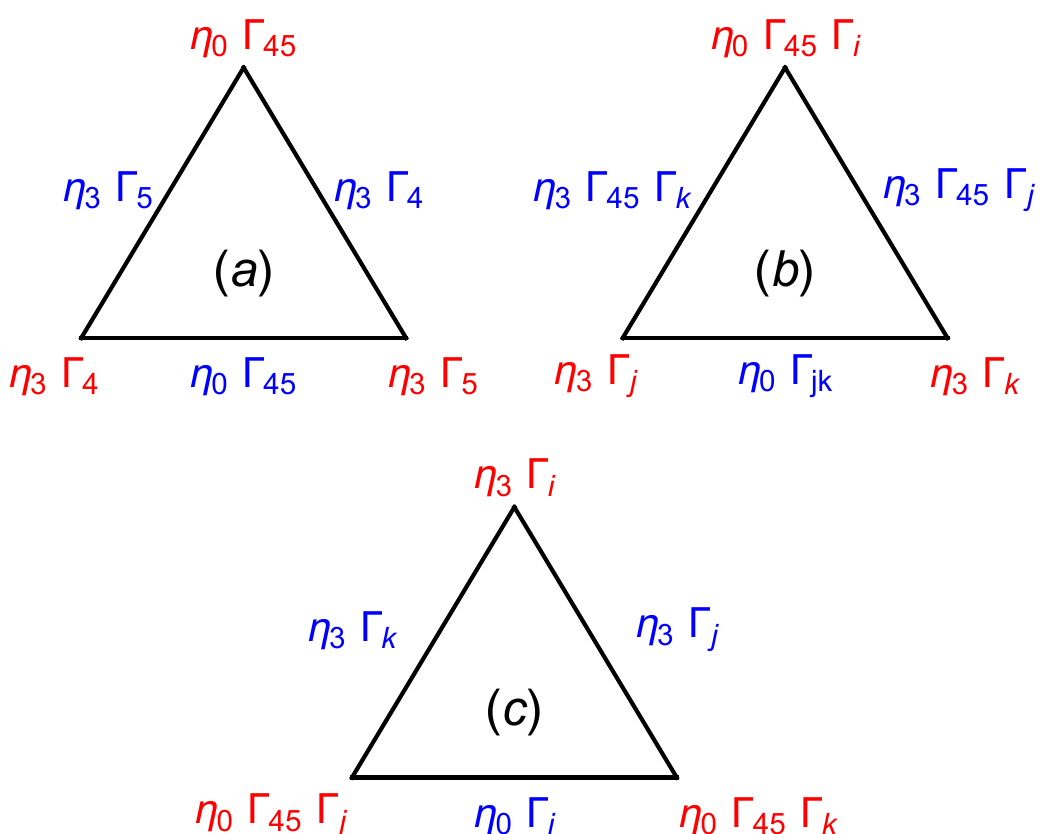}
\caption{Schematic representations of SU(2) symmetry among (a) $E_g$ nematic and $A_{2u}$ magnetic orders, and (b)-(c) components of $T_{2g}$ nematic and $T_{1u}$ magnetic orders (with $i \neq j \neq k =1,2,3$). Three vertices of each triangle are occupied by the order parameter matrices (in red), represented in the Nambu doubled basis ($\Psi_{\rm Nam}$). Three sides of each triangle represent SU(2) rotations. The generators of SU(2) rotations (also in Nambu doubled basis) are shown in blue.      
}~\label{Fig:Triangles_Excitons}
\end{figure}

    \subsection{Competing Orders and Selection Rule }~\label{SubSec:SelectionRule}

So far we have presented an extensive analysis of the role of electronic interactions among spin-3/2 fermions. We showed multiple cuts of the global phase diagram at zero and finite temperature and chemical doping (see Figs.~\ref{Fig:FiniteT_Summary}, \ref{Fig:swave_IsotropicLSM} and \ref{Fig:dwaves_IsotropicLSM}). In addition, we also addressed the imprint of the quadrupolar deformation (or the mass anisotropy parameter $\alpha$) on a such phase diagram (see Figs.~\ref{Fig:GlobalPD_Intro_FS}, \ref{Fig:T1uMagnetg4smallalpha} and \ref{Fig:GlobalPD_Intro_noFS}). Altogether we unearth a rich confluence of competing orders in this system. In this context an important question arises quite naturally: \emph{Is there a selection rule among short-range interactions in different channels (such as nematic and magnetic) and various ordered states (such as $s$- and $d$-wave pairings, magnetic phases and nematic orders) for interacting spin-3/2 fermions?} In this section we attempt to provide an affirmative (at least partially) answer to this question.

To this end it is convenient to express the quartic terms appearing in $H_{\rm int}$ [see Eq.~(\ref{Eq:Interaction_Hamiltonian})] in the Nambu doubled basis. For concreteness, we focus on the relevant interaction channels, namely $\lambda_{1,2,3,4}$. In the Nambu basis these four quartic terms take the form 
\begin{eqnarray}~\label{Eq:HInt_Nambu}
\left\{ 
\begin{array}{c}
  \sum^{3}_{j=1}\left( \Psi^\dagger \Gamma_j \Psi \right)^2 \\
  \sum^{5}_{j=4}\left( \Psi^\dagger \Gamma_j \Psi \right)^2  \\
  \left( \Psi^\dagger \Gamma_{45}\Psi \right)^2 \\
  \sum^{3}_{j=1}\left( \Psi^\dagger \Gamma_j \Gamma_{45} \Psi \right)^2 \\
\end{array}
\right\} \nonumber \\
\to 
\frac{1}{2}\left\{
\begin{array}{c}
 \sum^3_{j=1}\left( \Psi^\dagger_{\rm Nam} \eta_3 \Gamma_j \Psi_{\rm Nam} \right)^2 \\
 \sum^5_{j=4}\left( \Psi^\dagger_{\rm Nam} \eta_3 \Gamma_j \Psi_{\rm Nam} \right)^2 \\
 \left( \Psi^\dagger_{\rm Nam} \eta_0 \Gamma_{45} \Psi_{\rm Nam} \right)^2 \\
 \sum^3_{j=1}\left( \Psi^\dagger_{\rm Nam} \eta_0 \Gamma_j \Gamma_{45} \Psi_{\rm Nam} \right)^2 \\
\end{array}
\right\}.
\end{eqnarray} 
The factor of $1/2$ takes care of the artificial Nambu doubling. The above question can then be reformulated in the following way. When at least one of the coupling constants, say $\lambda_j$, diverges toward $+ \infty$, what is the nature of the resulting BSP; or which one of the source terms, appearing in Eq.~(\ref{Eq:Action_SourceTerm}), simultaneously diverges toward $+\infty$? This is our quest in this section.

\begin{figure}[t!]
\includegraphics[width=.95\linewidth]{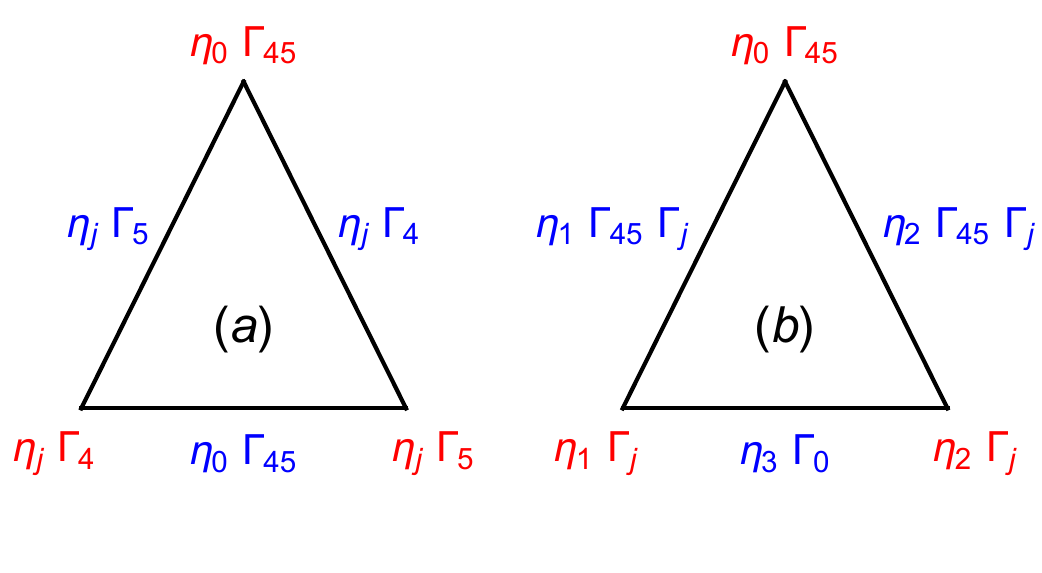}
\caption{Schematic representation of SU(2) symmetry between the $A_{2u}$ magnetic order and $d$-wave pairing belonging to the $E_g$ representation. Notations are the same as in Fig.~\ref{Fig:Triangles_Excitons}. In panel (a) $j=1,2$, whereas in panel (b) $j=4,5$.  
}~\label{Fig:Triangles_Pairing1}
\end{figure}

When the coupling constant $\lambda$ of the quartic interaction $\lambda \left( \Psi^\dagger_{\rm Nam} \; M^{\rm Int}_{{\rm Nam}} \; \Psi_{\rm Nam} \right)^2$ diverges toward $+\infty$ it nucleates a BSP, characterized by the order-parameter $\langle \Psi^\dagger_{\rm Nam} \; M_{\rm or} \; \Psi_{\rm Nam}\rangle$, only if one of the following two conditions is satisfied
\begin{equation}~\label{Eq:Selction_Rule}
{\rm (1)}\:\: M_{\rm or} = M^{\rm Int}_{{\rm Nam}} \quad 
{\rm or} \quad 
{\rm (2)} \:\: \left\{ M_{\rm or}, M^{\rm Int}_{{\rm Nam}} \right\}=0.
\end{equation}
If, on the other hand, $ M^{\rm Int}_{{\rm Nam}}$ and $M_{\rm or}$ have more than one component, then selection rule (2) requires a slight modification, see footnote~\ref{footnote-selectionrule}.

The above selection rule can be justified from the Feynman diagrams shown in Fig.~\ref{Fig:FeynDiag_Susceptibility}. A source term diverges toward $+\infty$ (indicating onset of a BSP) only when the net contribution from diagrams (b) and (c) is \emph{positive}. Feynman diagram (b) gives non-zero and \emph{positive} contribution only when condition (1) from Eq.~(\ref{Eq:Selction_Rule}) is satisfied (due to the ${\rm \bf Tr}$ arising from the fermion bubble). Even though diagram (c) then yields a negative contribution, all togther they still produce a \emph{positive definite} quantity, since the contribution from (b) dominates over (c), as the former one involves ${\rm \bf Tr}$. Hence, when condition (1) is satisfied, interaction $\lambda$ can enhance the propensity toward the formation of a BSP with $M_{\rm or} = M^{\rm Int}_{{\rm Nam}}$.

On the other hand, when condition (1) is not satisfied, only Feynman diagram (c) contributes [due to ${\rm \bf Tr}$ involved in (b)]. The contribution from this diagram is \emph{positive} only when condition (2) is satisfied. By contrast, when $M_{\rm or} \neq M^{\rm Int}_{{\rm Nam},j}$ and $\left[ M_{\rm or}, M^{\rm Int}_{{\rm Nam},j} \right]=0$ the net contribution from (b) and (c) is \emph{negative}. Interaction coupling $\lambda$ then does not support such an ordered phase. All phase diagrams presented in this work support one of these two selection rules (discussed below).

The above selection rule can be phrased in a slightly different fashion. Two ordered phases respectively breaking $O(M_1)$ and $O(M_2)$ symmetries,~\footnote{~\label{footnote:Z2OP}In this notation a $Z_2$ symmetry breaking order-parameter is denoted by an $O(1)$ vector.}
 can reside next to each other (when we tune the interaction strength in a particular channel, or temperature or chemical potential) if they constitute an $O(M)$ symmetric \emph{composite} order parameter, where 
\begin{equation}~\label{Eq:Selction_Rule_2}
M_1, M_2 \leq M \leq M_1+M_2.
\end{equation}
We invite the readers to verify the equivalence between Eqs.~(\ref{Eq:Selction_Rule}) and ~(\ref{Eq:Selction_Rule_2}). 
Note that at the transition between two competing phases the corresponding order-parameters do not need to be equal, as the emergent fermionic quasiparticle spectra inside the competing ordered phases, dictating the condensation energy gains inside the ordered phases, are typically distinct (see Sec.~\ref{SubSec:band-topology}). Furthermore, as the neighboring competing orders always mutually anticommute (at least partially), a coexistence between them with the two pure phases on either side of it is a very generic situation (see footnote~2), as in the coexistence regime two orders can then be rotated into each other. Such a coexistence between two competing phases can be demonstrated from standard mean-field or Ginzburg-Landau theory.
Next, we discuss some prototypical examples to support our claims.

\begin{figure}[t!]
\includegraphics[width=.95\linewidth]{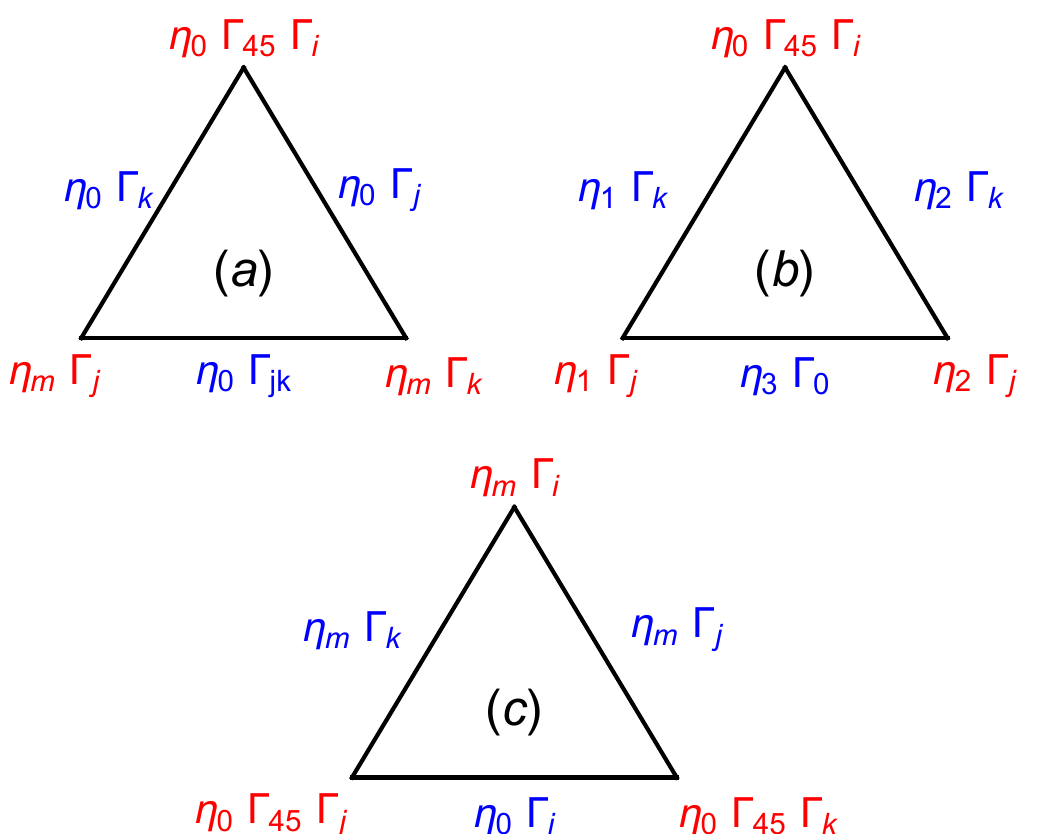}
\caption{Schematic representation of SU(2) symmetry between the components of the $T_{1u}$ magnetic order and $T_{2g}$ $d$-wave pairing, with $i \neq j \neq k =1,2,3$ [for (a), (b), (c)] and $m=1,2$ in (a) and (c). Notations are the same as in Fig.~\ref{Fig:Triangles_Excitons}. 
}~\label{Fig:Triangles_Pairing2}
\end{figure}

\begin{enumerate}

\item The $T_{2g}$ nematic order (an O(3) order-parameter) and $s$-wave pairing (an O(2) order-parameter) constitute an $O(5)$ vector [see Eq.~(\ref{Eq:masspiover2})], and these two ordered phases reside next to each other, see Figs.~\ref{Subfig:FiniteT_g1}, \ref{Fig:swave_IsotropicLSM} (left), \ref{Subfig:T2gswaveMU}, \ref{Subfig:T2gswave}, when we tune the strength of $g_{_1}$.  

\item The $E_{2g}$ nematicity (an O(2) order-parameter) and $s$-wave pairing constitute an $O(4)$ composite order parameter [see Eq.~(\ref{Eq:mass0})], and Figs.~\ref{Subfig:FiniteT_g1}, \ref{Fig:swave_IsotropicLSM} (right), \ref{Subfig:EgswaveMU}, \ref{Subfig:Egswave}  display a confluence of these two ordered phases, as one tunes the interaction $g_{_2}$. 

\item The $E_{g}$ nematicity and $A_{2u}$ magnet (described by an O(1) or $Z_2$ order-parameter) form an $O(3)$ vector [see Fig.~\ref{Fig:Triangles_Excitons}(a)], and these two ordered phases often (in particular, when $g_{_3}$ is tuned) reside next to each other, see Figs.~\ref{Subfig:FiniteT_g3} and \ref{Subfig:A2uEg}.

\item One can construct multiple copies of $O(3)$ composite order parameters by combining the components of $T_{2g}$ nematic and $T_{1u}$ (an O(3) order-parameter) magnetic orders [see Fig.~\ref{Fig:Triangles_Excitons}(b) and (c)], and these two phases can be realized by tuning the quartic interaction $g_{_4}$, see Figs~\ref{Subfig:FiniteT_g3} and \ref{Subfig:T1uT2g}.  

\item $A_{2u}$ magnetic order and $E_g$ $d$-wave pairing (an O(2) order-parameter) can be combined to form $O(3)$ vectors [see Fig.~\ref{Fig:Triangles_Pairing1}]. When the chemical potential is finite and we tune the strength of $g_{_3}$, the system accommodates a paired (magnetic) state at low (high) temperature, see Figs.~\ref{Fig:dwaves_IsotropicLSM} (top) and \ref{Subfig:A2udwaveMU}.     

\item Finally, note that multiple copies of an $O(3)$ vector can be formed by combining the components of $T_{1u}$ magnetic order and $T_{2g}$ $d$-wave pairing (an O(2) order-parameter), see Fig.~\ref{Fig:Triangles_Pairing2}. These two phases reside next to each other at finite chemical doping when we tune the quartic coupling $g_{_4}$, see Figs.~\ref{Fig:dwaves_IsotropicLSM} (bottom) and \ref{Subfig:T1udwaveMU}.  

\end{enumerate}

From the matrix representations of all quartic interactions [see Eq.~(\ref{Eq:HInt_Nambu})] and order parameters [see Eqs.~(\ref{Eq:excitonicOP}) and (\ref{Eq:superconductingOP})] the readers can convince themselves that the above examples are in agreement with our proposed selection rule from Eqs.~(\ref{Eq:Selction_Rule}) and ~(\ref{Eq:Selction_Rule_2}). It is admitted that we arrive at the conclusion from a leading order RG analysis. However, the alternative version of the selection rule [see Eq.~(\ref{Eq:Selction_Rule_2})] solely relies on the internal symmetry among competing orders. We therefore believe that our proposed selection rule is ultimately non-perturbative in nature, which can be tested in numerical experiments, for example. 


\section{Summary and Discussions}~\label{Sec:Conclusion}

To summarize we have presented a comprehensive analysis on the role of electron-electron interactions in a three-dimensional Luttinger system, describing a bi-quadratic touching of Kramers degenerate valence and conduction bands (in the absence of chemical doping) of effective spin-3/2 fermions at an isolated point in the Brillouin zone. This model can succinctly capture the low-energy physics of HgTe~\cite{hgte}, gray-Sn~\cite{gray-sn-1, gray-sn-2}, 227 pyrochlore iridates~\cite{Savrasov, Balents1, Exp:Nakatsuji-1, Exp:Nakatsuji-2, Exp:armitage} and half-Heuslers~\cite{Exp:cava, Exp:felser, binghai}. For concreteness, we focused only on the short-range components of the Coulomb interaction (such as the ones appearing in an extended Hubbard model), and neglected its long-range tail. Due to the vanishing density of states, namely $\varrho(E) \sim \sqrt{E}$, sufficiently weak, but generic local four-fermion interactions are irrelevant perturbations in this system. Consequently, any ordering sets in at an intermediate coupling through quantum phase transitions. We here address the instability of interacting Luttinger fermions at finite coupling within the framework of a renormalization group analysis, controlled by a ``\emph{small}" parameter $\epsilon$, where $\epsilon=d-2$. Notice that in two spatial dimensions a bi-quadratic band touching (similar to the situation in Bernal-stacked bilayer graphene) yields a \emph{constant} density of states, and local four-fermion interactions are \emph{marginal} in $d=2$. Hence, our renormalization group analysis is performed about the \emph{lower-critical} two spatial dimensions. In this framework all quantum phase transitions from a disordered Luttinger (semi)metal to any ordered phase take place at $g^\ast_{_j} \sim \epsilon$, where $g_{_j}$ is the dimensionless coupling constant, and for three dimensions $\epsilon=1$. We here restrict ourselves to repulsive (at the bare level) electron-electron interactions and present multiple cuts of the global phase diagram at zero and finite temperature (see Figs.~\ref{Fig:FiniteT_Summary} and ~\ref{Fig:GlobalPD_Intro_noFS}) and finite chemical doping (see Figs.~\ref{Fig:swave_IsotropicLSM}, ~\ref{Fig:dwaves_IsotropicLSM} and ~\ref{Fig:GlobalPD_Intro_FS}).

Using the renormalization group analysis, we show that in an isotropic system an $s$-wave pairing and two nematic orders are the prominent candidates for a broken symmetry phase at zero temperature and chemical doping. While the $s$-wave pairing only breaks the global $U(1)$ symmetry and produces a uniform mass gap, two nematic orders, transforming under the $T_{2g}$ and $E_g$ representations, produce lattice distortion along the $C_{3v}$ and $C_{4v}$ axes, respectively. Such ordered phases can describe either time-reversal symmetry preserving insulators or topological Dirac semimetals. However, a collection of strongly correlated gapless spin-3/2 fermions do not show any noticable propensity toward the nucleation of any magnetic order or $d$-wave pairings in an isotropic system at least when $t=0$. This is so, because the magnetic orders ($d$-wave pairings) produce gapless Weyl nodes (nodal loops) around which the density of states vanishes as $\varrho(E) \sim |E|^2 (|E|)$, while the former three orders support gapped spectra. Hence, the magnetic orders and $d$-wave pairings are energetically inferior to $s$-wave pairing and electronic nematicities. However, with increasing temperature, one finds a smooth crossover from nematic to magnetic phases, as shown in Fig.~\ref{Fig:FiniteT_Summary}. Hence, \emph{energetically superior orders are found at low temperatures, whereas at higher temperature broken symmetry phases possess larger entropy}. This energy-entropy competition is discussed in Sec.~\ref{SubSec:energy_Entropy} and summarized in Fig.~\ref{Fig:energyentropy}.

We identify that the mass anisotropy ($\alpha$), measuring the quadrupolar distortion in the Luttinger system, can be a useful non-thermal tuning parameter to further explore the territory of strongly interacting spin-3/2 fermions~\cite{goswami-roy-dassarma, Herbut-2}. In particular, we find that strong quadrupolar distortions can be conducive for various magnetic orderings even at zero temperature. Specifically, when the electronic dispersion along the $C_{3v}$ axes becomes almost flat (realized as $\alpha \to \frac{\pi}{2}$) the singlet $A_{2u}$ magnetic order stabilizes at $t=0$, see Figs.~\ref{Subfig:A2udwaveMU} and \ref{Subfig:A2uEg}. On the other hand, when Luttinger fermions are almost non-dispersive along the $C_{4v}$ axes (realized when $\alpha \to 0$), the system becomes susceptible toward the formation of a triplet $T_{1u}$ magnetic order, see Figs.~\ref{Fig:T1uMagnetg4smallalpha} and \ref{Subfig:T1uT2g}. For these two limiting scenarios the above two magnetic orders become (almost) \emph{mass} [see Secs.~\ref{SubSubSec:Susceptibility_piover2} and ~\ref{SubSubSec:Susceptibility_0}], and their nucleation becomes energetically beneficial even at zero temperature.

Irrespective of these details, we realize that all quantum phase transitions from the Luttinger semimetal to symmetry-breaking phases are continuous and controlled by various critical points, see Sec.~\ref{SubSec:QPTinLSM} and Table~\ref{Table:CriticalPoints}. To the leading order in the $\epsilon$-expansion, the universality class of these transitions is characterized by the (a) correlation length exponent $\nu^{-1}=\epsilon$ and (b) dynamic scaling exponent $z=2$ [see Sec.~\ref{SubSec:exponents}]. The presence of such quantum critical points manifests itself through the scaling of the transition temperature $t_c \sim |\delta|^{\nu z}$, yielding $t_c \sim |\delta|^2$ for $d=3$ or $\epsilon=1$, see Fig.~\ref{Fig:ScalingTc}, where $\delta$ is the reduced distance from the critical point.

Finally, we introduce (chemical) doping as another non-thermal tuning parameter to map out the global phase diagram of an interacting Luttinger metal, see Sec.~\ref{SubSec:LuttingerMetalRG}. Since any paired state maximally gaps the Fermi surface, its appearance at the lowest temperature is quite natural, at least when $|\mu|>0$. By contrast, excitonic phases (insulators or semimetals) become metallike (possessing a finite density of states) at finite chemical doping, according to the Luttinger theorem~\cite{Luttinger_Theorem}. Therefore, particle-hole orders are accompanied by higher entropy due to the presence of a Fermi surface (with a constant density of states). To demonstrate the energy-entropy competition in a metallic system, we choose the mass anisotropy parameter $\alpha$ such that at $\mu=0$ the system only supports excitonic orders. Upon raising (lowering) the chemical potential to the conduction (valence) band, we observe that a superconducting order develops at low temperature and the excitonic order gets pushed toward higher temperature and stronger interactions, see Figs.~\ref{Fig:dwaves_IsotropicLSM} and ~\ref{Fig:GlobalPD_Intro_FS}. Therefore, the overall structure of the global phase diagram is compatible with the energy-entropy competition, dictated by the emergent band topology of competing broken symmetry phases.

Furthermore, we also identify a definite ``selection rule" among competing phases [see Sec.~\ref{SubSec:SelectionRule}]. From multiple cuts of the global phase diagram of a collection of strongly interacting spin-3/2 fermions, we find that two phases can reside in close vicinity of each other if the order-parameters describing two distinct phases can be combined to form a composite order-parameter. In other words, two ordered phases, respectively described by $O(M_1)$ and $O(M_2)$ symmetric order-parameters [see footnote~\ref{footnote:Z2OP}], can reside next to each other only if one can construct an $O(M)$ symmetric composite-vector, where  $M_1,M_2 \leq M \leq M_1+M_2$, from the elements of two individual order parameters. This is so because when an interaction favors $O(M_1)$ symmetry breaking order, it also enhances the \emph{scaling dimension} of an $O(M_2)$ symmetry breaking order and vice-versa, when the above selection rule is satisfied. Therefore, by tuning a suitable parameter (such as temperature, mass anisotropy parameter, chemical potential) one can induce a transition between two competing phases. As an immediate outcome of this selection rule we realize that while repulsive interactions (short-range) in the nematic channels are conducive to $s$-wave pairing [see Figs.~\ref{Subfig:FiniteT_g1}, ~\ref{Subfig:FiniteT_g2}, ~\ref{Fig:swave_IsotropicLSM}, ~\ref{Subfig:T2gswaveMU}, ~\ref{Subfig:EgswaveMU}], magnetic interactions favor nucleation of $d$-wave pairings [see Figs.~\ref{Fig:dwaves_IsotropicLSM}, \ref{Subfig:A2udwaveMU}, \ref{Subfig:T1udwaveMU}] among Luttinger fermions.

A few specific cuts of the global phase diagram corroborate with ones extracted experimentally in Ln$_2$Ir$_2$O$_7$~\cite{takagi} and half-Heusler compounds~\cite{Exp:Paglione-1}. For exmaple, a finite temperature phase transition from $A_{2u}$ or all-in all-out magnetic order to a Luttinger semimetal has been observed in majority of 227 pyrochlore iridates (except for Ln=Pr)~\cite{takagi}, and Fig.~\ref{Subfig:FiniteT_g3} qualitatively captures this phenomena (for strong enough $g_{_3}$). By contrast, Pr$_2$Ir$_2$O$_7$ supports a large anomalous Hall effect below $1.5$K~\cite{AHE-1,AHE-2,AHE-3}. Note that a triplet $T_{1u}$ or 3-in 1-out magnetic order supports anomalous Hall effect due to the presence of Weyl nodes in the ordered state (see Sec.~\ref{SubSec:band-topology} and Ref.~\cite{goswami-roy-dassarma}) and the phase diagram from Fig.~\ref{Subfig:FiniteT_g4} shows the appearance of $T_{1u}$ magnetic order at finite temperature for strong enough interaction ($g_{_4}$). It is worth recalling that ARPES measurements strongly suggests that isotropic Luttinger semimetal describes the normal states of both Nd$_2$Ir$_2$O$_7$ and Pr$_2$Ir$_2$O$_7$~\cite{Exp:Nakatsuji-2,Exp:Nakatsuji-1}. On the other hand, half-Heusler compounds LnPdBi display a confluence of magnetic order and superconductivity~\cite{Exp:Paglione-1} and the phase diagrams shown in Figs.~\ref{Fig:dwaves_IsotropicLSM}, ~\ref{Subfig:A2udwaveMU} and ~\ref{Subfig:T1udwaveMU} capture such competition (at least qualitative). This connection can be further substantiated from a recent penetration depth ($\Delta \lambda$) measurement in YPtBi~\cite{Exp:Paglione-2}, suggesting $\Delta \lambda \sim T/T_c$ (roughly) at low enough temperature, where $T_c=0.78$K is the superconducting transition temperature. Such a T-linear dependence of the penetration depth can result from gapless BdG fermions yielding $\varrho(E) \sim |E|$. Any $d$-wave pairing, producing two nodal loops (see Table~\ref{Tab:Spectra_dwaves}), is therefore a natural candidate for the paired state in half-Heuslers (see also Refs.~\cite{brydon-1, Herbut-3, roy-nevidomskyy, savary-2}). It is admitted that more microscopic analysis is needed to gain further insights into the global phase diagram of strongly interacting spin-3/2 fermions in various materials, which we leave for future investigation.

The energy-entropy competition and the proposed selection rule among competing orders provide valuable insights into the overall structure of the global phase diagram of strongly interacting spin-3/2 fermions. Various cuts of the phase diagram, which we exposed by pursuing an unbiased renormalization group analysis, corroborate (at least qualitatively) the former two approaches. These approaches are not limited to interacting spin-3/2 fermions living in three dimensions. The methodology is applicable for a large set of strongly interacting multi-band systems, among which two dimensional Dirac (semi)metals, doped Bernal-stacked bilayer graphene (supporting bi-quadratic band touchings in two dimensions)~\cite{graphene-RMP, Balatsky, vafek}, twisted bilayer graphene near the so called magic angle~\cite{tblg-1, tblg-2, tblg-3}, three-dimensional Dirac or Weyl materials~\cite{Weyl-RMP}, doped topological (Kondo-)insulators (described by massive Dirac fermions)~\cite{TI-RMP-1, TI-RMP-2, TKI-Review} and nodal loop metals~\cite{roy-NLSM, Shouvik-NLSM} are the prominent and experimentally pertinent ones. In the future we will systematically study these systems, which should allow us to gain further insights into the global phase diagram of correlated materials, appreciate the role of emergent topology inside various broken symmetry phases, and search for possible routes to realize unconventional high temperature superconductors.


\acknowledgements

This work was in part supported by Deutsche Forschungsgemeinschaft under grant SFB~1143 and a start-up grant from Lehigh University (B.R.).


\appendix

\section{Organizing principle and selection rule in doped graphene}~\label{appendix:Dirac}

The low-energy Dirac excitations in monolayer graphene are captured by a sixteen component Nambu-doubled spinor $\Psi^\top_{\rm Nam}= \left(\Psi_{\vec{k}}, i \sigma_2 \tau_1 \Psi^\star_{-\vec{k}} \right)$, where $\Psi^\top_{\vec{k}}=\left( \Psi_{\vec{k},\uparrow}, \Psi_{\vec{k},\downarrow} \right)$, $\Psi^\top_{\vec{k},\sigma}=\left( \Psi_{\vec{K}+\vec{k},\sigma}, \Psi_{-\vec{K}+\vec{k},\sigma} \right)$, and
\begin{equation}
\Psi^\top_{\tau \vec{K}+\vec{k},\sigma}=\left( u_{\tau \vec{K} + \vec{k},\sigma}, v_{\tau \vec{K} + \vec{k},\sigma} \right). 
\end{equation}
Here, $u_{\tau \vec{K} + \vec{k},\sigma}$ and $v_{\tau \vec{K} + \vec{k},\sigma}$ are fermion annihilation operators with momentum $\vec{k}$ on the sublattices A and B of the honeycomb lattice, respectively, with the Fourier components near two inequivalent valleys at $ \tau \vec{K}$ with $\tau=\pm$, and spin projections $\sigma=\uparrow, \downarrow$. The two sets of Pauli matrices $\{ \sigma_\mu \}$ and $\{ \tau_\mu \}$ respectively operate on the spin and valley indices. In this Nambu-doubled basis the sixteen-dimensional Dirac Hamiltonian reads as 
\begin{equation}
\hat{h}^{\rm Nam}_{\rm D}= v \left( \eta_3 \sigma_0 \tau_3 \alpha_1 k_x -\eta_3 \sigma_0 \tau_0 \alpha_2 k_y \right).
\end{equation} 
Two sets of Pauli matrices $\{ \eta_\mu \}$ and $\{ \alpha_\mu \}$ respectively operate on the Nambu or particle-hole and sublattice indices, and $\vec{k}$ is measured from the respective valley.

In this basis the dominant component of the nearest-neighbor local four-fermion interactions takes the form $\big( \Psi^\dagger_{\rm Nam} \eta_3 \sigma_0 \tau_0 \alpha_3 \Psi_{\rm Nam} \big)^2$, and therefore $\hat{\rm I}=\eta_3 \sigma_0 \tau_0 \alpha_3$ (see Sec.~IIB3). It supports charge-density-wave and spin-triplet $f$-wave pairing respectively at zero and finite chemical doping, as recently found from a non-perturbative functional RG analysis~\cite{scherrerCDWfwave}. These two ordered states are respectively described by the fermion bilinears $\Psi^\dagger_{\rm Nam} \hat{\rm O}_{\rm CDW} \Psi_{\rm Nam}$ and $\Psi^\dagger_{\rm Nam} \hat{\rm O}_{f-{\rm wave}} \Psi_{\rm Nam}$, with 
\begin{eqnarray}
\hat{\rm O}_{\rm CDW}=\eta_3 \sigma_0 \tau_0 \alpha_3, \:\:
\hat{\rm O}_{f-{\rm wave}}=\eta_{j} {\boldsymbol \sigma} \tau_3 \alpha_0,
\end{eqnarray}
where $j=1,2$, and ${\boldsymbol \sigma}=(\sigma_1, \sigma_2, \sigma_3)$. Notice that nucleation of these two ordered states for nearest-neighbor interactions is consistent with the selection rule shown in Eq.~(\ref{Eq:SelectionRule_Intro}), as (1) $\hat{\rm I} \equiv \hat{\rm O}_{\rm CDW}$ and (2) $\{\hat{\rm I} , \hat{\rm O}_{f-{\rm wave}} \}=0$. Furthermore, the fact that by tuning chemical doping one can induce a transition from charge-density-wave to $f$-wave pairing is also consistent with the other selection rule that $\{ \hat{\rm O}_{\rm CDW}, \hat{\rm O}_{f-{\rm wave}}\}=0$ and these two order parameters can be combined to form an O(4) supervector $\{\eta_3 \sigma_0 \tau_0 \alpha_3, \eta_{j} {\boldsymbol \sigma} \tau_3 \alpha_0  \}$ for $j=1$ or $2$, see Sec.~\ref{SubSec:SelectionRule}.

On the other hand, the dominant component of the next-nearest-neighbor repulsion on honeycomb lattice is captured by the four-fermion term $\big( \Psi^\dagger_{\rm Nam} \eta_3 {\boldsymbol \sigma} \tau_3 \alpha_3 \Psi_{\rm Nam} \big)^2$, and therefore $\hat{\rm I}=\eta_3 {\boldsymbol \sigma} \tau_3 \alpha_3$ (see Sec.~IIB3). Quantum Monte Carlo simulation (non-perturbative) shows that such interaction respectively supports a quantum spin Hall insulator and spin-singlet $s$-wave pairing for zero and finite chemical doping, respectively~\cite{assaadQSHIswave}. These two ordered phases are described by the fermion bilinears $\Psi^\dagger_{\rm Nam} \hat{\rm O}_{\rm QSHI} \Psi_{\rm Nam}$ and $\Psi^\dagger_{\rm Nam} \hat{\rm O}_{s-{\rm wave}} \Psi_{\rm Nam}$, respectively, with 
\begin{eqnarray}
\hat{\rm O}_{\rm QSHI}=\eta_3 {\boldsymbol \sigma} \tau_3 \alpha_3, \:\:
\hat{\rm O}_{s-{\rm wave}}=\eta_{j} \sigma_0 \tau_3 \alpha_0,
\end{eqnarray}    
where $j=1,2$. Nucleation of these two ordered states is also compatible with the selection rule from Eq.~(\ref{Eq:SelectionRule_Intro}), since (1) $\hat{\rm I} \equiv \hat{\rm O}_{\rm QSHI}$ and (2) $\{\hat{\rm I} , \hat{\rm O}_{s-{\rm wave}} \}=0$. The fact that these two phases are neighbors as one tunes the chemical doping in the system is consistent with the fact that they can be be combined together to form an O(5) supervector $\{ \eta_3 {\boldsymbol \sigma} \tau_3 \alpha_3, \eta_1 \sigma_0 \tau_3 \alpha_0, \eta_2 \sigma_0 \tau_3 \alpha_0 \}$ and $\{ \hat{\rm O}_{\rm QSHI}, \hat{\rm O}_{s-{\rm wave}} \}=0$ (see Sec.~\ref{SubSec:SelectionRule}).

Therefore, our proposed selection rule among interaction channel and competing ordered phases is also operative in other correlated systems, such as monolayer graphene~\cite{scherrerCDWfwave, assaadQSHIswave}. Most importantly, these two non-perturbative and unbiased numerical works strongly suggest the non-perturbative nature of the selection rule, which we derive here by exploiting the internal symmetry among the competing orders. Also the fact that these two numerical analyses found insulators (superconductors) at zero (finite) chemical doping, is also compatible with the organizing principle from Sec.~IIB1, based on the energy and entropy argument. For detailed RG analysis, supporting these claims see Ref.~\cite{szabo-roy:dopedhoneycomb}.


\section{Specific heat and compressibility}~\label{Append:freeenergy}

This appendix is devoted to present the scaling of specific heat ($C_V$) and compressibility ($\kappa$) in a non-interacting Luttinger system, when the chemical potential ($\mu$) is placed away from the band touching point. We begin with the expression for the free-energy density in this system, given by (after setting $k_B=1$)
\begin{equation}
f=-D T \sum_{\tau=\pm} \int \frac{d^3 {\bf k}}{(2 \pi)^3} \; \ln \left[ 2 \cosh \left( \frac{E^\tau_{\bf k}}{2 T}\right) \right],
\end{equation}
where $D$ is the degeneracy of the valence and conduction band, hence $D=2$, and 
\begin{equation}
E^\tau_{\bf k}= \frac{k^2}{2 m} + \tau \mu.
\end{equation}
The chemical potential and momentum ($k$) are measured from the band touching point. One can rewrite the above expression for the free energy density as 
\begin{eqnarray}
f &=&-\frac{D}{2} \sum_{\tau=\pm} \int \frac{d^3 {\bf k}}{(2 \pi)^3} E^\tau_{\bf k} \nonumber \\
&-& D T \sum_{\tau=\pm} \int \frac{d^3 {\bf k}}{(2 \pi)^3} \ln \left[ 1+ \exp\left( -\frac{E^\tau_{\bf k}}{T} \right) \right].
\end{eqnarray}
The first term is independent of temperature and not important for the thermodynamic properties of the system. So we focus only on the second term. After proper rescaling of variables the free-energy density becomes 
\begin{eqnarray}
f &=& -D T^{5/2} \frac{(2m)^{3/2}}{4 \pi^2} \; \sum_{\tau=\pm} \int^\infty_0 dy \; \sqrt{y} \; \ln \left[ 1+ e^{-y-\tau \tilde{\mu}} \right], \nonumber \\
\end{eqnarray}
leading to Eq.~(\ref{Eq:freeenergy_Intro}), where $\tilde{\mu}=\mu/T$, and ${\rm Li}_s(z)$ represents the polylogarithm function of order $s$ and argument $z$. For $\tilde{\mu} \ll 1$ the free-energy density reads as 
\begin{equation}
f=-D T^{5/2} \frac{(2m)^{3/2}}{8 \pi^{3/2}} \left[ a + b \; \frac{\mu^2}{T^2} + c \; \frac{\mu^4}{T^4} + \; \cdots \right], 
\end{equation} 
where 
\begin{eqnarray}
a &=& \frac{1}{2} \left( 4-\sqrt{2} \right) \; \zeta \left( \frac{5}{2}\right) \approx 1.7344, \nonumber \\
b &=& \left( 1-\sqrt{2} \right) \; \zeta \left( \frac{1}{2}\right) \approx 0.6049, \nonumber \\
c &=& \frac{1}{12} \left( 1-4 \sqrt{2} \right) \; \zeta \left( -\frac{3}{2}\right) \approx 0.00989.
\end{eqnarray}
From the above expression of the free-energy density we arrive the expression for specific heat [see Eq.~(\ref{Eq:specificheat_Intro})] and compressibility [see Eq.~(\ref{Eq:compressibility_Intro})]. From these two expressions we finally arrive at the following universal ratio  
\begin{equation}
\frac{C_V/T}{\kappa}=\frac{15 \left( 4-\sqrt{2}\right)}{16 \left( 1-\sqrt{2}\right)} \times \frac{\zeta\left( 5/2\right)}{\zeta\left( 1/2\right)} \approx 5.37611,
\end{equation} 
reported in Eq.~(\ref{Eq:universalratio_Intro}). This number is a characteristic of a $z=2$ scale invariant fixed point in $d=3$.


\begin{figure}[t!]
\includegraphics[width=.85\linewidth]{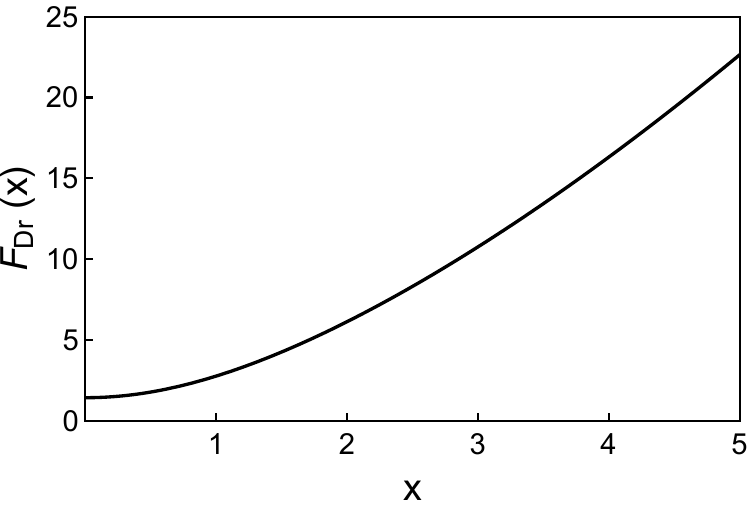}
\caption{Scaling of the universal function $F_{\rm Dr} (x)$ with its argument $x$, appearing in the expression for the Drude conductivity of Luttinger fermions, see Eq.~(\ref{Eq:Drude}). 
}~\label{Fig:Drude_LSM}
\end{figure}

\section{Dynamic conductivity}~\label{Append:dynamic conductivity}

This appendix is devoted to disclose some key steps of the computation of the dynamic conductivity in Luttinger system. We focus on an isotropic system (for the sake of simplicity) and explicitly compute the $z$-component of the conductivity ($\sigma_{zz}$). Due to the cubic symmetry $\sigma_{zz}=\sigma_{xx}=\sigma_{yy} \equiv \sigma$ (say). To this end we use the Kubo formula and compute the polarization bubble as a function of external (Matsubara) frequency
\begin{eqnarray}
\Pi_{zz}(i \omega_n) &=&- \frac{e^2}{\beta} \sum_{m} \int_{\bf k} \; {\bf Tr} \bigg[ \hat{j}_z \: G_0 \left(i p_m, {\bf k} \right) \: \hat{j}_z \nonumber \\ 
&\times& G_0 \left(i p_m+i \omega_n , {\bf k} \right) \bigg],
\end{eqnarray}
where $e$ is the electronic charge, $\beta$ is the inverse temperature and the current operator along the $z$ direction is given by 
\begin{equation}
\hat{j}_z=-\frac{1}{2m} \left[ \sqrt{3} \; k_y \Gamma_1 + \sqrt{3} \; k_x \Gamma_2 + 2 \; k_z \Gamma_5 \right].
\end{equation}
In the spectral representation the noninteracting Greens function reads as 
\begin{equation}
G_0 (i \omega_n, {\bf k})= \int^{\infty}_{-\infty} \: \frac{d \epsilon}{2 \pi} \; \frac{{\mathcal A} (\epsilon, {\bf k})}{i \omega_n-\epsilon},
\end{equation}
where 
\begin{eqnarray}
{\mathcal A} (\epsilon, {\bf k}) &=& \pi \left( 1 + \sum^5_{j=1} \Gamma_j \hat{d}_j \right) \: \delta\left( \omega+\mu- \frac{k^2}{2 m} \right) \\
&+& \pi \left( 1 - \sum^5_{j=1} \Gamma_j \hat{d}_j \right) \: \delta\left( \omega+\mu+ \frac{k^2}{2 m} \right). \nonumber 
\end{eqnarray}
Components of $\hat{\bf d}$ are shown in Eq.~(\ref{d-hats}). After performing the analytic continuation $i \omega \to \omega + i \eta$, we find the dynamic conductivity using the Kubo formula  
\begin{equation}
\sigma_{zz}(\omega)= \frac{\Im \: \Pi_{zz} \left( i \omega \to \omega + i \eta \right)}{\omega}.
\end{equation}
From the above expression we obtain both Drude and inter-band components of the dynamic conductivity, respectively shown in Eqs.~(\ref{Eq:Drude}) and (\ref{Eq:Inter-band}). The universal function $F_{\rm Dr} (x)$ appearing in the expression of the Drude component in Eq.~(\ref{Eq:Drude}) is given by 
\begin{equation}~\label{Eq:Drude_Function}
F_{\rm Dr} (x)= \frac{32}{3} \int^\infty_0 \; du \; u^{3/2} \: \sum_{\tau=\pm} \sech^2 \left( u + \tau \frac{\mu}{2 T}\right). 
\end{equation}
The scaling of this function is shown in Fig.~\ref{Fig:Drude_LSM}.


\section{Details of Luttinger model}~\label{Append:Luttingerdetails}

\begin{table}[t!]
\renewcommand{\arraystretch}{1.4}
\begin{tabular}{|c|c|c|}
\hline
FPs & $\left( g^\ast_0, g^\ast_1, g^\ast_2 \right) \times 10^{3}$ & EVs of SM  \\ 
\hline
Gaussian & $(0,0,0)$ & $(-1,-1,-1) \epsilon$ \\
\rowcolor{RowColor}
QCP$^1_{\frac{\pi}{4}}$ & $\left( 2.24, 2.03, 2.03 \right) \epsilon$ & $(1, -0.870, -0.704) \epsilon$ \\
QCP$^1_{\frac{\pi}{2}}$ & $\left( 1.49, 1.47, 1.26 \right) \epsilon$ & $(-1.075, 1, -0.654) \epsilon$ \\
\rowcolor{RowColor}
QCP$^2_{\frac{\pi}{2}}$ & $\left( -2.93, -2.84, 3.76 \right) \epsilon$ & $(-1.169, 1, -0.933) \epsilon$ \\
BCP$_{\frac{\pi}{2}}$ & $\left( -1.58, -1.33, 5.54 \right) \epsilon$ & $(1.271, 1, -0.527) \epsilon$ \\
\rowcolor{RowColor}
QCP$^1_{0}$ & $\left( 1.95, 1.71, 1.94 \right) \epsilon$ & $(1, -0.889, -0.642) \epsilon$ \\
QCP$^2_{0}$ & $\left( -5.98, 4.80, -5.87 \right) \epsilon$ & $(-1.131, 1, -0.134)\epsilon$ \\
\rowcolor{RowColor}
BCP$_{0}$ & $\left( -5.60, 4.97, -5.48 \right) \epsilon$ & $(-1.075, 1, 0.132 ) \epsilon$ \\
\hline
\end{tabular}
\caption{ Three eigenvalues (EVs) of the stability matrix (SM) $M\left( g_{_0}, g_{_1}, g_{_2} \right)$, defined in Eq.~(\ref{Eq:StabilityMatrix}) at various fixed points (FPs) located at $\left( g^\ast_{_0}, g^\ast_{_1}, g^\ast_{_2} \right)$, see Table~\ref{Table:CriticalPoints}. The trivial Guassian fixed point is found for any arbitrary value of $\alpha$. 
}~\label{Tab:StabilityMatrix}
\end{table}

In this appendix we present some essential details of the Luttinger model, introduced in Sec.~\ref{Sec:Luttinger_Model}. The components of a five-dimensional unit vector $\hat{\bf d}(\hat{\bf k})$, appearing in the Luttinger Hamiltonian [see Eq.~(\ref{Eq:Luttinger_Model})], are given by 
\begin{eqnarray}~\label{d-hats}
\hat{d}_1 &=&	\frac{i\left[Y^1_{2} + Y^{-1}_{2}\right]}{\sqrt{2}} = 	\frac{\sqrt{3}}{2} \sin 2 \theta \, \sin \phi = \sqrt{3} \; \hat{k}_y \hat{k}_z, \nonumber \\
\hat{d}_2 &=& \frac{\left[Y^{-1}_{2} + Y^{1}_{2}\right]}{\sqrt{2}} = \frac{\sqrt{3}}{2} \sin 2 \theta \cos \phi = \sqrt{3} \; \hat{k}_x \hat{k}_z, \nonumber \\
\hat{d}_3 &=& \frac{i\left[Y^{-2}_{2} + Y^{-2}_{2}\right]}{\sqrt{2}} = \frac{\sqrt{3}}{2} \sin^2 \theta \sin 2 \phi = \sqrt{3} \; \hat{k}_y \hat{k}_x, \nonumber \\
\hat{d}_4 &=& \frac{\left[Y^{-2}_{2} + Y^{2}_{2}\right]}{\sqrt{2}} = \frac{\sqrt{3}}{2} \sin^2 \theta \cos 2 \phi = \frac{\sqrt{3}}{2} \left[ \hat{k}^2_x-\hat{k}^2_y \right], \nonumber \\
\hat{d}_5 &=& Y^0_2 = \frac{1}{2} \left( 3 \cos^2\theta-1 \right)=\frac{1}{2} \left[ 2 \hat{k}^2_z-\hat{k}^2_x-\hat{k}^2_y \right],
\end{eqnarray}
where $Y^m_l \equiv Y^m_l (\theta, \phi)$, $\hat{d}_j \equiv \hat{d}_j(\hat{k})$, and $\theta$ and $\phi$ are the polar and azimuthal angles in the momentum space, respectively. Five mutually anti-commuting $\Gamma$ matrices are [see also Eq.~(\ref{Eq:Gammarepresentation})]
\begin{eqnarray}
\Gamma_1 &=& \frac{1}{\sqrt{3}} \{J_y,J_z\},
\Gamma_2 = \frac{1}{\sqrt{3}} \{J_x,J_z\},
\Gamma_3 = \frac{1}{\sqrt{3}} \{J_x,J_y\}, \nonumber\\
\Gamma_4 &=&\, \frac{1}{\sqrt{3}} \left[J_x^2 - J_y^2 \right],
\Gamma_5 = \frac{1}{3}\left[2 J_z^2 - J_x^2 - J_y^2 \right], 
\end{eqnarray}
where $\{ A,B\} \equiv A B+B A$, and ${\bf J}$ are three spin-3/2 matrices. Both $\hat{{\bf d}}$ and ${\bf \Gamma}$ transform as vectors under the cubic point group ($O_h$), and their scalar product yields the Luttinger Hamiltonian, an $A_{1g}$ quantity.

On the other hand, ten commutators $\Gamma_{jk}=[\Gamma_j,\Gamma_k]/(2i)$ with $j>k$ can be expressed in terms of the products of \emph{odd} number of spin-3/2 matrices as follows   
\allowdisplaybreaks[4]
\begin{eqnarray}
\Gamma_{45} &=& -\frac{2}{\sqrt{3}} \left[ J_x J_y J_z + J_z J_y J_x \right],
\Gamma_{12}  = \frac{1}{3} \left[ 7 J_z -4 J^3_z \right], \nonumber \\
\Gamma_{13} &=& -\frac{1}{3} \left[ 7 J_y -4 J^3_y \right],
\Gamma_{23}  = \frac{1}{3} \left[ 7 J_x -4 J^3_x \right], \nonumber \\ 
\Gamma_{34} &=& -\frac{1}{6} \left[ 13 J_z -4 J^3_z \right],
\Gamma_{35}  =\frac{1}{\sqrt{3}} \left\{ J_z, J^2_x - J^2_y \right\}, \nonumber \\
\Gamma_{14} &=& \frac{1}{12} \left[ 13 J_x -4 J^3_x \right] +\frac{1}{2} \left\{ J_x, J^2_y - J^2_z \right\}, \\
\Gamma_{15} &=& \frac{1}{4\sqrt{3}} \left[ 13 J_x -4 J^3_x \right] -\frac{1}{2\sqrt{3}} \left\{ J_x, J^2_y - J^2_z \right\}, \nonumber \\
\Gamma_{24} &=& \frac{1}{12} \left[ 13 J_y -4 J^3_y \right] -\frac{1}{2} \left\{ J_y, J^2_z - J^2_x \right\}, \nonumber \\
\Gamma_{25} &=& -\frac{1}{4\sqrt{3}} \left[ 13 J_y -4 J^3_y \right] -\frac{1}{2\sqrt{3}} \left\{ J_y, J^2_z - J^2_x \right\}. \nonumber 
\end{eqnarray}
The four-dimensional identity matrix can be written as 
\begin{equation}
\Gamma_0 = \frac{4}{15} \left( J^2_x + J^2_y +J^2_z \right).
\end{equation} 


\section{Fierz Identity}~\label{Append:Fierz}

In this appendix we present the Fierz reduction of the number of linearly independent quartic terms for the interacting Luttinger system. To perform this exercise for generic local density-density interactions we introduce a \emph{six} component vector 
\allowdisplaybreaks[4]
\begin{eqnarray}
X^\top &=& \bigg[ (\Psi^\dagger \Gamma_0 \Psi)^2, \sum^3_{j=1} (\Psi^\dagger \Gamma_j \Psi)^2, \sum^5_{j=4} (\Psi^\dag\Gamma_j \Psi)^2, \\ 
&& (\Psi^\dag\Gamma_{45}\Psi)^2, \sum^3_{j=1} (\Psi^\dag \Gamma_j \Gamma_{45} \Psi)^2,
\sum^{3}_{j=1} \sum^{5}_{k=4} (\Psi^\dagger \Gamma_{jk} \Psi)^2 \bigg], \nonumber 
\end{eqnarray}
constituted by the quartic terms appearing in $H_{\rm int}$, see Eq.~(\ref{Eq:Interaction_Hamiltonian}). The Fierz identity allows us to rewrite each quartic term appearing in $X$ as a linear combination of the remaining ones according to 
\begin{eqnarray}
\left[ \Psi^\dag(x) M \Psi(x)\right] \left[\Psi^\dag(y) N \Psi(y)\right]= \nonumber \\
-\frac{1}{16}\left( \mathrm{Tr} M\Gamma^a N \Gamma^b \right) \left[\Psi^\dag(x) \Gamma^b\Psi(y)\right] \left[ \Psi^\dag(y) \Gamma^a\Psi(x) \right],
\end{eqnarray}
where $M$ and $N$ are four-dimensional Hermitian matrices and for local interactions $x=y$. The set of sixteen matrices $\{ \Gamma_{a}, \: a=1, \cdots, 16 \}$ closes the $U(4)$ Clifford algebra of four-dimensional matrices, and we choose $\Gamma^\dagger_a=\Gamma_a=(\Gamma_a)^{-1}$. The above Fierz constraint then takes a compact form $F X=0$, where $F$ is the Fierz matrix given by 
\begin{equation}
F=
\begin{bmatrix}
5& 1& 1& 1& 1& 1 \\
3& 3& -3& 3& -1& 1 \\
2& -2& 4& -2& 2& 0 \\
1& 1& -1& 5& 1& -1 \\
3& -1& 3& 3& 3& -1 \\
6& 2& 0& -6& -2& 4 
\end{bmatrix}.
\end{equation}
The rank of $F$ is 3. Hence, the number of lienarly independent quartic terms is $3=6$ (dimensionality of $F$) -- $3$ (rank of $F$). We chose four-fermion interactions proportional to $\lambda_0$, $\lambda_1$ and $\lambda_3$ as \emph{three} linearly independent quartic terms. The remaining three quartic terms can then be expressed in terms of linear combinations of the above three according to 
\begin{eqnarray}~\label{EqAppend:FierzRelations}
(\Psi^\dag\Gamma_{45}\Psi)^2 &=& -\frac{1}{2} (\Psi^\dag  \Psi)^2 
-\frac{1}{2} \sum^{3}_{j=1} (\Psi^\dag\Gamma_j \Psi)^2 \nonumber \\
&+& \frac{1}{2} \sum^5_{j=4} (\Psi^\dag\Gamma_j \Psi)^2,  \nonumber \\
\sum^{3}_{j=1}(\Psi^\dag \Gamma_j \Gamma_{45}\Psi)^2 
&=& -\frac{3}{2} (\Psi^\dag \Psi)^2  + \frac{1}{2} \sum^{3}_{j=1} (\Psi^\dag\Gamma_j \Psi)^2 \nonumber \\
&-& \frac{3}{2} \sum^5_{j=4} (\Psi^\dag\Gamma_j \Psi)^2,  \nonumber \\
\sum^3_{j=1} \sum^{5}_{k=4} (\Psi^\dag\Gamma_{jk}\Psi)^2
&=& -3 (\Psi^\dag \Psi)^2 -\sum^{3}_{j=1} (\Psi^\dag\Gamma_j \Psi)^2.
\end{eqnarray}
Therefore, during the RG analysis whenever we generate any one of the above three quartic terms we can immediately express them in terms of the ones proportional to $\lambda_0$, $\lambda_1$ and $\lambda_2$. Therefore, the interacting model [see Eq.~(\ref{Eq:Action_Int})] remains closed under coarse garining to any order in the perturbation theory. 

\section{Details of RG flow equations}~\label{Append:RGdetails}

The RG flow equations, displayed in Eq.~(\ref{Eq:RGgeneral}), are expressed in terms of the functions $H^{j}_{km} (\alpha, t, \mu)$, where $j, k, m=0,1,2$. For brevity we here drop the explicit dependence of these functions on $\alpha, t$ and $\mu$. The functions appearing in the $\beta$ function of $g_{_0}$ are given by 
\begin{widetext}
\allowdisplaybreaks[4]
\begin{eqnarray}
H_{00}^0 &=& \frac{5}{4} \left[ -f_0-\tilde{f}_0-4 f_g-6 f_t \right], \:
H_{11}^0  = \frac{5}{4} \left[ -6 \tilde{f}_0 - 6 f_g + 6 \tilde{f}_g -3 f_t + 3 \tilde{f}_t \right], \nonumber \\
H_{22}^0 &=& \frac{5}{4} \left[ f_0 - 3 \tilde{f}_0 -3 f_t +3 \tilde{f}_t \right], \: 
H_{01}^0 = \frac{5}{4} \left[ 6 f_0 + 6 f_g + 6 \tilde{f}_g + 18 f_t + 12 \tilde{f}_t \right], \nonumber \\  
H_{02}^0 &=& \frac{5}{2} \left[ 2 f_0 + 5 f_g + 3 f_t +3 \tilde{f}_g + 3 \tilde{f}_t \right], \: 
H_{12}^0  = \frac{15}{2} \left[ -f_0 - \tilde{f}_0 - f_g + \tilde{f}_g - 2 f_t + 2 \tilde{f}_t \right]. 
\end{eqnarray}
The six functions appearing in the RG flow equation for $g_{_1}$ are given by
\allowdisplaybreaks[4]
\begin{eqnarray}
H_{00}^1 &=& \frac{5}{4} (f_t + \tilde{f}_t), \:
H_{11}^1 = \frac{5}{4} (-5 f_0 + \tilde{f}_0 + 10 f_g + 2 \tilde{f}_g + 10 f_t + 10 \tilde{f}_t), \:
H_{22}^1 = \frac{5}{4} ( -f_0 - \tilde{f}_0 + 3 f_t +\tilde{f}_t ),  \nonumber \\
H_{01}^1 &=& \frac{5}{2}( 2 f_0 - \tilde{f}_0 - 3 f_g + \tilde{f}_g - \tilde{f}_t ), \:
H_{02}^1 = -\frac{5}{2} ( f_g - \tilde{f}_g + f_t - \tilde{f}_t ), \: 
H_{12}^1 = -\frac{5}{2} ( 3 f_0 + \tilde{f}_0 - 7 f_g - \tilde{f}_g - 2 \tilde{f}_t ).
\end{eqnarray}
Finally, the flow equation for $g_{_2}$ is expressed in terms of the following six functions
\allowdisplaybreaks[4]
\begin{eqnarray}
H_{00}^2 &=& \frac{5}{4} (f_g + \tilde{f}_g ), \: 
H_{11}^2  = \frac{5}{4} ( -3 f_0 - 3 \tilde{f}_0 + 3 f_g + 3 \tilde{f}_g + 3 f_t - 3 \tilde{f}_t ), \:
H_{22}^2  = \frac{5}{4} ( -3 f_0 + \tilde{f}_0 + 4 f_g + 4 \tilde{f}_g + 9 f_t + 3 \tilde{f}_t ), \nonumber \\
H_{01}^2 &=& \frac{5}{4} (-6 f_t + 6 \tilde{f}_t ), \:
H_{02}^2  = \frac{5}{2} (2 f_0 - \tilde{f}_0 + f_g - \tilde{f}_g - 3 f_t ), \:
H_{12}^2  = \frac{15}{2} (-f_0 - f_g +\tilde{f}_g + 4 f_t + \tilde{f}_t ). 
\end{eqnarray} 
Note that in the above expression $f_j \equiv f_j(\alpha, t, \mu)$ and $\tilde{f}_j=\tilde{f}_j(\alpha, t, \mu)$s for $j=0,t,g$, and 
\allowdisplaybreaks[4]
\begin{eqnarray}
f_0 (\alpha, t, \mu) &=& -\frac{1}{2}\int \mathrm{d}\Omega \: \frac{1}{2t} \sum_{\tau=\pm 1}\left[ \mathrm{sech}^2\left(\frac{f+\tau\mu}{2t}\right) + \frac{2t}{f}\tanh\left(\frac{f+\tau\mu}{2t}\right)\right], \\
f_t (\alpha, t, \mu) &=& -\frac{1}{2}\int \mathrm{d}\Omega \: \frac{\cos^2\alpha }{2t} \sum_{\tau=\pm 1}\left[ \frac{1}{f^2}\mathrm{sech}^2\left(\frac{f+\tau\mu}{2t}\right) - \frac{2t}{f^3} \tanh\left(\frac{f+\tau\mu}{2t}\right)\right]\frac{\hat{d}_1^2+\hat{d}_2^2+\hat{d}_3^2}{3},  \\
f_g (\alpha, t, \mu) &=& -\frac{1}{2}\int \mathrm{d}\Omega \: \frac{\sin^2\alpha}{2t} \sum_{\tau=\pm 1}\left[ \frac{1}{f^2}\mathrm{sech}^2\left(\frac{f+\tau\mu}{2t}\right) - \frac{2t}{f^3} \tanh\left(\frac{f+\tau\mu}{2t}\right)\right]\frac{\hat{d}_4^2+\hat{d}_5^2}{2},  \\
\tilde{f}_0 (\alpha, t, \mu) &=& -\frac{1}{2}\int \mathrm{d}\Omega \: \frac{1}{\mu \left(1-(\frac{\mu}{f})^2\right)}  \sum_{\tau=\pm 1} \tanh\left(\frac{f+\tau\mu}{2t} \right)\left[\tau+\frac{\mu}{f}-2\tau \left(\frac{\mu}{f}\right)^2\right],  \\ 
\tilde{f}_t (\alpha, t, \mu) &=& \frac{1}{2}\int \mathrm{d}\Omega \: \frac{\cos ^2 \alpha }{f^2 \mu \left( 1-(\frac{\mu}{f})^2 \right)} \sum_{\tau=\pm 1}\tanh\left( \frac{f+\tau\mu}{2t} \right) \left[ -\tau+\frac{\mu}{f} \right]\frac{\hat{d}_1^2+\hat{d}_2^2+\hat{d}_3^2}{3}, \\
\tilde{f}_g (\alpha, t, \mu) &=& \frac{1}{2}\int \mathrm{d}\Omega \: \frac{\sin^2 \alpha }{f^2 \mu \left( 1-(\frac{\mu}{f})^2 \right)} \sum_{\tau=\pm 1}\tanh\left( \frac{f+\tau\mu}{2t} \right) \left[ -\tau+\frac{\mu}{f} \right]\frac{\hat{d}_4^2+\hat{d}_5^2}{2},
\end{eqnarray}     
\end{widetext}
where $d\Omega=\sin \theta d\theta d\phi$, $f \equiv f (\alpha, \theta, \phi)$ and 
\begin{equation}
f(\alpha,\theta,\phi)= \sqrt{\cos^2\alpha (\hat{d}_1^2+\hat{d}_2^2+\hat{d}_3^2) + \sin^2\alpha(\hat{d}_4^2+\hat{d}_5^2)}.
\end{equation}
The definition of $\hat{\bf d}$ is already provided in Eq.~(\ref{d-hats}).

In terms of $f_j$s and $\tilde{f}_j$s we can express the leading order RG flow equations for all source terms [see Eq.~(\ref{Eq:Action_SourceTerm})] as
\allowdisplaybreaks[4]
\begin{eqnarray}
\frac{d \ln \Delta_0}{dl}-2 &=& -\frac{5}{4} ( f_0 + 2 f_g + 3 f_t )( 3 g_0 - 3 g_1 -2 g_2 ), \nonumber  \\
\frac{d \ln \Delta_1}{dl}-2 &=& \frac{5}{4} ( f_0 - 2 f_g -f_t ) (g_0 - 5 g_1 - 2 g_2 ), \nonumber  \\
\frac{d \ln \Delta_2}{dl}-2 &=& \frac{5}{4} (f_0 - 3f_t) (g_0 - 3 g_1 - 4 g_2 ), \nonumber  \\
\frac{d \ln \Delta_3}{dl}-2 &=& \frac{5}{4} (f_0 - 2 f_g + 3 f_t ) (g_0 + 3 g_1 - 2 g_2 ),\nonumber  \\
\frac{d \ln \Delta_4}{dl}-2 &=& \frac{5}{4} (f_0 + 2 f_g - f_t ) (g_0 - g_1 + 2 g_2 ),\nonumber  \\
\frac{d \ln \Delta_5}{dl}-2 &=& \frac{5}{4} (f_0 + f_t )( g_0 + g_1 ), \nonumber  \\
\frac{d \ln \Delta^{\rm p}_{A_{1g}}}{dl}-2 &=& -\frac{5}{4} (\tilde{f}_0 - 2 \tilde{f}_g - 3 \tilde{f}_t ) (g_0 + 3 g_1 + 2 g_2 ), \nonumber  \\
\frac{d \ln \Delta^{\rm p}_{T_{2g}}}{dl}-2 &=& -\frac{5}{4} (\tilde{f}_0 + 2 \tilde{f}_g + \tilde{f}_t) (g_0 - g_1 -2 g_2), \nonumber  \\
\frac{d \ln \Delta^{\rm p}_{E_g}}{dl}-2 &=& -\frac{5}{4} (\tilde{f}_0 + 3 \tilde{f}_t )(g_0 - 3 g_1). 
\end{eqnarray}
These flow equations are schematically shown in Eq.~(\ref{Eq:sourceRG}).

\section{Stability matrix analysis}~\label{Append:Stabilitymatrix}

   To gain insight into the local structure of the RG flow trajectories in the close proximity to any fixed point, we perform the \emph{stability} analysis around them. To proceed with this analysis we introduce a $3 \times 3$ stability matrix $M \left(g_{_0},g_{_1}, g_{_2} \right)$ as follows
	\begin{equation}~\label{Eq:StabilityMatrix}
	M \left( g_{_0}, g_{_1}, g_{_2} \right) = \left[ \begin{array}{ccc}
	\frac{d\beta_{g_{_0}}}{dg_{_0}} & \frac{d\beta_{g_{_1}}}{dg_{_0}} & \frac{d\beta_{g_{_2}}}{dg_{_0}} \\ 
	\frac{d\beta_{g_{_0}}}{dg_{_1}} & \frac{d\beta_{g_{_1}}}{dg_{_1}} & \frac{d\beta_{g_{_2}}}{dg_{_1}} \\
	\frac{d\beta_{g_{_0}}}{dg_{_2}} & \frac{d\beta_{g_{_1}}}{dg_{_2}} & \frac{d\beta_{g_{_2}}}{dg_{_2}}
	\end{array}
	\right].
	\end{equation}
We classify the fixed points, located at $(g^\ast_{_0}, g^\ast_{_1}, g^\ast_{_2})$, according to the number of positive and negative eignevalues of $M \left(g^\ast_{_0}, g^\ast_{_1}, g^\ast_{_2} \right)$.

1. There exists only one fixed point at $\left(g^\ast_{_0}, g^\ast_{_1}, g^\ast_{_2} \right)=(0,0,0)$ with three negative eigenvalues $(-1,-1,-1) \epsilon$ of the stability matrix. This fixed point, referred as \emph{Gaussian} in Table~\ref{Tab:StabilityMatrix}, is stable from all directions and represents the noninteracting LSM for sufficiently weak but generic short-range interactions.

2. Fixed points possessing one positive and two negative eigenvalues for the stability matrix are the QCPs. Such fixed points control QPTs from LSM to BSPs. Since such QPTs can be triggered by tuning only one parameter (the interaction strength along the relevant direction, determined by the eigenvector associated with the positive eigenvalue of the stability matrix), they are continuous in nature. Concomitantly, a single parameter scaling emerges in the vicinity of all QCPs. The positive eigenvalue of the stability matrix also determines the correlation length exponent ($\nu$) at the QCPs [see Sec.~\ref{SubSec:exponents}].

3. Fixed points with two positive and one negative eigenvalues for the stability matrix are the BCPs. Typically BCPs are found when there exists more than one QCP in the multi-dimensional coupling constant space. For example, only around $\alpha=\frac{\pi}{2}$ and $0$, when the coupled RG flow equations support two QCPs, there exists one BCP (respectively denoted by BCP$_\frac{\pi}{2}$ and BCP$_0$). Existence of such a BCP is necessary to ensure the continuity of the RG flow trajectories. A BCP also separates the basin of attractions of two QCPs. 

A summary of the eigenvalues for the stability matrix is presented in Table~\ref{Tab:StabilityMatrix}. We did not find any fixed point with three positive eigenvalues of the corresponding stability matrix.



\end{document}